\documentclass[11pt]{article} 

\usepackage{natbib}
\usepackage{graphicx}
\usepackage{amsmath,wrapfig,amssymb,multirow}
\usepackage[margin=1in]{geometry}

\usepackage{chngcntr}
\counterwithout{figure}{section}

\usepackage[parfill]{parskip}

\usepackage{url}
\usepackage{authblk}

\newcommand{\mbf}{\mathbf}

\usepackage{easymat}
\usepackage{bigstrut}
\usepackage{threeparttable} 
\usepackage[format=hang,labelfont=bf]{caption}
\usepackage{appendix}
\usepackage{rotating}
\usepackage{tensor}

\DeclareMathOperator{\Tr}{Tr}

\begin{document}
\pagenumbering{roman}

\title{\vspace*{\fill}{\bf {\sc A Singular Value Decomposition-based Factorization and Parsimonious Component Model of Demographic Quantities Correlated by Age\\[10pt] \large{Predicting Complete Demographic Age Schedules with Few Parameters}}}
\vspace{2cm}}

\author[1,2,3,4,*]{Samuel J. Clark}

\affil[1]{Department of Sociology, University of Washington}
\affil[2]{MRC/Wits Rural Public Health and Health Transitions Research Unit (Agincourt), \authorcr School of Public Health, Faculty of Health Sciences, University of the Witwatersrand}
\affil[3]{ALPHA Network, London School of Hygience and Tropical Medicine, London, UK}
\affil[4]{INDEPTH Network, Accra, Ghana}
\affil[*]{Correspondence to: \texttt{work@samclark.net}}


\date{May 8, 2014 \vspace*{\fill}}

\maketitle

\newpage

\centerline{{\sc Abstract}}

\noindent {\scshape {\bfseries Background.}} Formal demography has a long history of building simple models of age schedules of demographic quantities, e.g. mortality and fertility rates.  These are widely used in demographic methods to manipulate whole age schedules using few parameters.

\noindent {\scshape {\bfseries Objective.}} The Singular Value Decomposition (SVD) factorizes a matrix into three matrices with useful properties including the ability to reconstruct the original matrix using many fewer, simple matrices.  This work demonstrates how these properties can be exploited to build parsimonious models of whole age schedules of demographic quantities that can be further parameterized in terms of arbitrary covariates.

\noindent {\scshape {\bfseries Methods.}} The SVD is presented and explained in detail with attention to developing an intuitive understanding.  The SVD is used to construct a general, component model of demographic age schedules, and that model is demonstrated with age-specific mortality and fertility rates.  Finally, the model is used (1) to predict age-specific mortality using HIV indicators and summary measures of age-specific mortality, and (2) to predict age-specific fertility using the total fertility rate (TFR).

\noindent {\scshape {\bfseries Results.}} The component model of age-specific mortality and fertility rates succeeds in reproducing the data with two inputs, and acting through those two inputs, various covariates are able to accurately predict full age schedules.

\noindent {\scshape {\bfseries Conclusions.}} The SVD is a potentially useful as a way to summarize, smooth and model age-specific demographic quantities.  The component model is a general method of relating covariates to whole age schedules.

\noindent {\scshape {\bfseries Comments.}} The focus of this work is the SVD and the component model.  The applications are for illustrative purposes only.

\medskip 


\vfill

\centerline{{\sc Acknowledgments}}
This work was supported by grants K01 HD057246 and R01 HD054511 and from the Eunice Kennedy Shriver National Institute of Child Health and Human Development (NICHD). The funders had no role in study design, data collection and analysis, decision to publish, or preparation of the manuscript. The work described below was conducted in the {\texttt R} programming language and run in the {\texttt R} statistical software package \citep{R-url}.  David Sharrow and Adrian Raftery have been colleagues working on related topics for some time, and their advice and feedback were important to writing this paper.  Jon Wakefield and Tyler McCormick read drafts of this article and provided helpful comments.

\normalsize

\newpage

\tableofcontents

\newpage

\pagenumbering{arabic}

\section{Introduction}

A long-standing pursuit of formal demography is the construction of empirical models of quantities that vary predictably by age \citep{coale1996development}.  Model life tables \citep[e.g.][]{coale1966,un1982,murray2003,wang2013} and various mortality models \cite[e.g.][]{wilmoth2012,heligman1980} aim to identify and parsimoniously express the regularity of mortality with age, and some fertility models \citep[e.g.][]{coale1974model} do the same.  Because age is a strong predictor of these quantities, age profiles of these quantities from different populations, times and places are correlated.  This fact allows us to take advantage of the singular value decomposition (SVD), a classic and long-standing result in mathematics \citep{stewart1993early}, to construct a general, parsimonious, weighted sum-based model of the age pattern of demographic quantities that can incorporate covariates and be used to make predictions.

The SVD has been used before in demography. Building on earlier work by \cite{wilmoth1989quand}, \cite{lee1992modeling} use the SVD to generate a rank-1 approximation of the residual produced by subtracting the mean from a matrix of log mortality rates, effectively yielding a least-squares solution for an under-determined part of their model \citep[][ and Section \ref{sec:sumRank1Mats}, Equation \ref{eq:XsumRank-1}]{wilmoth1993computational,good1969some}.  This is similar in spirit to what we develop below but solves a different, specific problem and does not identify or develop the  generalizable features of the SVD of age-correlated quantities or in a general sense exploit the properties of the reduced-rank form of the SVD.  Later \cite{wilmoth2012} again use the SVD in a similar way to characterize the age-pattern of residuals in the their Log-Quad model.  In earlier work \citep{clark2009,clarkINDEPTHMx,clarkPhD,Sharrow2014} we have developed and used a precursor of the weighted sum, component model that we fully develop below, and in the most recent iteration we use the SVD to construct the components - without fully exploiting its capabilities.   Finally, \cite{fosdick2012separable} develop a separable covariance model and demonstrate it using age-specific mortality.  This work focuses on the covariance model of the mean-subtracted mortality rates and like the others does not identify or develop the general implications of the generic SVD for age-correlated quantities.

The purpose of this work is to discuss the SVD in detail and demonstrate how it can be used to develop a general, parsimonious model of demographic quantities correlated by age.  The intended audience is demographers who might want to use the model.  With that in mind the presentation is intuitive with an emphasis on geometric interpretations rather than mathematically rigorous, and it is supported by several fully worked examples.  There are many mathematical presentations of the SVD elsewhere, e.g. \cite{good1969some, kalman1996singularly, strang2009introduction}.

The remainder of the article begins with a detailed presentation of what the SVD is and how it is related to principal components analysis (PCA).  That is followed by a re-expression of the SVD of a data matrix $\mbf{X}$ in a form that expresses each column vector of $\mbf{X}$ as a weighted sum of increasingly less consequential terms. That result is used to develop a general, parsimonious model of demographic age profiles, and finally, that model is explored and demonstrated thoroughly using mortality and fertility data from the Agincourt health and demographic surveillance system (HDSS) in South Africa \citep{kahn2012agin}.

\section{The Singular Value Decomposition}

This section presents the SVD drawing on Strang's presentation \citep{strang2009introduction} and can be skipped by those who do not want to know the origin of the SVD or how it relates to PCA.

\subsection{Sketch of a Linear Algebra Derivation of the SVD}

Imagine an arbitrary $m \times n$ linear transformation (or data matrix) $\mbf{X}$.  The row-space of $\mbf{X}$ contains vectors with $n$ elements in $\text{R}^n$, and the column-space vectors with $m$ elements in $\text{R}^m$.  The SVD results from the identification of an orthonormal basis $\mbf{V}$ in the row-space of $\mbf{X}$ that when transformed by $\mbf{X}$ yields another orthonormal basis $\mbf{U}$ in the column-space of $\mbf{X}$, potentially stretched by a diagonal matrix with positive entries $\mbf{S}$,%
\begin{align}
\mbf{XV} &= \mbf{U S} \ , \\
\mbf{X} &= \mbf{U S V}^{-1} \ , \nonumber
\end{align}
and because $\mbf{V}$ is orthogonal,
\begin{align}
\mbf{X} &= \mbf{U S V}^{\text{T}} \ . \label{eq:XdecompPrimative}
\end{align}%
Equation \ref{eq:XdecompPrimative} is the SVD of $\mbf{X}$. The column vectors of $\mbf{U}$ are the `left singular vectors', the column vectors of $\mbf{V}$ are the singular vectors, and $\mbf{S}$ is a diagonal matrix containing the `singular values'.%

\subsection{The Singular Value Decomposition and Principal Component Analysis}

The natural question now is how to identify $\mbf{U}$, $\mbf{V}$, and $\mbf{S}$.  Beginning with $\mbf{U}$, we examine the SVD of $\mbf{X}^{\text{T}}\mbf{X}$. $\mbf{X}^{\text{T}}\mbf{X}$ is special because it is positive and semi-definite, all of which means it has a well-behaved eigen decomposition with real, pairwise orthogonal eigenvectors (when their eigenvalues are different) and positive or null eigenvalues.  Using Equation \ref{eq:XdecompPrimative} and the fact that $\mbf{U}^{\text{T}}\mbf{U}$ is the identity matrix (because $\mbf{U}$ is orthogonal),%
\begin{align}
\mbf{X}^{\text{T}}\mbf{X} &= (\mbf{V S U}^{\text{T}})(\mbf{U S V}^{\text{T}}) \ , \nonumber \\
&= \mbf{V S}^2\mbf{V}^{\text{T}} \ . \label{eq:xtxEigen}
\end{align}%
The right-hand side of Equation \ref{eq:xtxEigen} is the eigen decomposition of $\mbf{X}^{\text{T}}\mbf{X}$ which means that $\mbf{V}$ contains the eigenvectors of $\mbf{X}^{\text{T}}\mbf{X}$ and the elements of $\mbf{S}^2$ are the eigenvalues of $\mbf{X}^{\text{T}}\mbf{X}$. We can identify $\mbf{U}$ in a similar way by examining the SVD of $\mbf{X}\mbf{X}^{\text{T}}$, which is again positive and semi-definite,%
\begin{align}
\mbf{X}\mbf{X}^{\text{T}} &= (\mbf{U S V}^{\text{T}})(\mbf{V S U}^{\text{T}}) \ , \nonumber \\
&= \mbf{U S}^2\mbf{U}^{\text{T}} \ . \label{eq:xxtEigen}
\end{align}%
Equation \ref{eq:xxtEigen} says that $\mbf{U}$ contains the eigenvectors of $\mbf{X}\mbf{X}^{\text{T}}$, and again the elements of $\mbf{S}^2$ are the eigenvalues of $\mbf{X}\mbf{X}^{\text{T}}$, the same as the eigenvalues of  $\mbf{X}^{\text{T}}\mbf{X}$.  Together Equations \ref{eq:xtxEigen} and \ref{eq:xxtEigen} give us a way to identify all the components of the SVD\footnote{In practice this is not how the SVD is calculated; it is generally numerically easier and less uncertain to calculate the SVD directly \citep{kalman1996singularly} and use the SVD to calculate eigen decompositions.}.  

Those same equations explain how the SVD is related PCA.  PCA is typically conducted by taking the eigen decomposition of either the covariance or correlation matrix.  In both cases the first eigenvector lines up with the axis along which there is most variation in the cloud of data points, and subsequent eigenvectors point in orthogonal directions and capture in decreasing quantities the remaining variation in the data cloud, see \cite{abdi2010principal}.  

Both the SVD and PCA identify the same dominant dimensions in preprocessed versions of the data matrix $\mbf{X}$ \citep{abdi2010principal,wall2003singular}. Both the covariance and correlation matrices can be calculated as $\mbf{X}_{*}^{\text{T}}\mbf{X}_{*}$ where $\mbf{X}_{*}$ is a preprocessed version of $\mbf{X}$. The covariance matrix is $\mbf{X}_{\text{ms}}^{\text{T}}\mbf{X}_{\text{ms}}$, a centered and scaled version of $\mbf{X}^{\text{T}}\mbf{X}$, computed from  $\mbf{X}_{\text{ms}}$ (mean subtracted) which is created by subtracting the column mean from each column of $\mbf{X}$ and multiplying each element of the result by either $\tfrac{1}{\sqrt{N}}$ or $\tfrac{1}{\sqrt{N-1}}$. The correlation matrix is formed by additional processing of the variables. In addition to centering and scaling by $\tfrac{1}{\sqrt{N-1}}$, each variable is normalized (to account for different scales) by dividing by its norm, the square root of the sum of its squared elements, to form $\mbf{X}_{\text{msn}}$ (mean subtracted, normalized).  The correlation matrix is then $\mbf{X}_{\text{msn}}^{\text{T}}\mbf{X}_{\text{msn}}$. The eigenvectors of the covariance matrix are the right singular vectors $\mbf{V}$ of the SVD of $\mbf{X}_{\text{ms}}$, and the eigenvectors of the correlation matrix are the right singular vectors of the SVD of $\mbf{X}_{\text{msn}}$.  In both cases the singular values are the square roots of the eigenvalues.

Both the covariance and correlation matrices are effectively centered and scaled versions of the original data, and hence both the SVD of $\mbf{X}_{*}$ and the eigen decomposition of $\mbf{X}_{*}^{\text{T}}\mbf{X}_{*}$ yield primary dimensions of the centered data cloud, and because the cloud is centered, these primary dimensions will line up with the orthogonal dimensions of greatest variation in the cloud.  As we saw above in Equation \ref{eq:xtxEigen} the SVD of $\mbf{X}$ and the eigen decomposition of $\mbf{X}^{\text{T}}\mbf{X}$ identify the same primary dimensions, but these will not necessarily line up with the dimensions of greatest variation in the cloud of data points because the first primary dimension simply points from the origin to the data cloud.  Consequently, the first primary dimension will correspond to the dimension of maximum variation in the cloud only if that dimension happens to lie on the line from the origin to the center of the cloud.  The characteristics of the SVD and PCA are displayed graphically in Figure \ref{fig:SVDPCAdims}.

\newsavebox{\smlmat}
\savebox{\smlmat}{$\left[\begin{smallmatrix} 200 & 100 \\ 100 & 75 \end{smallmatrix}\right]$}
\begin{figure}[htbp]
\begin{center}
\begin{tabular}{cc}
A & B \\
\includegraphics[width=0.48\textwidth]{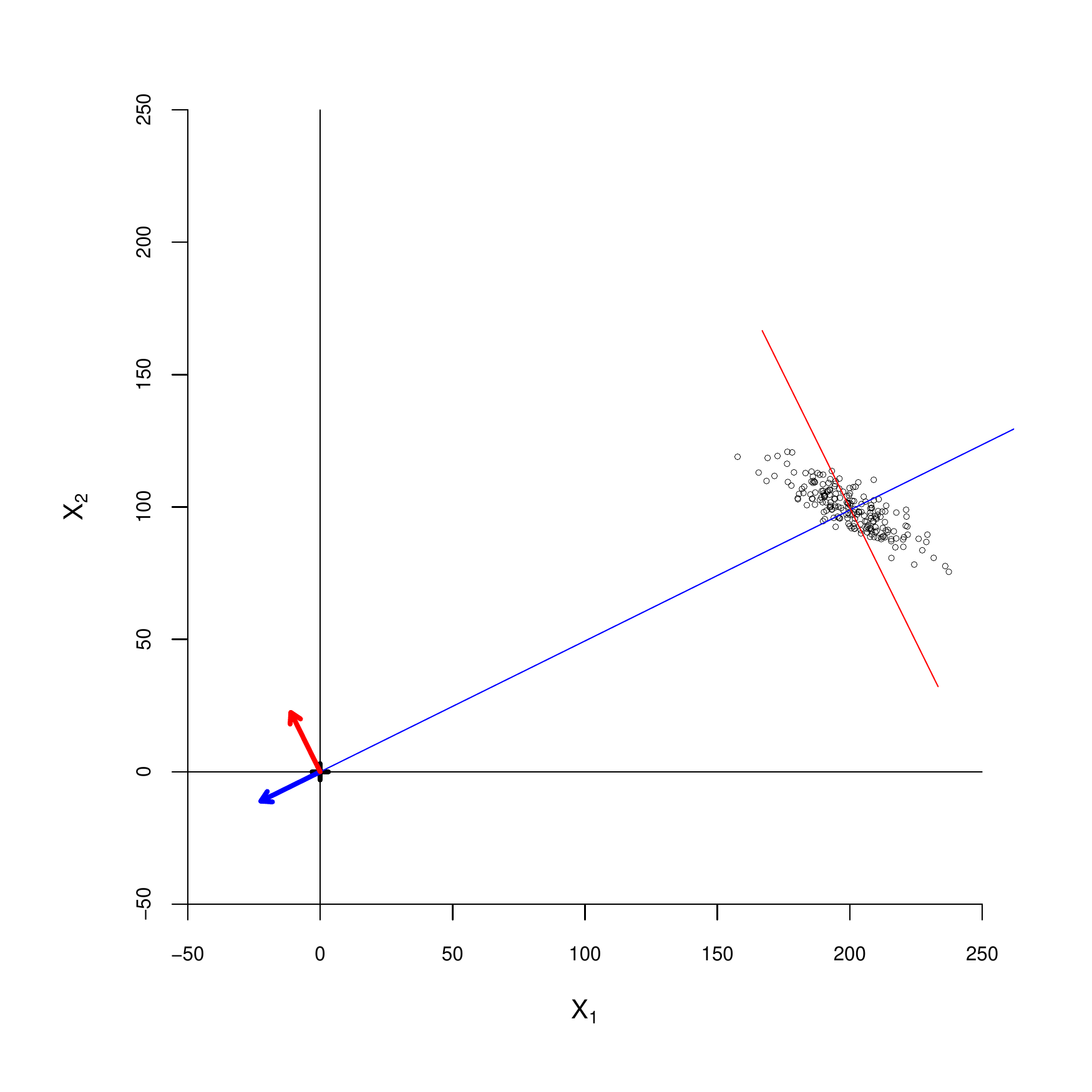} &
\includegraphics[width=0.48\textwidth]{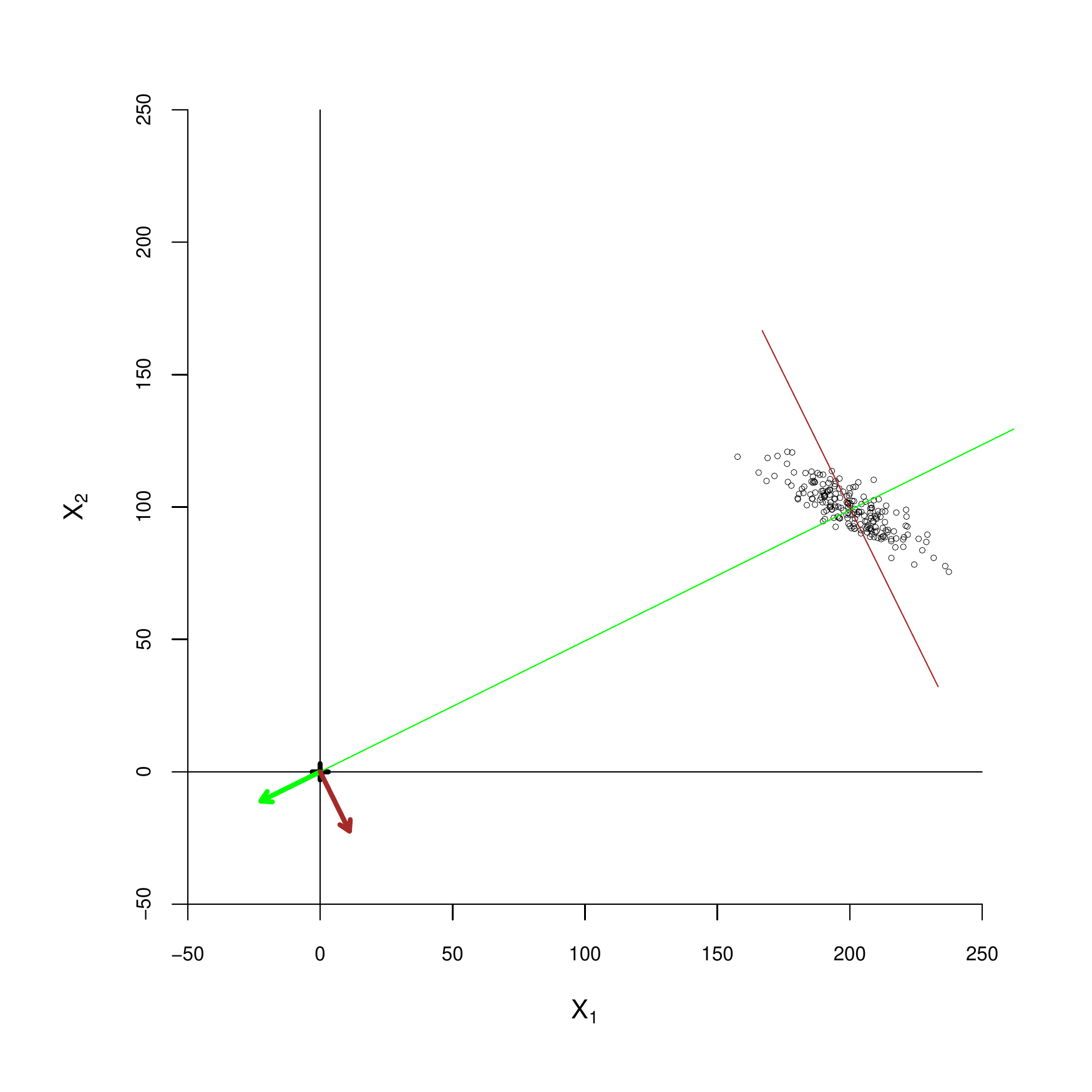} \\
C & D \\
\includegraphics[width=0.48\textwidth]{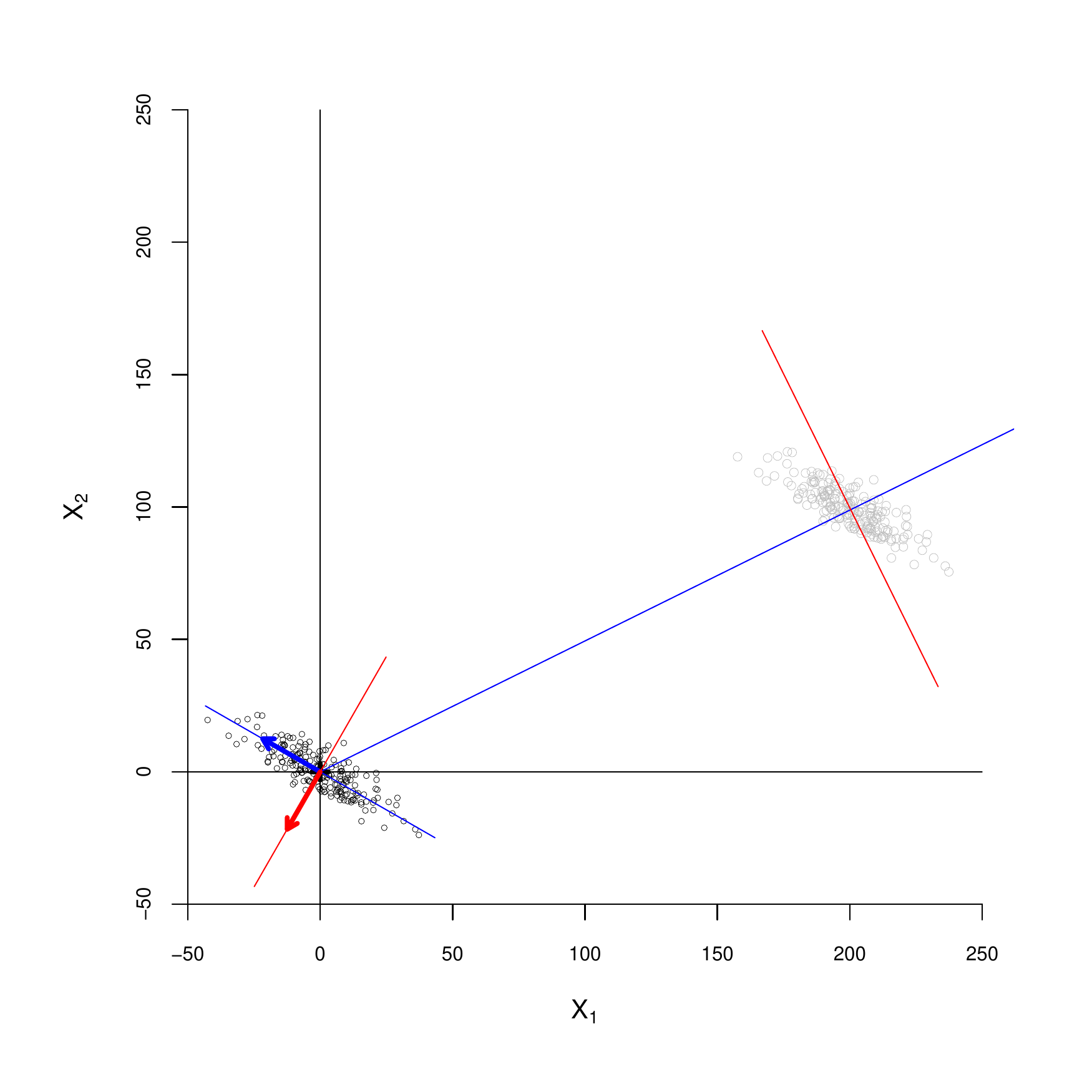} &
\includegraphics[width=0.48\textwidth]{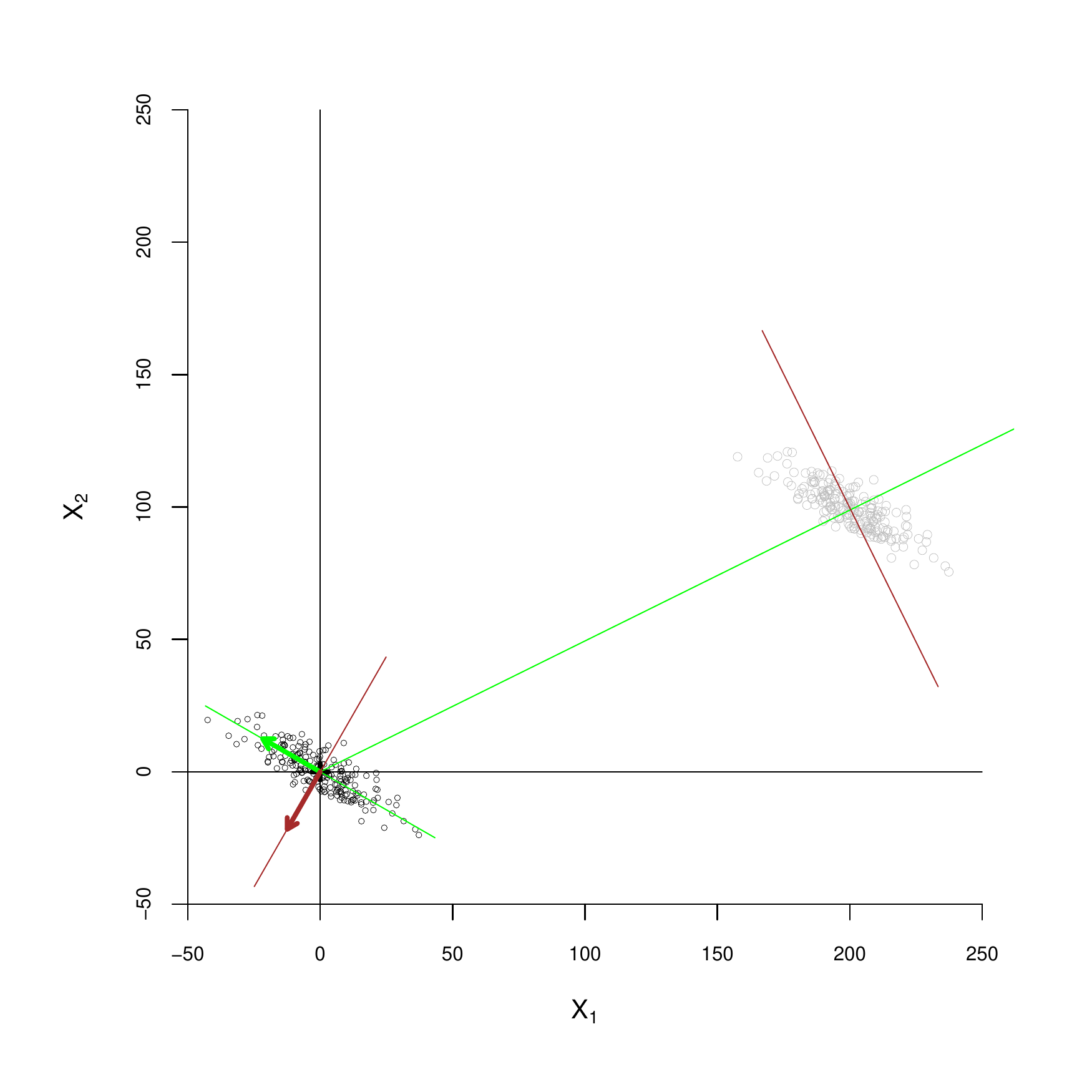} 
\end{tabular}
\captionsetup{format=plain,format=plain,font=small,margin=0.5cm,justification=justified}
\caption{{\bf General Geometry of SVD and PCA.} The data $\mathbf{X}$ are the cloud of points: 200 points distributed in bivariate Normal $\mathcal{N}(\mu,\Sigma), \mu=(200,100), \Sigma =$~\usebox{\smlmat}. \textsc{Panel} \textbf{(A)}: New dimensions identified by SVD. Blue vector is first right singular vector $\mathbf{v}_{1}$. Red vector is second right singular vector $\mathbf{v}_{2}$. \textsc{Panel} \textbf{(B)}: New dimensions identified by eigen decomposition of $\mathbf{X^\text{T}X}$. Green vector is first eigenvector, and brown vector is second eigenvector. \textsc{Panel} \textbf{(C)}: Same as Panel (A) adding centered cloud and new dimensions identified by SVD of the centered cloud. \textsc{Panel} \textbf{(D)}: Same as Panel (B) adding centered cloud and new dimensions identified by Eigen decomposition of the centered cloud. Notice (1) that the SVD of $\mathbf{X}$ and the eigen decomposition of $\mathbf{X^\text{T}X}$ produce \textit{exactly} the same new dimensions (net of sign), (2) that the new dimensions of the `raw' cloud \textbf{do not} line up with the primary dimensions of the raw cloud, and (3) that the new dimensions of the centered cloud \textbf{do} line up with the primary dimensions of the centered cloud.}
\label{fig:SVDPCAdims}
\end{center}
\end{figure}

\section{Matrix Approximation and a useful Geometric Interpretation of the SVD}

For our purposes there is another equivalent and more useful way of understanding the SVD as a sum of rank-1 matrices that can be rearranged to express each column of $\mbf{X}$ as a weighted sum of the column vectors of $\mbf{U}$.  Below we derive this re-expression and then discuss why it is useful.

\subsection{The SVD as a Sum of Rank-1 Matrices}\label{sec:sumRank1Mats}

Let the data matrix $\mbf{X}$ be an arbitrary $K \times L$ matrix or real values with $\text{rank}(\mathbf{X}) = \rho$; $\rho \le \text{min}(K,L)$. The rank of a matrix is the number of independent rows and columns that it has; the number of columns and rows that cannot be expressed as multiplies of others.  Intuitively this is the number of dimensions defined or actually occupied by the column and row vectors in the matrix.  Now, let the factors of the SVD be:%
\begin{itemize}
\item $\mbf{U}$, a $K \times \rho$ matrix whose column vectors are the left singular vectors.%
\item $\mbf{S}$, a $\rho \times \rho$ diagonal (square) matrix containing the singular values.%
\item $\mbf{V}$, a $L \times \rho$ matrix whose column vectors are the right singular vectors.
\end{itemize}%
Then the SVD of $\mathbf{X}$ is:%
\begin{align}
\mbf{X}
&=
\mbf{USV}^\text{T} 
\\
\left[ 
	\begin{matrix}
	| & & | \\
	\mbf{x}_{1} & \ldots  & \mbf{x}_{L}  \\
	| & & | \\
	\end{matrix}
\right] 
&=
\left[ 
	\begin{matrix}
	| & & | \\
	\mbf{u}_{1} & \ldots  & \mbf{u}_{\rho}  \\
	| & & |  \\
	\end{matrix}
\right]  
\left[ 
	\begin{matrix}
		s_{1} & \ldots & 0  \\
		\vdots & \ddots & \vdots \\
		0 & \ldots & s_{\rho}  \\
	\end{matrix}
\right]
\left[ 
	\begin{matrix}
		\text{---} & \mbf{v}_1 & \text{---} \\
		& \vdots  & \\
		\text{---} & \mbf{v}_\rho & \text{---} \\
	\end{matrix}
\right] 	 
\nonumber \\
&=
\left[ 
	\begin{matrix}
		| & & | \\
	\mbf{u}_{1} & \ldots  & \mbf{u}_{\rho}  \\
		| & & | \\
	\end{matrix}
\right]  
\left[ 
	\begin{matrix}
		\text{---} & s_{1} \mbf{v}_1 & \text{---} \\
		& \vdots & \\
		\text{---} & s_{\rho} \mbf{v}_\rho & \text{---} \\
	\end{matrix}
\right] 	 
\nonumber \\
&=
\left[ 
	\begin{matrix}
		\sum_{i=1}^{\rho} u_{1i} s_{i} v_{1i} & \ldots & \sum_{i=1}^{\rho} u_{1i} s_{i} v_{Li} \\
		\vdots & \ddots  & \vdots \\		
		\sum_{i=1}^{\rho} u_{Ki} s_{i} v_{1i} & \ldots & \sum_{i=1}^{\rho} u_{Ki} s_{i} v_{Li} \\
	\end{matrix}
\right]  
\nonumber \\
&=
\left[ 
	\begin{matrix}
		| & & | \\
		\sum_{i=1}^{\rho} s_{i} v_{1i} \mbf{u}_{i}  & \ldots & \sum_{i=1}^{\rho} s_{i} v_{Li} \mbf{u}_{i}  \\
		| & & | \\
	\end{matrix}
\right] 
\label{eq:genComponents} \\ 
&=
\sum_{i=1}^{\rho} \left[ 
	\begin{matrix}
		| & & | \\
		s_{i} v_{1i} \mbf{u}_{i}  & \ldots  & s_{i} v_{Li} \mbf{u}_{i}  \\
		| & & | \\
	\end{matrix}
\right] 
\nonumber \\
&=
\sum_{i=1}^{\rho} \left[ 
	\begin{matrix}
		s_{i} v_{1i} u_{1i} & \ldots & s_{i} v_{Li} u_{1i} \\
		\vdots & \ddots & \vdots \\
		s_{i} v_{1i} u_{Ki}  & \ldots  & s_{i} v_{Li} u_{Ki} \\
	\end{matrix}
\right] 
\nonumber \\
&=
\sum_{i=1}^{\rho}
s_{i} 
\left[ 
	\begin{matrix}
		u_{1i} \\
		\vdots \\
		u_{Ki} \\
	\end{matrix}
\right]  
\left[ 
	\begin{matrix}
		v_{1i} \ldots v_{Li} \\
	\end{matrix}
\right] 
\label{eq:XsumRank-1Detail} \\
\mathbf{X} &=
\sum_{i=1}^{\rho} s_{i} \mbf{u}_{i} \mbf{v}_{i}^\text{T}
\label{eq:XsumRank-1}
\end{align}%
Equation \ref{eq:XsumRank-1} expresses $\mbf{X}$ as a sum of rank-1 matrices (each term contains a matrix constructed from a single column vector $\mbf{u}_i$, hence rank-1).  The \textit{Eckart-Young-Mirsky} matrix approximation theorem \citep{golub1987generalization} describes the fact that the matrices in this sum have the special property that they account for successively less and less of the overall variability in $\mbf{X}$.  For each $i<\rho$ in Equation \ref{eq:XsumRank-1} we can form a partial sum that is an approximation of $\mbf{X}$, and each such sum obeys the constraint that it produces the best possible rank-$i$ approximation of $\mbf{X}$ in a perpendicular, sum-of-squares sense, i.e. its row vectors are points that are as close as possible in a Euclidean sense to the corresponding points (row vectors) in $\mbf{X}$ as can be achieved using $i$ dimensions. Formally, the Euclidean difference between $\mbf{X}$ and the rank-$i$ approximation $\mbf{X}^{[i]}$ is as small as possible.  This can be expressed as \citep{abdi2010principal,golub1987generalization},%
\begin{align}
\Big\lVert \mbf{X} - \mbf{X}^{[i]} \Big\rVert ^2 = \Tr \left\{(\mbf{X} - \mbf{X}^{[i]})(\mbf{X} - \mbf{X}^{[i]})^{\text{T}} \right\} =  \min\limits_{\mbf{X}^{[\le i]}} \ \Big\lVert \mbf{X} - \mbf{X}^{[\le i]} \Big\rVert ^2
\label{eq:approxCondition}
\end{align}%
where $\mbf{X}^{[\le i]}$ are matrices of rank less than or equal to $i$, and $\lVert \cdot \rVert$ is the square root of the sum of squared elements of the row vectors of matrix $\mbf{A}$:%
\begin{align}
\lVert \mbf{A} \rVert = \sqrt{\Tr \left( \mbf{A}\mbf{A}^{\text{T}} \right)} \nonumber
\end{align}%
Intuitively, what this means is that the first term in the sum in Equation \ref{eq:XsumRank-1} produces `predicted' points that are as close as possible to the actual points (in a perpendicular sense) using just one dimension, i.e. a line.  The predicted points are vectors that are multiples of the first right singular vector $\mbf{v}_1$, and the extension/contraction factors that define them are $s_1\mbf{u}_1$, easiest to see in Equation \ref{eq:XsumRank-1Detail}.  The second term in the sum produces vectors that are multiples of the second right singular vector $\mbf{v}_2$, similarly with weights equal to $s_2\mbf{u}_2$, that when added to the vectors defined by the first term produce predicted points that are a little closer to the actual points.  This pattern continues through the sum until the last term adds multiples of the last right singular vector and finally completes a vector sum that reproduces the actual points exactly. 

\begin{figure}[htbp]
\begin{center}
\includegraphics[width=1.0\textwidth]{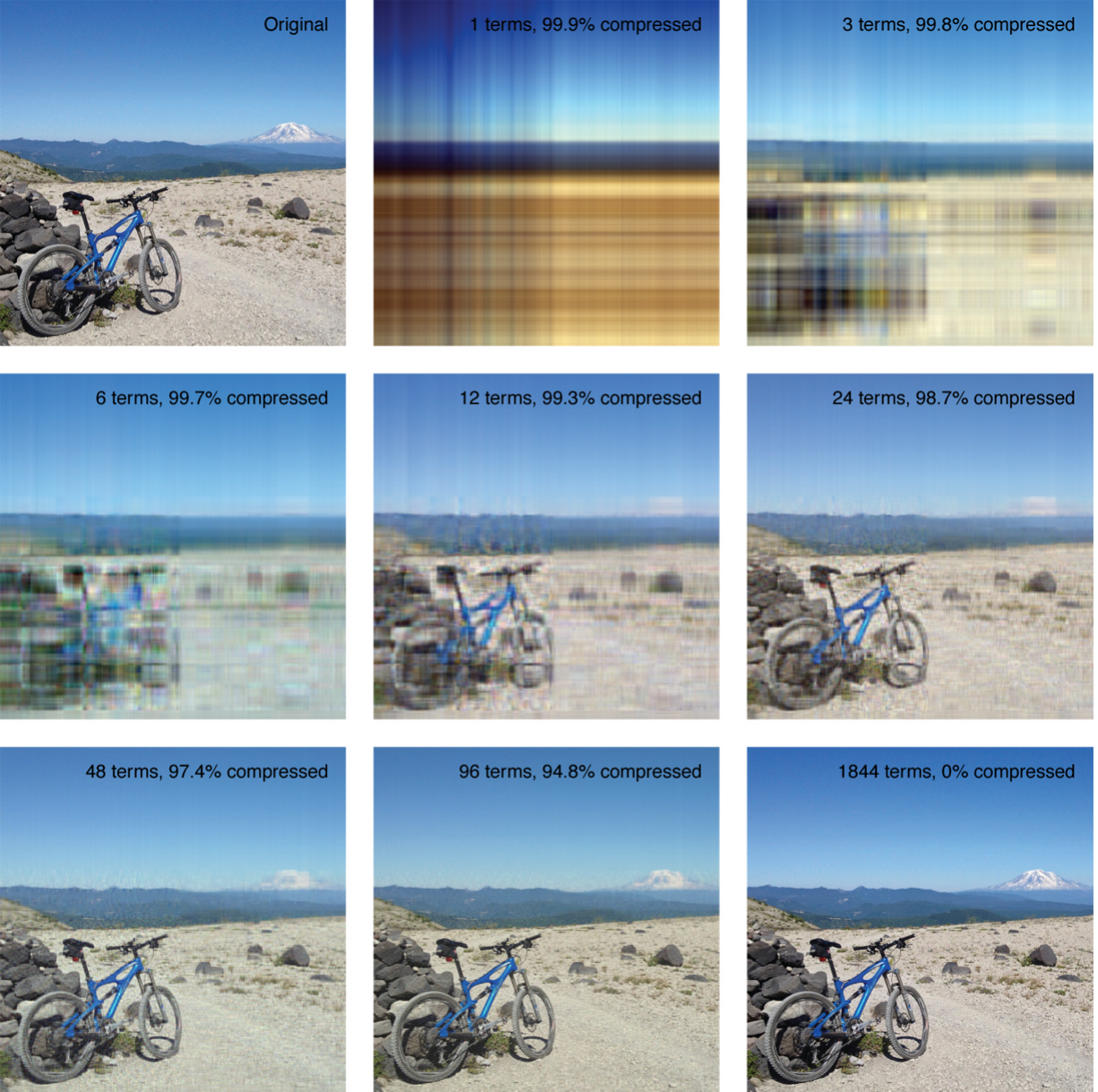}
\captionsetup{format=plain,font=small,margin=0cm,justification=justified}
\caption{{\bf Illustration of the Reduced-rank Approximation using an Image.} The top left panel contains the original $1,844 \times 1,852$ color image of Mt. Adams from the east flank of Mt. St. Helens.  The image is a three-dimension array containing one $1,844 \times 1,852$ matrix of $\left[0,1\right]$ values (one value for each pixel) for each primary color red, blue and green.  The SVD was used to decompose each of the three matrices, and then they are reconstructed using the number of terms in Equation \ref{eq:XsumRank-1} indicated in their label.  The image in the lower right panel uses the maximum number of components (1,844) and reproduces the original exactly. With just 6 components (0.3\% of the information in the original, or 99.7\% compressed) all fundamental aspects of the image are in place, the colors are close to their correct values, and it is possible to detect that there is a bicycle in the photo; with just 24 components (98.7\% compressed) the entire scene is interpretable with high confidence, and with 96 components (94.8\% compressed) you would not know you are missing most of the original.}
\label{fig:imageSVD}
\end{center}
\end{figure}

Because each successive approximation gets as close as possible to the actual points using the orthonormal basis $\mbf{V}$, the first few terms cover most of the distance from the origin to the actual points, and the remaining terms make comparatively small and often negligible contributions.  It is this property of the sum that makes it useful -- \textit{the original data matrix $\mbf{X}$ can be closely approximated by a sum with (possibly many) fewer terms than the rank of $\mbf{X}$}.  Figure \ref{fig:imageSVD} is a convincing visual display of this property of Equation \ref{eq:XsumRank-1}.

There are two final points to note.  The approximation condition defined in Equation \ref{eq:approxCondition} implies that each term in Equation \ref{eq:XsumRank-1} is associated with a fraction of the overall `perpendicular' squared (Euclidean) distance from the origin to the points.  The first term minimizes the difference between the points and their best approximation along a single line, or conversely, maximizes the share of the overall perpendicular squared distance from the origin to the points accounted for by the one-term approximation.  The remaining terms account for smaller and smaller fractions of this overall perpendicular squared distance.  The perpendicular squared distance accounted for by each term is equal to the square of the singular value corresponding to that term.  This fact ensures that the singular values are always arranged in a monotonically decreasing list: $s_{1} \ge s_{2} \ldots \ge s_{\rho}$, and unless the cloud of data points is centered, $s_{1}$ is much larger than the rest, and the first few are much larger than the others.  The singular values are the fraction of the total squared distance from the origin to all of the points that is `accounted for' by each new dimension $\mbf{v}_i$.  This contribution can be quantified by calculating the square of each singular value and dividing that by the sum of the squares of the singular values. 

Finally, it is worth reiterating and stating explicitly that Equation \ref{eq:approxCondition} ensures that the right singular vectors $\mbf{V}$ point in the directions of maximum variation in the original cloud of points defined by $\mbf{X}$, subject to the constraint that they remain perpendicular, and that the overall perpendicular distance starts from the origin.  The last condition is why data clouds must be centered in order for the right singular vectors to point in directions that correspond to the dominant orthogonal dimensions of the cloud and why PCA is conducted on centered data. 

\subsection{Each Column of the Data Matrix as a Sum of Left Singular Vectors}

Interpreting the SVD as a sum of rank-1 matrices leads directly to a similar and very useful expression for each of its column vectors.  In the derivation above, just before converting to a sum of matrices, we express the SVD as a matrix of sums, one for each column.  Equation \ref{eq:genComponents} says that each column vector $\mbf{x}_{\ell}$ in $\mbf{X}$ can be written%
\begin{align}
\mbf{x}_{\ell} = \sum_{i=1}^{\rho} s_{i} v_{\ell,i} \mbf{u}_{i} \ .
\label{eq:goldenEq}
\end{align}%
\textit{This is the key equation} that we have been building up to.  Equation \ref{eq:goldenEq} says that we can write all the columns in $\mbf{X}$ as weighted sums of the left singular vectors scaled by their corresponding singular values, and it tells us what the weights are, namely the $\ell^{\text{th}}$ elements of each corresponding right singular vector.  Moreover, because of the nature of the reduced-rank approximations described above, we know that these sums have the property of concentrating most of the variation in the first few terms, and consequently we only need the first few terms to produce predicted values that are very close to the actual values.  This allows us to closely approximate the columns of $\mbf{X}$ with (potentially very) few effective parameters -- just the first few weights.

\subsection{A Parsimonious Model and Smoother for Vectors Similar to the Columns of the Data Matrix}

Equation \ref{eq:goldenEq} suggests the form of a parsimonious model for an arbitrary column vector of the same length and similar to the columns of $\mbf{X}$, namely%
\begin{align}
\widehat{\mbf{x}} &= \sum_{i = 1}^{c} \beta_{i} \cdot s_{i}\mbf{u}_{i} \ , \label{eq:genModel} \\
\mbf{x} &= \widehat{\mbf{x}} + \mbf{r} \ . \nonumber
\end{align}%
where the $\beta_i$ are chosen to minimize the magnitude of the residual $\lVert \mbf{r} \rVert$; $\mbf{r}$ is the difference between $\mbf{x}$ and its predicted value $\widehat{\mbf{x}}$ using $c \le \rho$ components.

In addition to serving as a reduced-dimension, compact model for $\mbf{x}$, Equation \ref{eq:genModel} can also be thought of as a \textit{smoother}.  By not including the higher-order, small magnitude terms in the sum, it is possible to eliminate the small, less systematic, mostly stochastic differences between the \textit{elements in $\mbf{x}$}.  What is left when only the first few terms are included are the systematic differences between the elements of $\mbf{x}$ that are shared by all or the majority of the column vectors in the data matrix whose SVD produced the singular values and left singular vectors used in Equation \ref{eq:genModel} to produce the approximation of $\mbf{x}$.  This is a common use of SVD that has been applied in many fields for many purposes, e.g.  signal processing, image compression and clustering.

\subsection{A Simple $3 \times 2$ Example SVD with Geometric Interpretation}

\begin{figure}[htbp]
\begin{center}
\includegraphics[width=1.0\textwidth]{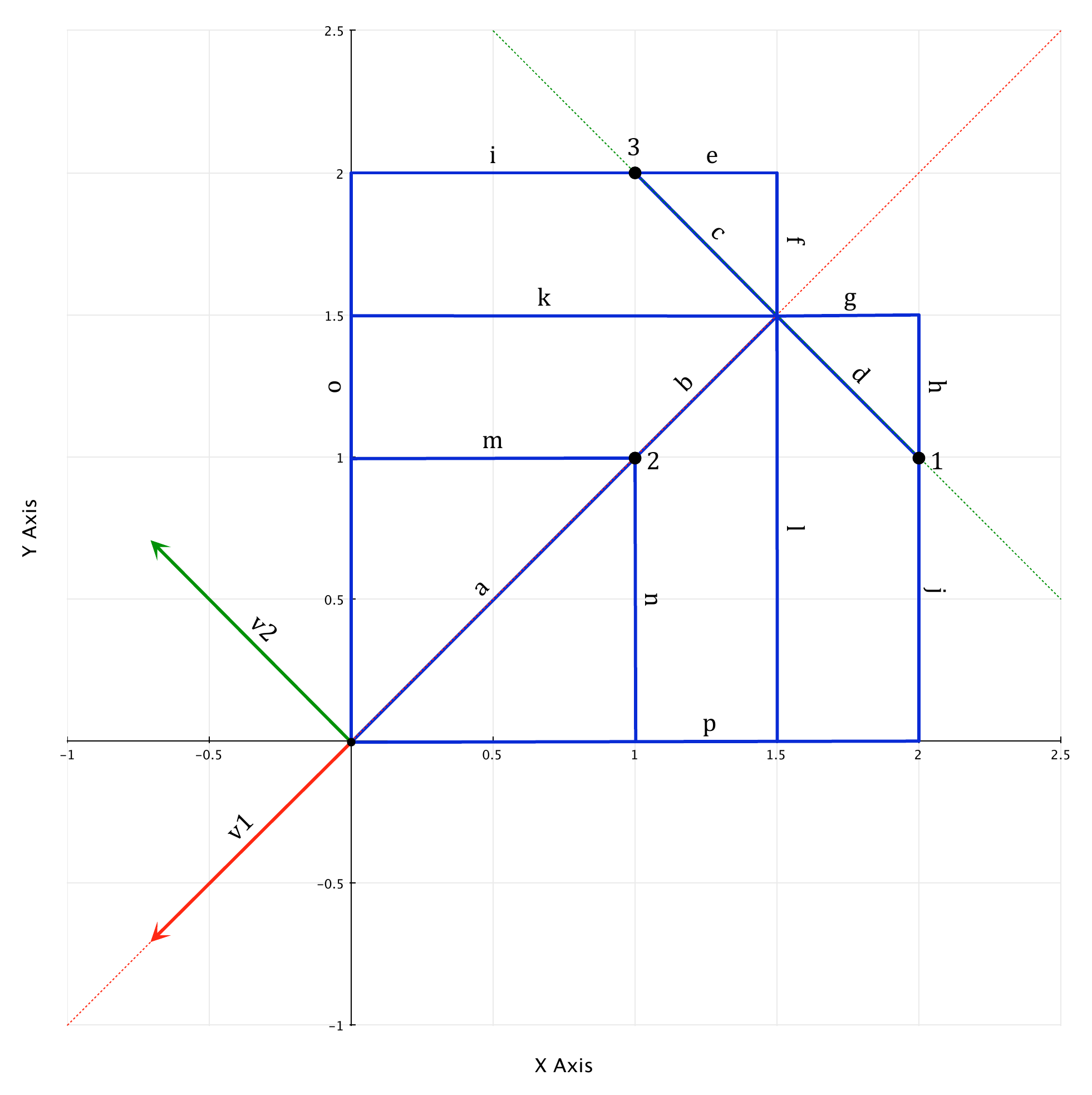}
\captionsetup{format=plain,font=small,margin=0cm,justification=justified}
\caption{{\bf $3\times2$ Example.} The `data matrix' consists of the three points labeled `1', `2' and `3'.  The right singular vectors $\mbf{V}$ are red and green and labeled `v1' and `v2'; they define the new orthogonal basis associated with the SVD of the three points.  The blue line segments mark distances that are used in the calculation of the SVD and are labeled with letters from the alphabet.  Segments `a', `a+b', `c' and `d' are proportional to the left singular vectors of the SVD (scaled by the singular values).}
\label{fig:32Example}
\end{center}
\end{figure}

Equation \ref{eq:32svdDetail} defines the general SVD of a simple $3\times2$ matrix $\mbf{X}$ in detail, and Equations \ref{eq:32comps} (detailed) and \ref{eq:32compact} (compact) describe the component form of the SVD for this particular case.  Let:
\begin{align}
\mathbf{X} 
&=%
\left[ 
	\begin{matrix}
		x_{11} & x_{12}  \\
		x_{21} & x_{22}  \\
		x_{31} & x_{32}  \\
	\end{matrix}
\right],
&\mbf{x}_{i} 
&=%
\left[ 
	\begin{matrix}
		x_{1i} \\
		x_{2i} \\
		x_{3i} \\
	\end{matrix}
\right]
\nonumber \\	
\mathbf{U} 
&=%
\left[ 
	\begin{matrix}
		u_{11} & u_{12}  \\
		u_{21} & u_{22}  \\
		u_{31} & u_{32}  \\
	\end{matrix}
\right], 
&\mathbf{u}_{i} 
&=%
\left[ 
	\begin{matrix}
		u_{1i} \\
		u_{2i} \\
		u_{3i} \\
	\end{matrix}
\right]
\nonumber \\
\mathbf{S} 
&=%
\left[ 
	\begin{matrix}
		s_{1} & 0  \\
		0 & s_{2}  \\
	\end{matrix}
\right] 
\nonumber \\
\mathbf{V} 
&=%
\left[ 
	\begin{matrix}
		v_{11} & v_{12}  \\
		v_{21} & v_{22}  \\
	\end{matrix}
\right],
&\mathbf{v}_{i} 
&=%
\left[ 
	\begin{matrix}
		v_{1i} \\
		v_{2i} \\
	\end{matrix}
\right] \nonumber
\end{align}%
Then the SVD of $\mbf{X}$ is%
\begin{align}
\mbf{X} 
&=%
\mbf{USV}^\text{T} 
\nonumber \\
\left[ 
	\begin{matrix}
		x_{11} & x_{12}  \\
		x_{21} & x_{22}  \\
		x_{31} & x_{32}  \\
	\end{matrix}
\right] 
&=%
\left[ 
	\begin{matrix}
		u_{11} & u_{12}  \\
		u_{21} & u_{22}  \\
		u_{31} & u_{32}  \\
	\end{matrix}
\right]  
\left[ 
	\begin{matrix}
		s_{1} & 0  \\
		0 & s_{2}  \\
	\end{matrix}
\right]
\left[ 
	\begin{matrix}
		v_{11} & v_{21}  \\
		v_{12} & v_{22}  \\
	\end{matrix}
\right]
\label{eq:32svdDetail} 	 
\\
&=%
\left[ 
	\begin{matrix}
		u_{11} & u_{12}  \\
		u_{21} & u_{22}  \\
		u_{31} & u_{32}  \\
	\end{matrix}
\right]  
\left[ 
	\begin{matrix}
		s_{1} v_{11} & s_{1} v_{21}  \\
		s_{2} v_{12} & s_{2} v_{22}  \\
	\end{matrix}
\right] 
\nonumber \\
&=%
\left[ 
	\begin{matrix}
		u_{11} s_{1} v_{11} + u_{12} s_{2} v_{12} & 
			u_{11}  s_{1} v_{21} + u_{12} s_{2} v_{22} \\
		u_{21} s_{1} v_{11} + u_{22} s_{2} v_{12} & 
			u_{21}  s_{1} v_{21} + u_{22} s_{2} v_{22} \\
		u_{31} s_{1} v_{11} + u_{32} s_{2} v_{12} & 
			u_{31}  s_{1} v_{21} + u_{32} s_{2} v_{22} \\
	\end{matrix}
\right]  
\nonumber \\
&=%
\left[ 
\begin{matrix}
	s_{1} v_{11} 
\left[ 
	\begin{matrix}
		u_{11} \\
		u_{21} \\
		u_{31} \\
	\end{matrix}
\right]  +
	s_{2} v_{12} 
\left[ 
	\begin{matrix}
		u_{12} \\
		u_{22} \\
		u_{32} \\
	\end{matrix}
\right] 
	& s_{1} v_{21} 
\left[ 
	\begin{matrix}
		u_{11} \\
		u_{21} \\
		u_{31} \\
	\end{matrix}
\right]  +
	s_{2} v_{22} 
\left[ 
	\begin{matrix}
		u_{12} \\
		u_{22} \\
		u_{32} \\
	\end{matrix}
\right] \\
\end{matrix}
\right]
\label{eq:32comps} \\
&=%
\left[ 
\begin{matrix}
	s_{1} v_{11} \left[ 
	\begin{matrix}
		u_{11} \\
		u_{21} \\
		u_{31} \\
	\end{matrix}
\right]  
	& s_{1} v_{21} 
\left[ 
	\begin{matrix}
		u_{11} \\
		u_{21} \\
		u_{31} \\
	\end{matrix}
\right]  \\
\end{matrix}
\right] +
\left[ 
\begin{matrix}
	s_{2} v_{12} \left[ 
	\begin{matrix}
		u_{12} \\
		u_{22} \\
		u_{32} \\
	\end{matrix}
\right] 
	& s_{2} v_{22} \left[ 
	\begin{matrix}
		u_{12} \\
		u_{22} \\
		u_{32} \\
	\end{matrix}
\right] \\
\end{matrix}
\right]
\label{eq:32ScaledVs} \\
&=%
\left[
	\begin{matrix}
	s_{1} v_{11} u_{11} & s_{1} v_{21} u_{11} \\
	s_{1} v_{11} u_{21} & s_{1} v_{21} u_{21} \\
	s_{1} v_{11} u_{31} & s_{1} v_{21} u_{31} 
	\end{matrix}
\right] +
\left[
	\begin{matrix}
	s_{2} v_{12} u_{12} & s_{2} v_{22} u_{12} \\
	s_{2} v_{12} u_{22} & s_{2} v_{22} u_{22} \\
	s_{2} v_{12} u_{32} & s_{2} v_{22} u_{32} 
	\end{matrix}
\right]
\nonumber \\
&=%
s_{1} 
\left[ 
	\begin{matrix}
		u_{11} \\
		u_{21} \\
		u_{31} \\
	\end{matrix}
\right]  
\left[ 
	\begin{matrix}
		v_{11} & v_{21} \\
	\end{matrix}
\right] 
	+
	s_{2} \left[ 
	\begin{matrix}
		u_{12} \\
		u_{22} \\
		u_{32} \\
	\end{matrix}
\right]  
\left[ 
	\begin{matrix}
		v_{12} & v_{22} \\
	\end{matrix}
\right] 
\nonumber \\
\mbf{X} 
&=%
\sum_{ i = 1}^2 s_ i \mbf{u}_{i} \mbf{v}_{i}^\text{T}
\label{eq:32compact}
\end{align}%
From (\ref{eq:32comps})
\begin{align}
\mbf{x}_{1} 
&=%
s_{1} v_{11} 
\left[ 
	\begin{matrix}
		u_{11} \\
		u_{21} \\
		u_{31} \\
	\end{matrix}
\right]  +
	s_{2} v_{12} \left[ 
	\begin{matrix}
		u_{12} \\
		u_{22} \\
		u_{32} \\
	\end{matrix}
\right] 
\nonumber \\
\mbf{x}_{1} 
&= s_{1}v_{11}\mbf{u}_{1} + 	s_{2}v_{12}\mbf{u}_{2}
\label{eq:32x1comp}
\end{align}%
and
\begin{align}
\mbf{x}_{2} 
&=%
s_{1} v_{21} 
\left[ 
	\begin{matrix}
		u_{11} \\
		u_{21} \\
		u_{31} \\
	\end{matrix}
\right]  +
	s_{2} v_{22} \left[ 
	\begin{matrix}
		u_{12} \\
		u_{22} \\
		u_{32} \\
	\end{matrix}
\right] 
\nonumber \\
\mbf{x}_{2} 
&= s_{1}v_{21}\mbf{u}_{1} + 	s_{2}v_{22}\mbf{u}_{2}
\label{eq:32x2comp}
\end{align}%
Following this general form, below we work an example with three specific points in two dimensions.  Figure \ref{fig:32Example} displays the three example points: (2,1) labeled `1', (1,1) labeled `2' and (1,2) labeled `3'.  

The line segments necessary to demonstrate the SVD are labeled $a$ -- $p$, and the lengths of each are:
\begin{align}
a &= \sqrt{2} \ ,					&  c 	&= \tfrac{\sqrt{2}}{2} \ ,	&	e	&= -0.5 \ ,	
	&	g	&= \phantom{-}0.5	\ ,		&	i	&= 1 \ , 	&	k	&= 1.5 \ , 	&	m	&= 1 \ , 	&	o	&= 2 \nonumber \\ 
b &= \tfrac{\sqrt{2}}{2} \ ,	&  d 	&= \tfrac{\sqrt{2}}{2} \ ,	& 	f	&= \phantom{-}0.5 \ ,		
	&	h	&= -0.5 \ ,	&	j	&= 1 \ , 	&	l	&= 1.5 \ , 	&	n	&= 1 \ , 	&	p	&= 2  \nonumber 
\end{align}%
First, we identify the singular vectors $\mbf{v}_1$ and $\mathbf{v}_2$.  The requirement is that the first singular vector point in a direction that maximizes the sum of squared distances \textit{along this dimension} to all of the points, or conversely, minimizes the sum of squared differences between the points and their predicted values, i.e. their projections onto this new dimension.  Given the simplicity and symmetry of this example, we can readily see that $\mbf{v}_1$ must be in the direction of the red vector labeled `v1' in Figure \ref{fig:32Example}. The projections of point 1 (2,1) and 3 (1,2) onto this new dimension are both the point where line segments b, c, d, f, g, k \& l meet, and point 2 (1,1) actually lies on this dimension already.  A final requirement is that the length of the singular vectors be 1, so we choose the point $\big(-\sqrt{0.5},-\sqrt{0.5} \big)$ to define $\mbf{v}_1$, whether it points down and the left or up and to the right does not matter, and it is easier to see in the figure if we define it like this.  

$\mbf{v}_2$ must be perpendicular to $\mbf{v}_1$, and the only option in this two-dimension example is along the direction of the green vector labeled `v2' in Figure \ref{fig:32Example}.  We choose $\mbf{v}_2$ defined by $\big(-\sqrt{0.5},\sqrt{0.5} \big)$, again to ensure its magnitude is 1.

Second, we calculate the singular values $s_1$ and $s_2$ and demonstrate how these are related to the total sum of squares.  The total sum of squares is the sum of squared perpendicular distances from the origin along each original dimension $X,Y$ to the three points, in order 1--3:
\begin{align}
SS_{tot}& = \bigl( p^2 + j^2 \bigr) + \bigl( m^2 + n^2 \bigr) + \bigl(i^2 + o^2 \bigr)  \\
& = 4+1 + 1+1 + 1 + 4 = 12 \nonumber
\end{align}
The first singular value $s_1$ is the square root of the sum of squared distances from the origin to each point on the dimension defined by the first singular vector $\mbf{v}_1$, again in order 1--3:
\begin{align}
s_{1}^2 = SS_1& = \bigl(a+b\bigr)^2 + a^2 + \bigl(a+b\bigr)^2  \\
& = \big(\tfrac{3}{2}\sqrt{2}\big)^2 + \bigl(\sqrt{2}\bigr)^2 +  \big(\tfrac{3}{2}\sqrt{2}\big)^2 \nonumber \\
& = \tfrac{9}{2} + \tfrac{4}{2} + \tfrac{9}{2} \nonumber \\
& = \tfrac{22}{2} = 11 \nonumber \\
\rightarrow s_1 &= \sqrt{11} = 3.3166 \nonumber
\end{align}
The second singular value $s_2$ is the square root of the sum of squared distances from the projection of the points onto the first new dimension ($\mbf{v}_1$) to each point along the dimension defined by the second singular vector $\mbf{v}_2$ (i.e. segment $d$ for point 1, 0 for point 2 and segment $c$ for point 3), in order 1--3:
\begin{align}
s_{2}^2 = SS_2& =d^2 + 0^2 + c^2  \\
& = \Big(\tfrac{\sqrt{2}}{2}\Big)^2 + 0+ \Big(\tfrac{\sqrt{2}}{2}\Big)^2 \nonumber \\
& = \tfrac{1}{2} + \tfrac{1}{2} = 1 \nonumber \\
\rightarrow s_2 &= \sqrt{1} = 1 \nonumber
\end{align}
We can verify that the sum of the squares of the singular values is equal to the total sum of squares.
\begin{align}
SS_1 + SS_2 &= s_1^2 + s_2^2  \\
&= 11 + 1 = 12 \ \checkmark \nonumber \\
&= SS_{tot}
\end{align}
Using the original $X,Y$ dimensions, the sum of squared perpendicular distances associated with either the $X$ or $Y$ dimension is $\left(p^2+m^2+i^2\right)$ or $\left(j^2+n^2+o^2\right) = \left(1^2+1^2+2^2\right) = 6$, or 12 for both dimensions combined. The SVD identifies a new set of dimensions such that a majority of the squared distance to the points is along the `primary' new dimension and the remainder along the secondary new dimension(s), in this case 11 along the first and 1 along the second.

Finally, we identify the left singular vectors $\mbf{u}_1$ and $\mbf{u}_2$.  From the first term in Equation \ref{eq:32ScaledVs} we see that the product of $\mbf{u}_1$ and the first singular value $s_1$ multiplies (extends or contracts) the $X$ and $Y$ elements of $\mbf{v}_1$, and similarly, the product of $\mbf{u}_2$ and the second singular value $s_2$ multiplies the $X$ and $Y$ elements of $\mbf{v}_2$.  To calculate the values of the elements of $\mbf{u}_1$ we take the $X$ and $Y$ distances to the projections of the points along the the first right singular vector $\mbf{v}_1$ and divide them by $s_1v_{11}$ in the case of the $X$ dimension or $s_1v_{21}$ for the $Y$ dimension.  Using values from the example along the $X$ dimension:%
\begin{align}
u_{11} &= \frac{k}{s_1v_{11}} = \frac{1.5}{\sqrt{11} \times -\sqrt{0.5}} = -0.6396 \nonumber \\
u_{21} &= \frac{m}{s_1v_{11}} = \frac{1}{\sqrt{11} \times -\sqrt{0.5}} = -0.4264 \nonumber \\
u_{31} &= \frac{k}{s_1v_{11}} = \frac{1.5}{\sqrt{11} \times -\sqrt{0.5}} = -0.6396 \nonumber 
\end{align}%
or, using the $Y$ dimension:%
\begin{align}
u_{11} &= \frac{\ell}{s_1v_{21}} = \frac{1.5}{\sqrt{11} \times -\sqrt{0.5}} = -0.6396 \nonumber \\
u_{21} &= \frac{n}{s_1v_{21}} = \frac{1}{\sqrt{11} \times -\sqrt{0.5}}= -0.4264 \nonumber \\
u_{31} &= \frac{\ell}{s_1v_{21}} = \frac{1.5}{\sqrt{11} \times -\sqrt{0.5}} = -0.6396 \nonumber 
\end{align}%
The result is the same in both cases, as it must be.

Similarly from Equation \ref{eq:32ScaledVs} we see that $\mbf{u}_2$ multiplies the $X$ and $Y$ elements of $\mbf{v}_2$ to create vectors in the $\mbf{v}_2$ direction that when added to the vectors produced by extending $\mbf{v}_1$ by $\mbf{u}_1$ yield the original points.  This time the numerators may have a negative sign because we are projecting the vectors obtained by subtracting the extended versions of $\mbf{v}_1$ described just above from the original points, and the resulting vectors are not all in the 1$^{\text{st}}$ quadrant. Again using values from the example in the $X$ dimension:%
 \begin{align}
u_{12} &= \frac{g}{s_2v_{12}} = \frac{0.5}{1 \times -\sqrt{0.5}} = -0.7071 \nonumber \\
u_{22} &= \frac{0}{s_2v_{12}} = \frac{0}{1 \times -\sqrt{0.5}} = \phantom{-}0 \nonumber \\
u_{32} &= \frac{e}{s_2v_{12}} = \frac{-0.5}{1 \times -\sqrt{0.5}} = \phantom{-}0.7071 \nonumber 
\end{align}%
or, using the $Y$ dimension:%
\begin{align}
u_{12} &= \frac{h}{s_2v_{22}} = \frac{-0.5}{1 \times \sqrt{0.5}} = -0.7071 \nonumber \\
u_{22} &= \frac{0}{s_2v_{22}} = \frac{0}{1 \times \sqrt{0.5}}= \phantom{-}0 \nonumber \\
u_{32} &= \frac{f}{s_2v_{22}} = \frac{0.5}{1 \times \sqrt{0.5}} = \phantom{-}0.7071 \nonumber 
\end{align}%
Again, the result is the same either way.  

Summarizing our intuitive, \textit{geometric} calculation of the SVD of $\mbf{X}$:%
\begin{align}
\mathbf{X} 
&=%
\left[ 
	\begin{matrix}
		2 & 1  \\
		1 & 1  \\
		1 & 2  \\
	\end{matrix}
\right]
& 
\mathbf{U} 
&=%
\left[ 
	\begin{matrix}
		-0.6396 & -0.7071\\
		-0.4264 & 0\\
		-0.6396 & \phantom{-}0.7071\\
	\end{matrix}
\right]
\nonumber \\
\mathbf{S} 
&=%
\left[ 
	\begin{matrix}
		3.3166 & 0  \\
		0 & 1  \\
	\end{matrix}
\right] 
&
\mathbf{V} 
&=%
\left[ 
	\begin{matrix}
		-0.7071 & -0.7071  \\
		-0.7071 & \phantom{-}0.7071  \\
	\end{matrix}
\right]
\nonumber
\end{align}%
Using the statistical package \texttt{R} to calculate the SVD of $\mbf{X}$ yields:%
\begin{align}
\mathbf{X} 
&=%
\left[ 
	\begin{matrix}
		2 & 1  \\
		1 & 1  \\
		1 & 2  \\
	\end{matrix}
\right]
& 
\mathbf{U} 
&=%
\left[ 
	\begin{matrix}
		-0.6396021 & \phantom{1}7.071068\text{e-01}\\
		-0.4264014 & \phantom{1}2.775558\text{e-17}\\
		-0.6396021 & -7.071068\text{e-01}\\
	\end{matrix}
\right]
\nonumber \\
\mathbf{S} 
&=%
\left[ 
	\begin{matrix}
		3.316625 & 0  \\
		0 & 1.000000  \\
	\end{matrix}
\right] 
&
\mathbf{V} 
&=%
\left[ 
	\begin{matrix}
		-0.7071068 & \phantom{-}0.7071068  \\
		-0.7071068 & -0.7071068  \\
	\end{matrix}
\right]
\nonumber
\end{align}%
The results are effectively identical; the SVD algorithm in \texttt{R} chose to point $\mbf{v}_2$ in the opposite direction which caused the signs on the members of $\mbf{u}_2$ to flip; equivalent to our result.

Looking more closely at $\mbf{U}$, we can begin to see how the SVD can be useful in demography.  $\mbf{u}_1$ in the direction of the points, i.e. $-1 \times \mbf{u}_1 = \left[\begin{smallmatrix}
0.64\\
0.43\\
0.64\end{smallmatrix}\right]$, encodes the `shape' of the points along $\mbf{v}_1$.  Starting with point 1, we go \textit{out}, then come \textit{back in} for point 2, and finally back \textit{out} again for point 3.  If the dimension defined by $\mbf{v}_1$ is age-specific, then \textit{$\mbf{u}_1$ is a scaled age pattern}.

Finally, turning our attention back to Equations \ref{eq:32x1comp} and \ref{eq:32x2comp}, we see that the column vectors of $\mbf{X}$ are expressed as weighted sums of the $\mbf{u}$'s scaled by their corresponding singular values, and the weights are the elements of the row vectors of $\mbf{V}$.  Each term in these sums is a fraction of the distance along the original $X$ and $Y$ axes from the origin to our three points.  Because the first left singular vector is associated with the new basis direction along which there is most `squared distance' from the origin, the first term in these weighted sums represents the largest fraction, with subsequent terms accounting for smaller and smaller fractions.  This allows us to approximate the column vectors in $\mbf{X}$ with a subset of the terms in these weighted sums.  In our two-dimension example, we have just two terms; with the first we have a reasonable approximation of the column vectors of $\mbf{X}$, and with both we reproduce them exactly.  In this re-expression, the first column vector $\mbf{x}_1$ is
\begin{align}
{\mbf{x}_1}
&=%
\left[ 
	\begin{matrix}
		2 \\
		1 \\
		1 \\
	\end{matrix}
\right]
=%
s_{1} v_{11} 
\left[ 
	\begin{matrix}
		u_{11} \\
		u_{21} \\
		u_{31} \\
	\end{matrix}
\right]
+
s_{2} v_{12} 
\left[ 
	\begin{matrix}
		u_{12} \\
		u_{22} \\
		u_{32} \\
	\end{matrix}
\right]
\nonumber \\
&=
3.3166 \times -0.7071 \left[ 
	\begin{matrix}
		-0.6396 \\
		-0.4264 \\
		-0.6396 \\
	\end{matrix}
\right] 
+
1 \times -0.7071 \left[ 
	\begin{matrix}
		-0.7071 \\
		\phantom{-0.707}0 \\
		\phantom{-}0.7071 \\
	\end{matrix}
\right] 
\nonumber \\
&=
\left[ 
	\begin{matrix}
		1.5 \\
		\phantom{1.}1 \\
		1.5 \\
	\end{matrix}
\right]
+
\left[ 
	\begin{matrix}
		\phantom{-}0.5 \\
		\phantom{-0.}0 \\
		-0.5 \\
	\end{matrix}
\right]
=
\left[ 
	\begin{matrix}
		2 \\
		1 \\
		1 \\
	\end{matrix}
\right] \ \checkmark
\nonumber
\end{align}%
The first term in this sum $\left[ 
	\begin{smallmatrix}
		1.5 \\
		\phantom{1.}1 \\
		1.5 \\
	\end{smallmatrix}
\right]$ approximates $\mbf{x}_1$ quite well, and the second $\left[ 
	\begin{smallmatrix}
		\phantom{-}0.5 \\
		\phantom{-0.}0 \\
		-0.5 \\
	\end{smallmatrix}
\right]$ makes the small refinement necessary to reproduce $\mbf{x}_1$ exactly.  In this example the situation is very similar for the second column vector $\mbf{x}_2$, the only difference being in the second term.%
\begin{align}
{\mbf{x}_2}
&=%
\left[ 
	\begin{matrix}
		1 \\
		1 \\
		2 \\
	\end{matrix}
\right]
=%
s_{1} v_{21} 
\left[ 
	\begin{matrix}
		u_{11} \\
		u_{21} \\
		u_{31} \\
	\end{matrix}
\right]
+
s_{2} v_{22} 
\left[ 
	\begin{matrix}
		u_{12} \\
		u_{22} \\
		u_{32} \\
	\end{matrix}
\right]
\nonumber \\
&=
3.3166 \times -0.7071 \left[ 
	\begin{matrix}
		-0.6396 \\
		-0.4264 \\
		-0.6396 \\
	\end{matrix}
\right] 
+
1 \times 0.7071 \left[ 
	\begin{matrix}
		-0.7071 \\
		\phantom{-0.707}0 \\
		\phantom{-}0.7071 \\
	\end{matrix}
\right] 
\nonumber \\
&=
\left[ 
	\begin{matrix}
		1.5 \\
		\phantom{1.}1 \\
		1.5 \\
	\end{matrix}
\right]
+
\left[ 
	\begin{matrix}
		-0.5 \\
		\phantom{-0.}0 \\
		\phantom{-}0.5 \\
	\end{matrix}
\right]
=
\left[ 
	\begin{matrix}
		1 \\
		1 \\
		2 \\
	\end{matrix}
\right] \ \checkmark
\nonumber
\end{align}%
In both cases we are almost all the way there with just the first term.

\section{The SVD and Demographic Quantities Correlated by Age}

In this section we turn to a practical application of the SVD in demography.

\subsection{Demographic Quantities Correlated by Age -- \textit{Age Schedules}}

Various demographic quantities including age-specific mortality and fertility are correlated by age.  The data used in the examples in the Section \ref{sec:examples} is displayed in Figure \ref{fig:aginData} and reveal the very strong age-dependance that is typical for both mortality and fertility. The pair-wise correlation coefficients for the two-sex (female joined to male in one 38-element vector) age-specific log mortality schedules plotted in Figure \ref{fig:aginData} are $\ge 0.90$ in all cases.  Likewise the log age-specific fertility schedules have pair-wise correlation coefficients that are no less than 0.99.  

This feature of many age-specific demographic quantities makes them amenable to direct decomposition using the SVD.  Imagine organizing the age schedules into a matrix $\mbf{A}$ so that each age schedule is a column and each age (group) is a row; a $G \times H$ matrix of age-specific quantities.  Geometrically, each of the $G$ rows in this matrix corresponds to a point in $H$-dimension space, and together the points defined by the rows make up the cloud of data points that we want to characterize using the SVD.  In order to make the SVD interpretable \textit{and} to retain the original scale of our data so that we can model them without translation (e.g. centering) or rescaling (e.g. normalizing), we need to be sure that the primary dimension identified by the SVD -- the first right singular vector $\mbf{v}_1$ -- is lined up with the dimension of maximum variability in the cloud of points representing our age schedules.  Remember that the first right singular vector points from the origin to the center of the cloud, which means that the primary dimension of the cloud must \textit{be} or be very similar to the first right singular vector in order to make the SVD of the uncentered cloud useful.  

The fact that demographic age schedules are correlated by age and none are offset, or moved, along the age axis means that the distance from the origin to each point along each of the $H$ axes is similar, that is, \textit{the points in $\mbf{A}$ all lie roughly on a line that intersects the origin and has `slope' $\approx$ 1}.  Of course when the number of columns exceeds three, one cannot visualize this line, but the fact still holds for an arbitrary number of dimensions (columns). 

\begin{figure}[htbp]
\begin{center}
\begin{tabular}{cc}
A: Log Mortality Rate Cloud & B: Geometry of Female Point 7\\
\includegraphics[width=0.5\textwidth]{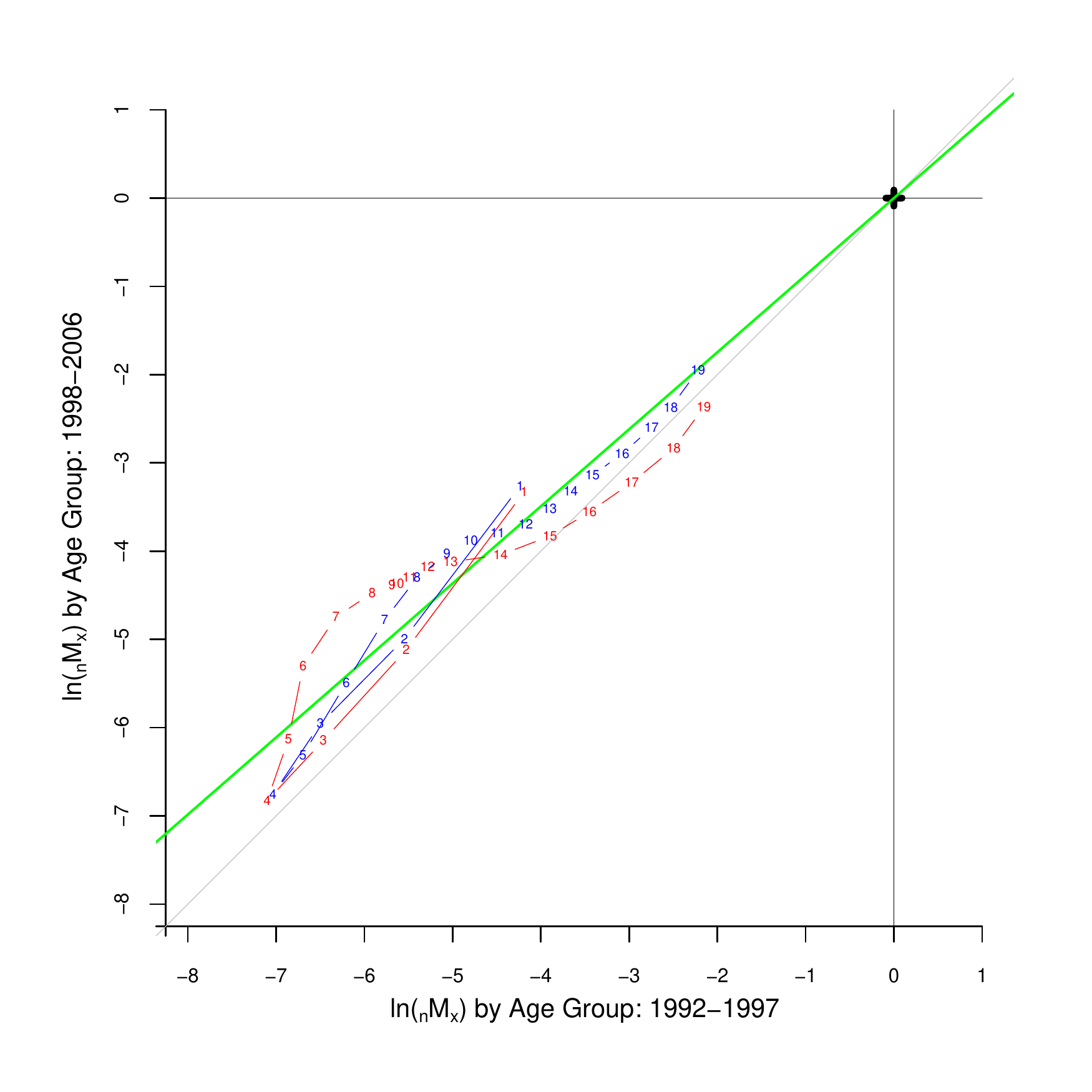} &
\includegraphics[width=0.5\textwidth]{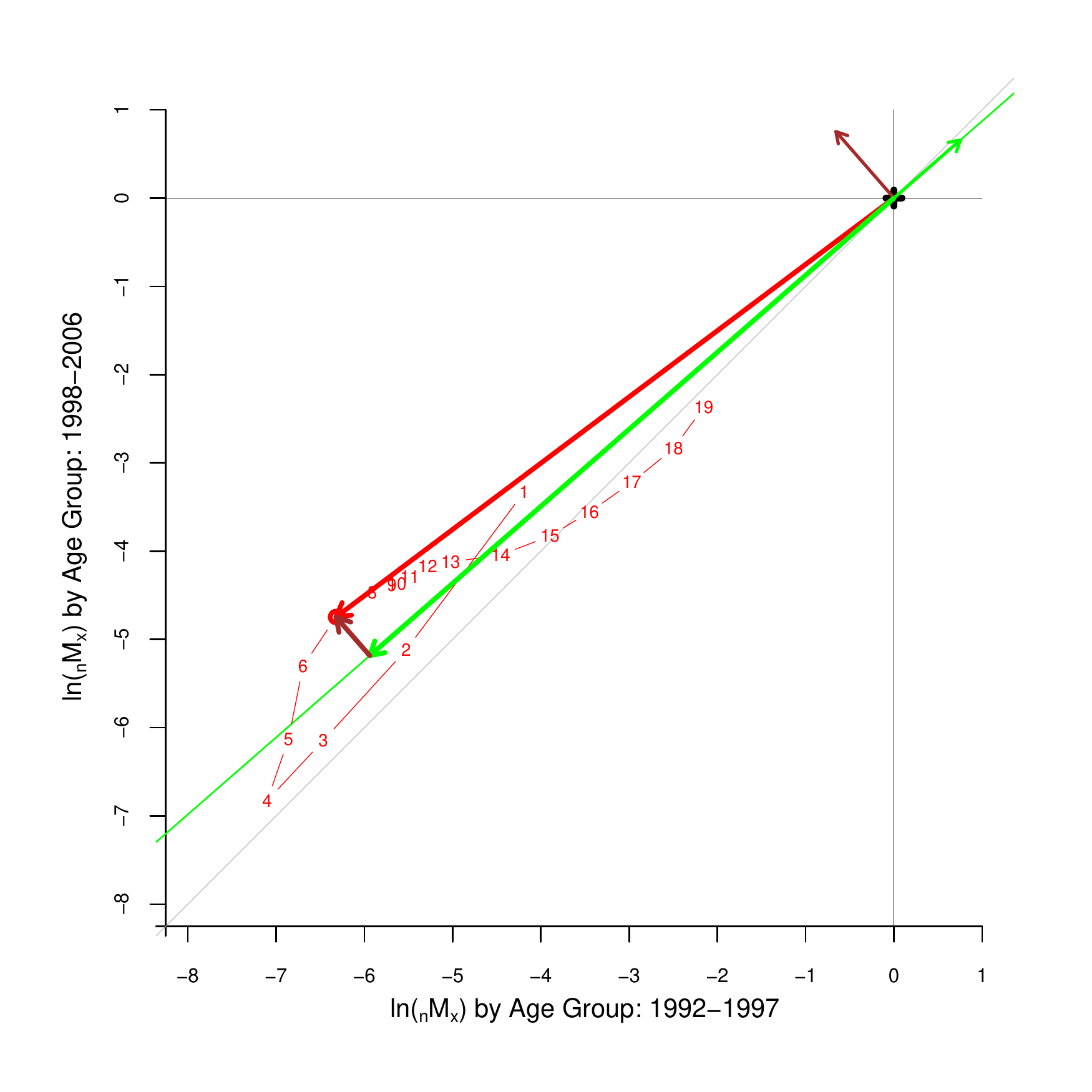}
\end{tabular}
\captionsetup{format=plain,format=plain,font=small,margin=0.5cm,justification=justified}
\caption{{\bf Two-dimension Example -- Geometry of SVD.} \textsc{Panel} \textbf{(A)}: Scatterplot of 1998--2006 life table by 1992--1997 life table. Red = female, Blue = male. Points numbered from youngest to oldest age group. Grey line is $y=x$.  Green line is direction along which there is most variation in the cloud of points. \textsc{Panel} \textbf{(B)}: SVD-defined vectors that reconstruct female point number 7.  Small green vector pointing up and to right from origin is first right singular vector $\mathbf{v}_1$; small brown vector pointing up and to left from origin is second right singular vector $\mathbf{v}_2$.  Long green vector pointing down and to left is projection along $\mathbf{v}_1$ corresponding to female point 7, and short brown vector from the tip of the long green vector to female point 7 is the projection along $\mathbf{v}_2$ corresponding to female point 7.  Adding the two projections of the right singular vectors produces the long red vector from the origin down and to the left that defines female point 7.}
\label{fig:2dExampleSVD}
\end{center}
\end{figure}

This can be seen easily by imagining just two age schedules.  In that case we have a simple two-column, two-dimension data set that corresponds to a cloud of $G$ points on a familiar two-dimension $X$, $Y$ plot.  Now imagine plotting the points that correspond to the age schedules.  Because the values are similar for each age group, the distance along both axes is similar for every point, and all the points  cluster around a line with slope $\approx$ 1, see \textsc{Panel} \textbf{(A)} in Figure \ref{fig:2dExampleSVD}.   

It is also generally true that the primary dimension identified by the first right singular vector also captures the \textit{vast} majority of the variation in the cloud of points.  This is obvious when one notes that this dimension effectively captures the level or magnitude of the indicator by age, the differences that are on the vertical axis if each age schedule were plotted by age.  The remaining orthogonal axes identified by the remaining right singular vectors capture the remaining variability in the cloud; variability that is age-specific but largely unrelated to the overall level or magnitude of the age schedule.  

\begin{figure}[htbp]
\begin{center}
\includegraphics[width=1.0\textwidth]{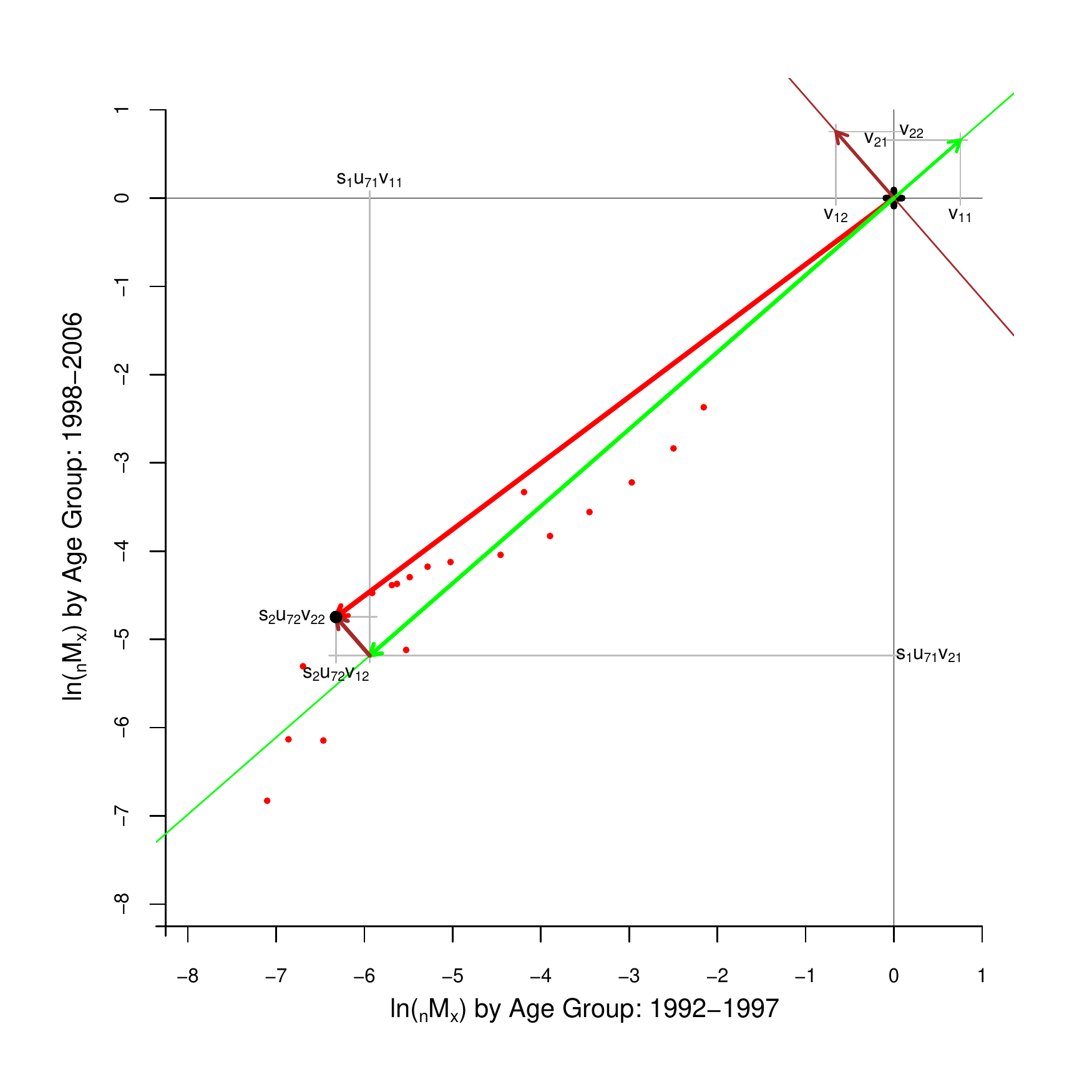} 
\captionsetup{format=plain,font=small,margin=0cm,justification=justified}
\caption{{\bf Two-dimension Example -- Geometry of SVD, continued.}  Scatterplot of 1998--2006 life table by 1992--1997 life table. Red = female, Blue = male. Green line is direction along which there is most variation in the cloud of points. SVD-defined vectors that reconstruct female point number 7: small green vector pointing up and to right from origin is first right singular vector $\mathbf{v}_1$; small brown vector pointing up and to left from origin is second right singular vector $\mathbf{v}_2$.  Long green vector pointing down and to left is projection along $\mathbf{v}_1$ corresponding to female point 7, and short brown vector from the tip of the long green vector to female point 7 is the projection along $\mathbf{v}_2$ corresponding to female point 7.  Adding the two projections of the right singular vectors produces the long red vector from the origin down and to the left that defines female point 7.  Each vector defined in terms of the original $X$, $Y$ coordinate system.  This makes clear that the projections `stretch' the right singular vectors by mulitplicative factors specified in the left singular vectors $\mathbf{u}$.}
\label{fig:2dExampleSVDDetail}
\end{center}
\end{figure}

Figures \ref{fig:2dExampleSVD} and \ref{fig:2dExampleSVDDetail} display an example of the geometry of the SVD using two mortality age schedules from the example data listed in Appendix \ref{app:exampleMx} Table \ref{app:exampleMx}.\ref{tab:aginMx3}.   

In summary, the first right singular vector $\mbf{v}_1$ will always identify a dimension that is close to the primary dimension of the cloud of points defined by the age schedules, and this dimension will be associated with the overall level of the indicator by age.  Additional orthogonal dimensions identified by the remaining right singular vectors will capture the remaining variability, most of which will be age-specific but not closely related to the overall level of the age schedules.

\subsection{A General, Parsimonious, SVD-derived Model for Demographic Quantities Correlated by Age}\label{sec:genModel}

Demography is largely about understanding the structure and dynamics of populations, and a key underlying dimension of `structure' is age.  Consequently demographers measure and manipulate age schedules of various quantitates -- e.g. mortality, fertility, nuptially, migration, etc. -- in many ways.  To make these tasks easier, and to make it possible to relate age schedules \textit{as a whole} (i.e. an indicator across all ages) to various covariates or predictors, it is desirable to have parsimonious models of complete age structures that can incorporate covariates.  These can then be used to:%
\begin{itemize}\addtolength{\itemsep}{-0.4\baselineskip}
\item smooth noisy age schedules,
\item fill-in or extend incompletely measured age schedules, and/or
\item produce full age schedules using one or a small number of parameters and/or predictors of those parameters.
\end{itemize}%
The objective of the model we develop below is to provide a general modeling framework for whole age structures that requires very few (usually just two or three) parameters.  A framework of this type can then be used to summarize the empirical regularities in any collection of age schedules correlated by age, and further the parameter values that replicate the observed age schedules can be themselves modeled as functions of covariates.  Those models can then be used to generate parameter values from the covariates, which can in turn can be turned into full age schedules by the model. The final result is an empirical model that can be driven either directly by the parameters themselves or by the covariates used to generate parameter values. This provides a mechanism by which to predict full age schedules from the covariates.  

\subsubsection{Data and Model Objectives}

As above, the data consist of a $G \times H$ matrix $\mbf{A}$ of age-specific quantities.  The columns of $\mbf{A}$ correspond to the age schedules, and the rows contain the values of the indicator for each age or age group.  Each row is a point in $\text{R}^H$ corresponding to an age group; the number of points equals to the number of ages or age groups.  The objective of our model is to summarize the shape of the cloud of points as parsimoniously as possible using well-behaved and interpretable parameters.  A secondary objective is to be able to remove random `noise,' i.e. stochastic variation, that is not systematically related to either age or anything else.  Finally, we would like to be able to identify natural groups or clusters of these points, if they exist.  Cleanly separated clusters would indicate that there are groups of age schedules that are similar to one another but systematically different from all the others.  If true, this is an important feature of the empirical data that likely results from some underlying mechanism (that could be explained) and can be exploited to improve our ability to both fit and predict age schedules using the model.

\subsubsection{The Model -- Summarizing Empirical Regularities}\label{sec:theModel}

Given a $G \times H$ matrix $\mbf{A}$ of age schedules, the SVD of $\mbf{A}$ yields a set of age-varying components and corresponding weights that can be used to reconstruct the $h \in \{1 \ldots H \}$ individual age schedules in $\mbf{A}$ to within arbitrary precision using a weighted sum based on the  Eckart-Young-Mirsky formula,
\begin{align}
\widehat{\mbf{a}}_h &= \sum_{i = 1}^{c} {v}{_{hi}} \cdot {s}{_i} {\mbf{u}}{_i} \ , \label{eq:genModelVs} \\
\mbf{a}_h &= \widehat{\mbf{a}}_h + \mbf{r}_h \ . \nonumber
\end{align}%
where $\mbf{r}_h$ is a residual; ${v}{_{hi}}$, ${s}{_i}$ and ${\mbf{u}}{_i}$ come from the SVD of $\mbf{A}$; and $c$ is chosen so that $\lVert \mbf{r}_h \rVert$ is small enough to satisfy the desired level of precision.  Taking the age-varying components $\left( {s}{_i} {\mbf{u}}{_i} \right)$ as fixed, Equation \ref{eq:genModelVs} is a $c$-parameter model for the age schedules in $\mbf{A}$.  Because of the concentration of variation in the first few new dimensions $\left(\mbf{V}\right)$ of the SVD, $\widehat{\mbf{a}}_h$ is a smoothed or de-noised version of $\mbf{a}_h$, and experience indicates that in most cases $c \le 3$ is sufficient to adequately reproduce all of the $\mbf{a}_h$. For a given matrix of age schedules $\mbf{A}$ Equation \ref{eq:genModelVs} will have very high `within sample' validity, i.e. it will be able to reproduce all of the $\mbf{a}_h$ in $\mbf{A}$ to within arbitrary precision (by adjusting $c$). 

Equation \ref{eq:genModelVs} is also useful in an `out of sample' sense to represent age schedules that are not included in $\mbf{A}$, as long as they are similar to those in $\mbf{A}$. Geometrically a new age schedule adds a new dimension to our cloud of $H$ points, and to be similar to the age schedules in $\mbf{A}$, this new dimension has to preserve the overall shape of the cloud rather than pulling or pushing it in a new way, i.e. having lower or higher indicator values at a given age compared to the schedules in $\mbf{A}$.  This is unlikely to be a problem if $\mbf{A}$ is large and diverse with respect to age schedules.  

When used in this out-of-sample way, Equation \ref{eq:genModelVs} needs to be modified to specify that the age-varying components come from the SVD of a particular matrix of age schedules $\mbf{A}$, and further that although the weights do not come from the SVD of $\mbf{A}$ they are defined with respect to the age-varying components from $\mbf{A}$,
\begin{align}
\tensor[_A]{\widehat{\mbf{a}}}{} &= \sum_{i = 1}^{c} \tensor[_A]{\beta}{_{i}} \cdot \tensor[_A]{\mbf{\Lambda}}{_i} \ , \label{eq:genModel1} \\
\mbf{a} &= \tensor[_A]{\widehat{\mbf{a}}}{} + \tensor[_A]{\mbf{r}}{} \ . \nonumber
\end{align}
where $\tensor[_A]{\mbf{r}}{}$ is a residual; the $\tensor[_A]{\mbf{\Lambda}}{_i} = \tensor[_A]{s}{_i} \tensor[_A]{\mbf{u}}{_i}$ and come from the SVD of $\mbf{A}$; and the $\tensor[_A]{\beta}{_{i}}$ are chosen to minimize $\lVert \tensor[_A]{\mbf{r}}{} \rVert$ for a given $c$.  For an arbitrary age schedule $\mbf{a}$, a reasonable set of $\tensor[_A]{\beta}{_{i}}$ can be identified easily through OLS regression of the age schedule on the $c$ age-varying components $\tensor[_A]{\mbf{\Lambda}}{_i}$, subject to the constraint that the intercept is 0.  Taking the $\tensor[_A]{\mbf{\Lambda}}{_i}$ as fixed parameters, the age schedule $\mbf{a}$ is represented by the small number of effective parameters $\tensor[_A]{\beta}{_{i}}, \ i \in \left\{1 \ldots c \right\} $, where $c$ is typically much smaller than the number of age groups in the age schedule.  This feature of Equation \ref{eq:genModel} generally makes it a very parsimonious representation of complete age schedules.

\begin{figure}[htbp]
\begin{center}
\begin{tabular}{cc}
A: Data and SVD Decomposition & B Reconstruction of Schedule -- 1 \\
\includegraphics[width=0.48\textwidth]{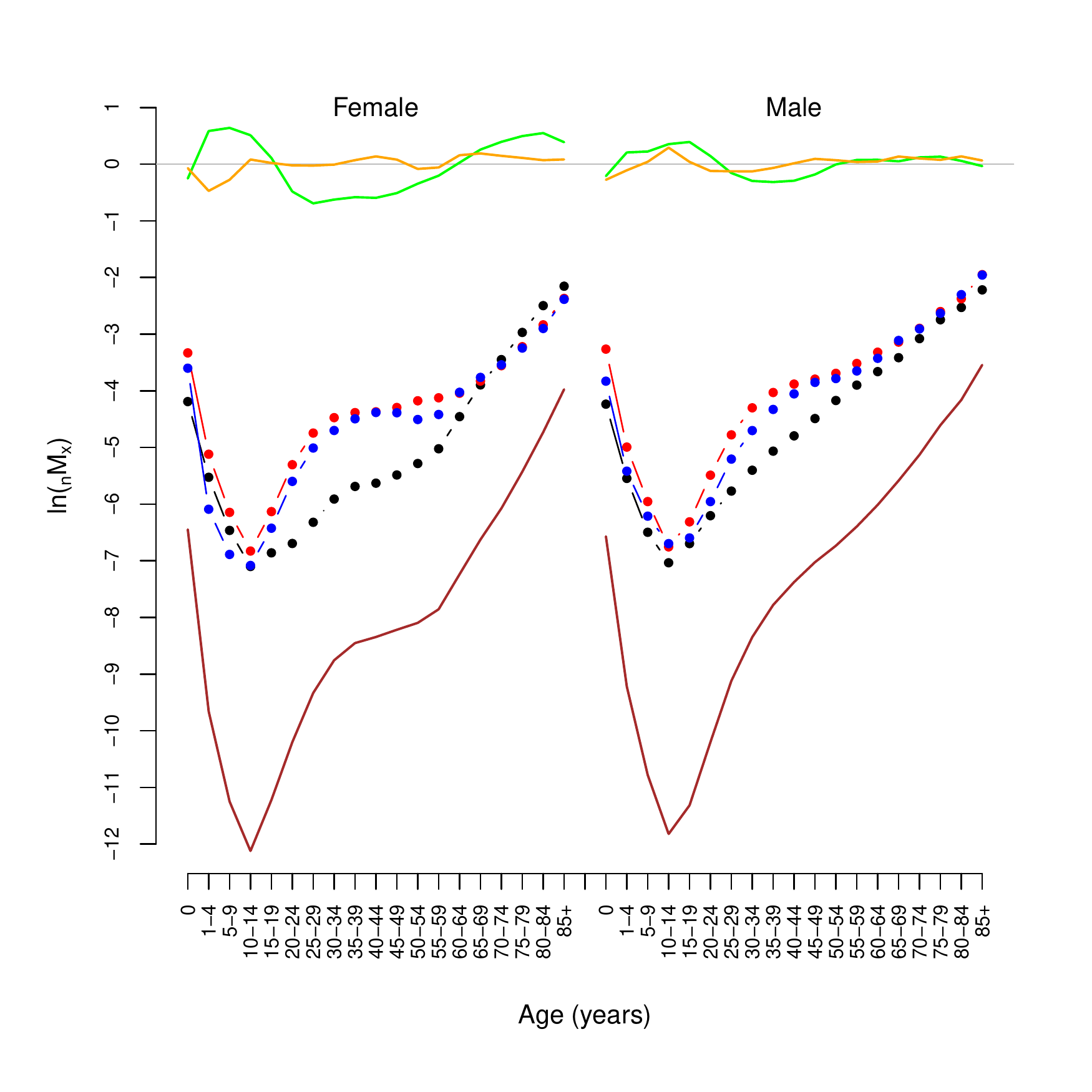} &
\includegraphics[width=0.48\textwidth]{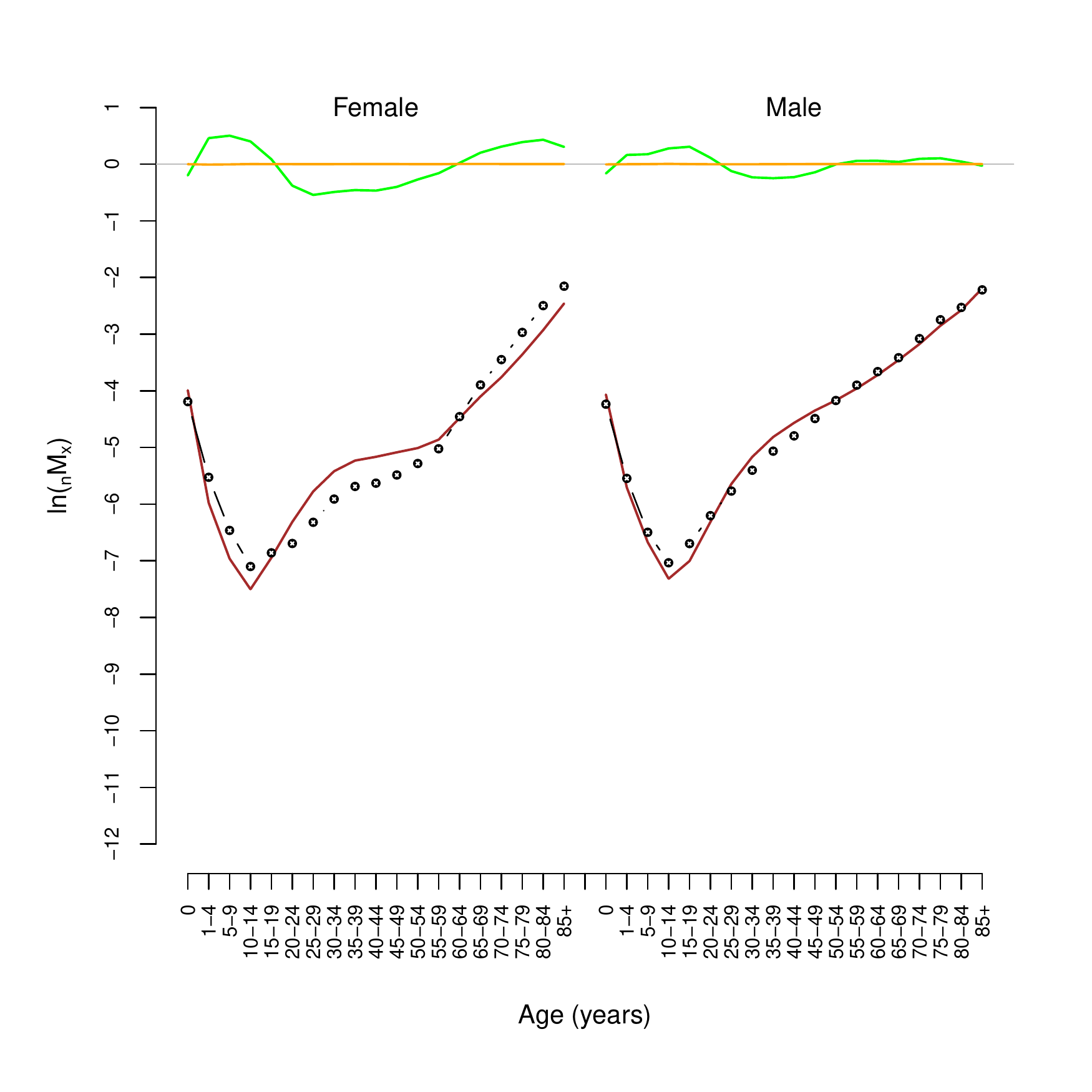} \\
C: Reconstruction of Schedule 2 & D: Reconstruction of Schedule 3 \\
\includegraphics[width=0.48\textwidth]{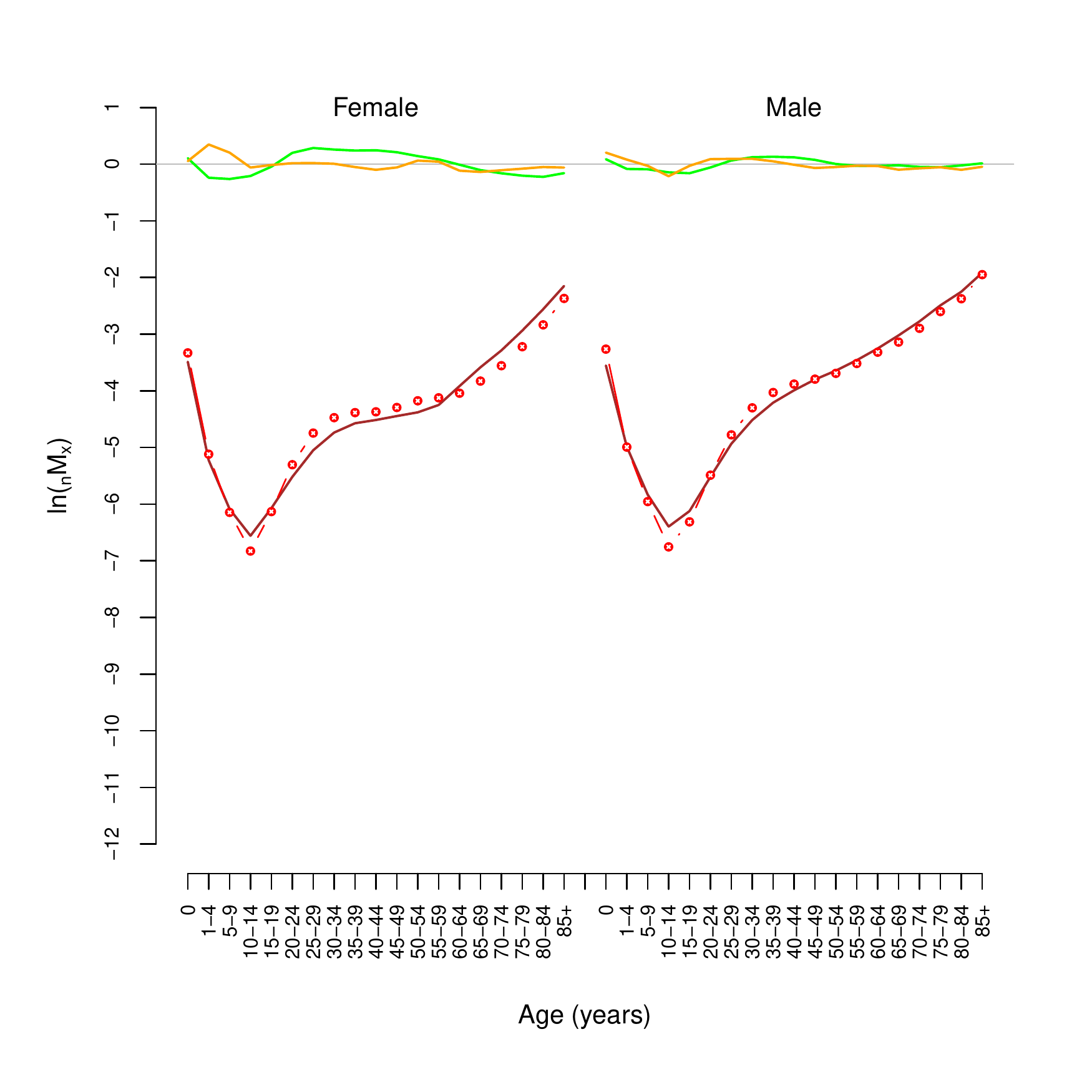} &
\includegraphics[width=0.48\textwidth]{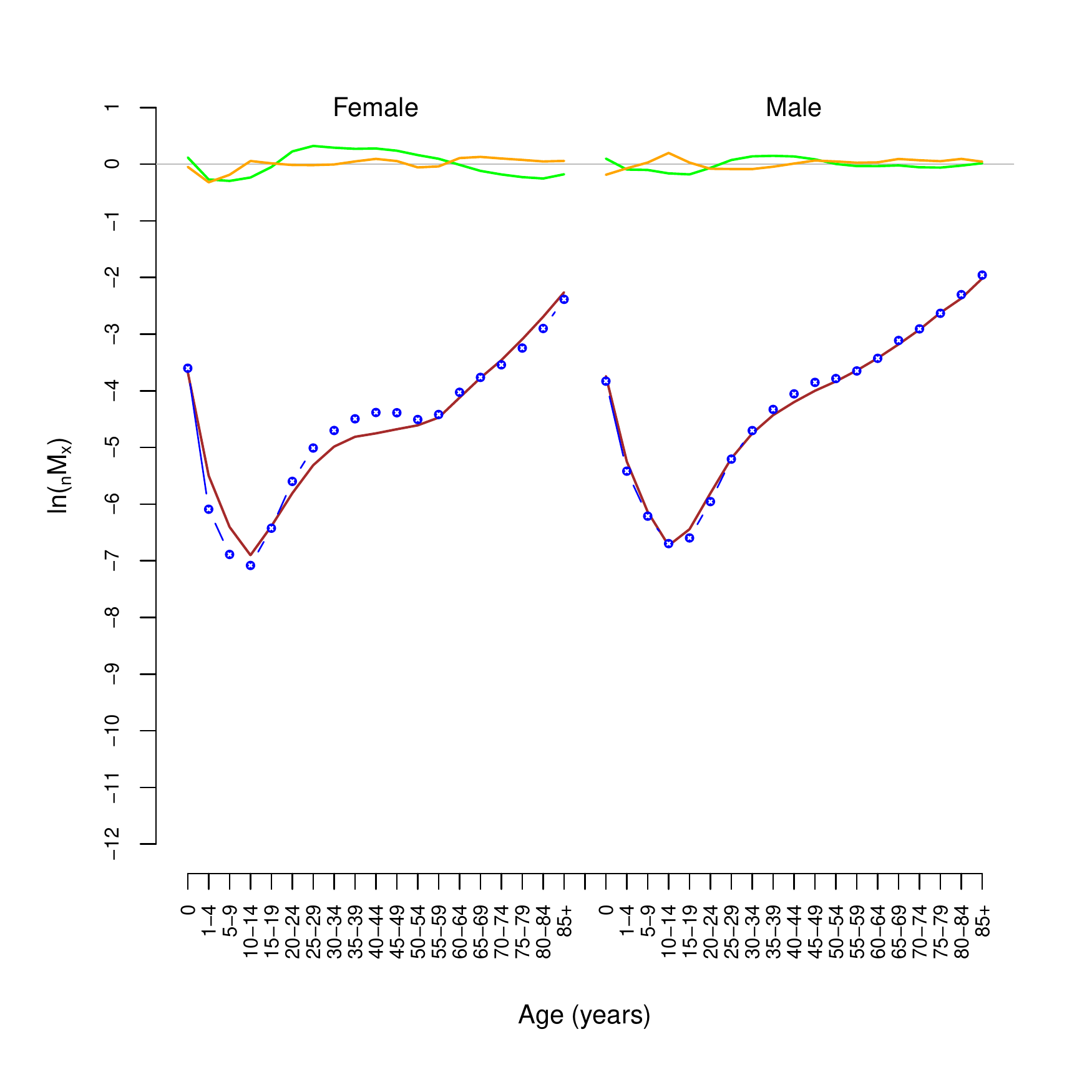}
\end{tabular}
\captionsetup{format=plain,font=small,margin=0cm,justification=justified}
\caption{{\bf Three-dimension Example with Log Mortality Schedules from Agincourt -- Reconstructions.} The mortality schedules cover periods 1: 1992-97, 2: 1998-06, and 3: 2007-12; data are in Table \ref{app:exampleMx}.\ref{tab:aginMx3}. \textsc{Panel} \textbf{(A)}: Three-dimension example components $\tensor[_3]{\mbf{\Lambda}}{_i} \ i \in \{1,2,3\}$; brown = first component $s_1\mathbf{u}_1$; green = second component  $s_2\mathbf{u}_2$; and orange = third component $s_3\mathbf{u}_3$. Black dots are data values for mortality schedule 1; red dots are data values for mortality schedule 2; and blue dots are data for mortality schedule 3. \textsc{Panel} \textbf{(B)}: Reconstruction of mortality schedule 1; brown = weighted component 1 $v_{11}\tensor[_3]{\mbf{\Lambda}}{_1}$; green = weighted second component $v_{12}\tensor[_3]{\mbf{\Lambda}}{_2}$; and orange = weighted third component $v_{13}\tensor[_3]{\mbf{\Lambda}}{_3}$.  Black dots are data values for mortality schedule 1, and white x's are reconstructed values (all at the center of the data dots) -- the sum of weighted components 1--3. \textsc{Panels} \textbf{(C--D)}: Same as Panel (B) for mortality schedules 2--3.  Notice that the the second (green) component creates the HIV-related `hump' by \textit{subtracting} it from the primary component for the first period (\textsc{Panel} \textbf{(A)}) and adding small amounts for the other two periods.}
\label{fig:3dimMxExample}
\end{center}
\end{figure}

\begin{figure}[htbp]
\begin{center}
\begin{tabular}{cc}
A: Mortality Schedule 1 (1992-97) & B: Mortality Schedule 2 (1998-06)\\
\includegraphics[width=0.48\textwidth]{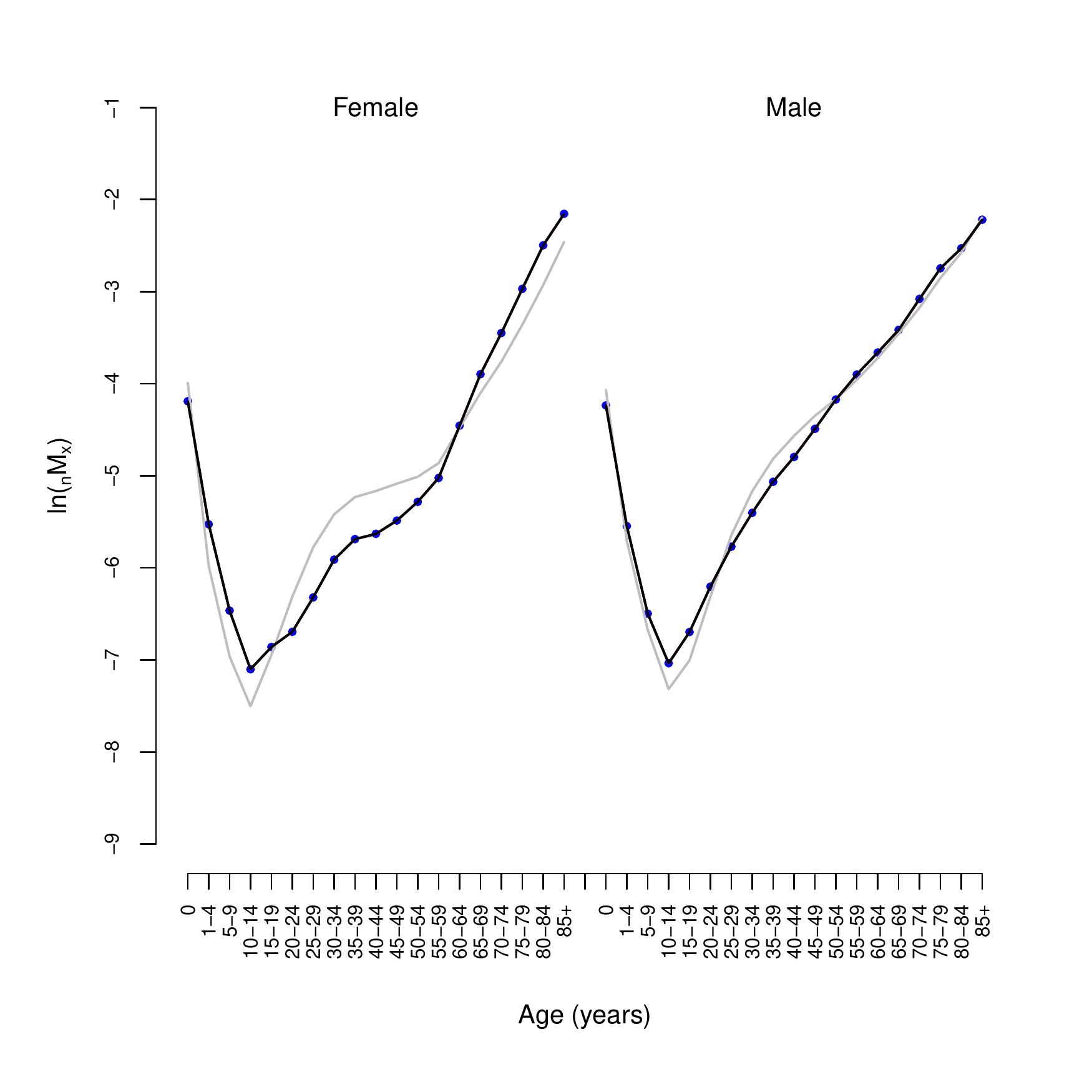} &
\includegraphics[width=0.48\textwidth]{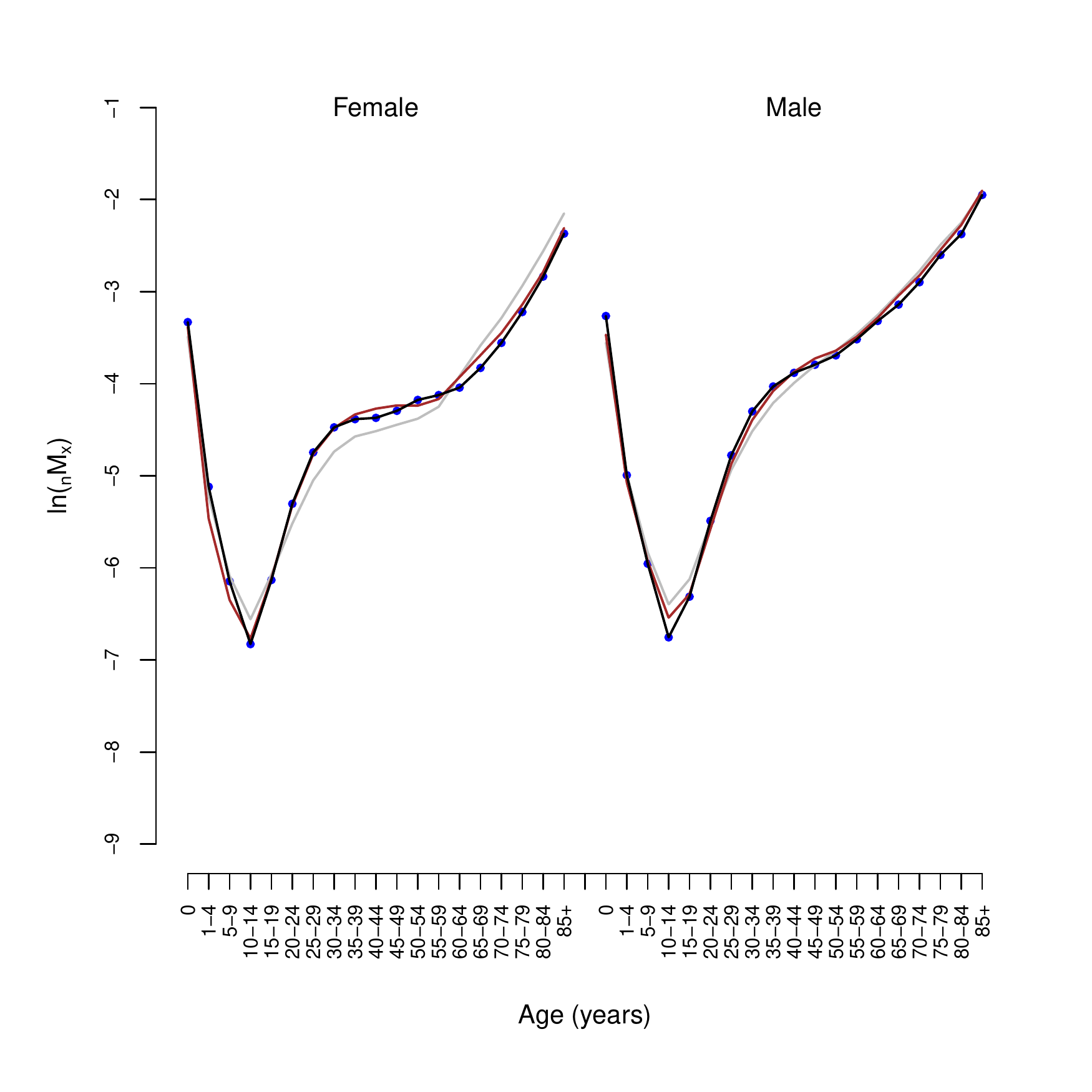} \\
C: Mortality Schedule 3 (2007-12) & C -- Weights\\
\includegraphics[width=0.48\textwidth]{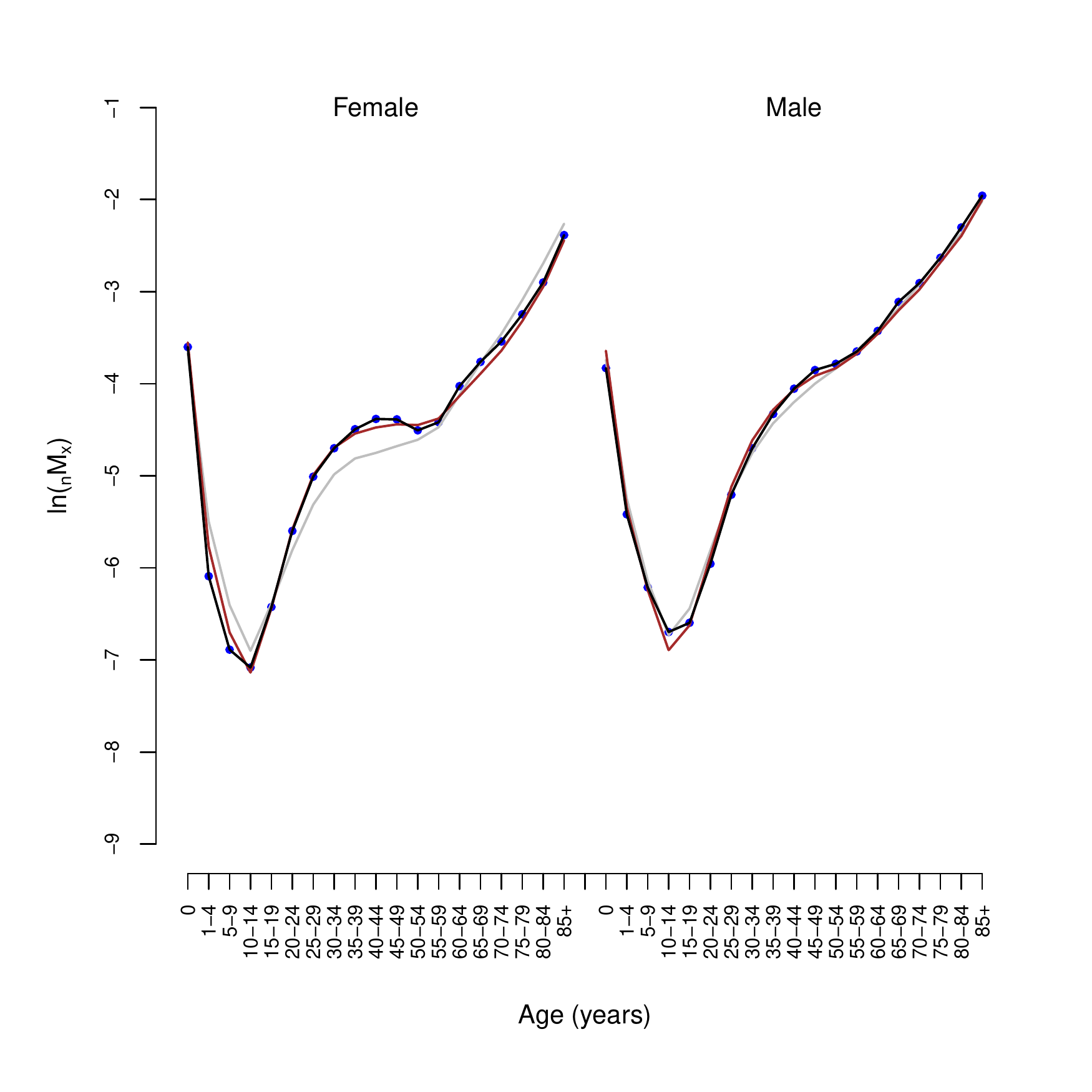} &
\includegraphics[width=0.48\textwidth]{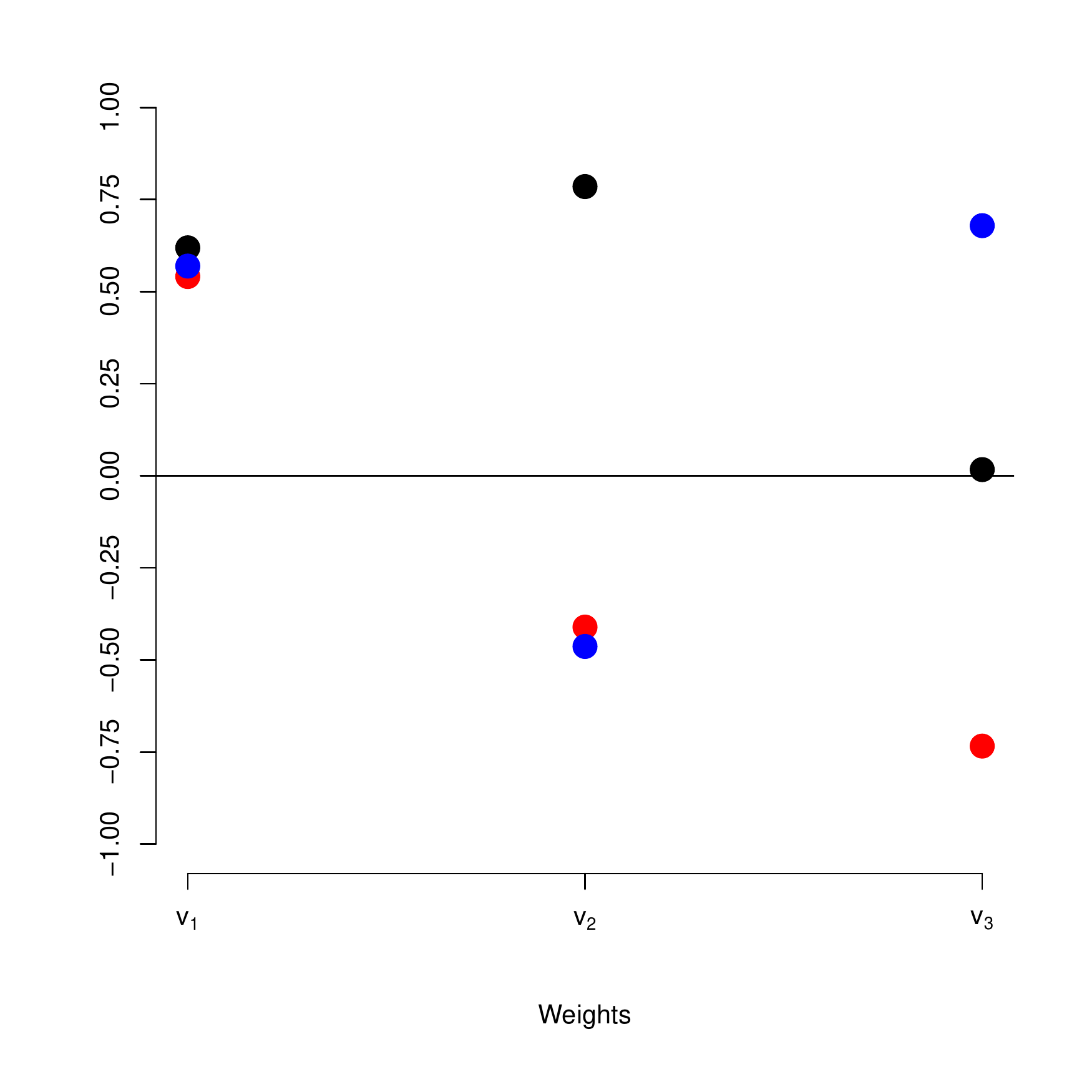} 
\end{tabular}
\captionsetup{format=plain,font=small,margin=0cm,justification=justified}
\caption{{\bf Three-dimension Example with Log Mortality Schedules from Agincourt -- Dimension Reduction.} The mortality schedules cover periods 1: 1992-97, 2: 1998-06, and 3: 2007-12; data are in Table \ref{app:exampleMx}.\ref{tab:aginMx3}. \textsc{Panel} \textbf{(A)}: Reconstruction of mortality schedule 1 using 1 (grey), 2 (brown) and 3 (black) components.  Blue dots are data values. \textsc{Panels} \textbf{(B--C)}: Similar reconstructions for mortality schedules 2--3. Notice that reconstructions with the first component (grey) capture the basic shape of the schedules; adding the second component (brown) produces schedules that are very close to the data (or match the data as for the first period), and adding the third component (black) matches the data perfectly. \textsc{Panel} \textbf{(D)}: The weights applied to each component to reconstruct the original values. Black is period 1, red is period 2, and blue is period 3.  Notice that each period contains a unique `pattern' of weights.}
\label{fig:3dimRecons}
\end{center}
\end{figure}

The model is illustrated in Figures \ref{fig:3dimMxExample} and \ref{fig:3dimRecons} using the smoothed three-period example data listed in Table \ref{app:exampleMx}.\ref{tab:aginMx3} in appendix \ref{app:exampleMx}.  \textsc{Panel} \textbf{(A)} of Figure \ref{fig:3dimMxExample} displays the log mortality schedules along with the three components $\tensor[_A]{\mbf{\Lambda}}{_i}$ calculated from the SVD of the schedules.  Based on our understanding of the SVD of uncentered data clouds, we expect the first component to `locate' the cloud, i.e. to correspond to the distance from the origin to the cloud and therefore to contain values that are well clear of zero, and in the case of log mortality rates, all negative.  Further we expect the remaining dimensions to capture variability within the cloud, and therefore to contain values of generally smaller magnitude that fall on either side of zero. The remaining panels of Figure \ref{fig:3dimMxExample} contain reconstructions of the three mortality schedules and display each of the weighted components $\tensor[_A]{\beta}{_{i}} \cdot \tensor[_A]{\mbf{\Lambda}}{_i}$, their sum, and finally the data that their sum reconstructs.  Figure \ref{fig:3dimRecons} displays the partial reconstructions of each of the three mortality schedules using 1, 2, and 3 components.  The last panel of Figure \ref{fig:3dimRecons} graphically displays the weights applied to each component and makes clear that the pattern of weights clearly differentiates the three schedules. It is this fact that is used to categorize age schedules into clusters by identifying common patterns of weights using a clustering algorithm, see Section \ref{sec:clustering}.

\subsubsection{Parameters as Functions of Covariates -- Predicting Age Schedules}\label{sec:paramsCovars}

If there is a systematic relationship between age schedules and an interesting covariate, then  Equation \ref{eq:genModel} indicates that the relationship will also hold between the $\tensor[_A]{\beta}{_{i}}, \ i \in \left\{1 \ldots c \right\}$ (hereafter $\tensor[_A]{\beta}{_{i}^{\prime}}$) and the covariate.  Quantifying the relationship between the covariate(s) and the $\tensor[_A]{\beta}{_{i}^{\prime}}$ allows the $\tensor[_A]{\beta}{_{i}^{\prime}}$ to be predicted from the covariate, and then the age schedule from the resulting $\tensor[_A]{\beta}{_{i}^{\prime}}$.  This is an efficient way to characterize the relationship between whole age schedules and interesting covariates, and perhaps more usefully, to be able to predict whole age schedules from covariates, even just one.  We have applied this idea to an earlier version of the component model of mortality in the context of HIV-related mortality \citep{Sharrow2014}.  We will demonstrate it in several examples with both mortality and fertility below.

\subsubsection{Identifying Clusters in Collections of Empirical Age Schedules}\label{sec:clustering}

Now we turn again to the cloud of points associated with a matrix $\mbf{A}$ of age schedules.  As we mentioned above, if there are groups of age schedules in $\mbf{A}$ that are similar to each other and largely different from the other age schedules in $\mbf{A}$, then there will be clusters of points in the cloud defined by $\mbf{A}$.  Because of the strong age dependence of all of the age schedules, geometrically these clusters will all be `long and thin', lying close and roughly parallel to a line through the origin and the center of the cloud.  Even if we could visualize things in 4+ dimensions, it would be hard to identify and separate these clusters. 

The SVD of $\mbf{A}$ helps solve this problem.  Just as the calculated weights did just above when we were thinking about predicting age schedules from covariates, the first few age-schedule- ($h$) and component- ($i$) specific weights $v_{hi}$ in Equation \ref{eq:genModelVs} capture most of information necessary to define each individual age schedule in $\mbf{A}$, and moreover, they quantify the contribution of orthogonal components to each age schedule.  Together these properties make them ideal inputs to clustering algorithms such as Mclust \citep{fraley2002,fraley2009mclust} that automatically identify clusters and label their members.  We demonstrate this below.

\section{Examples}\label{sec:examples}

\subsection{Example Data: The Agincourt HDSS, South Africa}

\begin{figure}[htbp]
\begin{center}
\begin{tabular}{cc}
A: Mortality Rates & B: Fertility Rates \\
\includegraphics[width=0.48\textwidth]{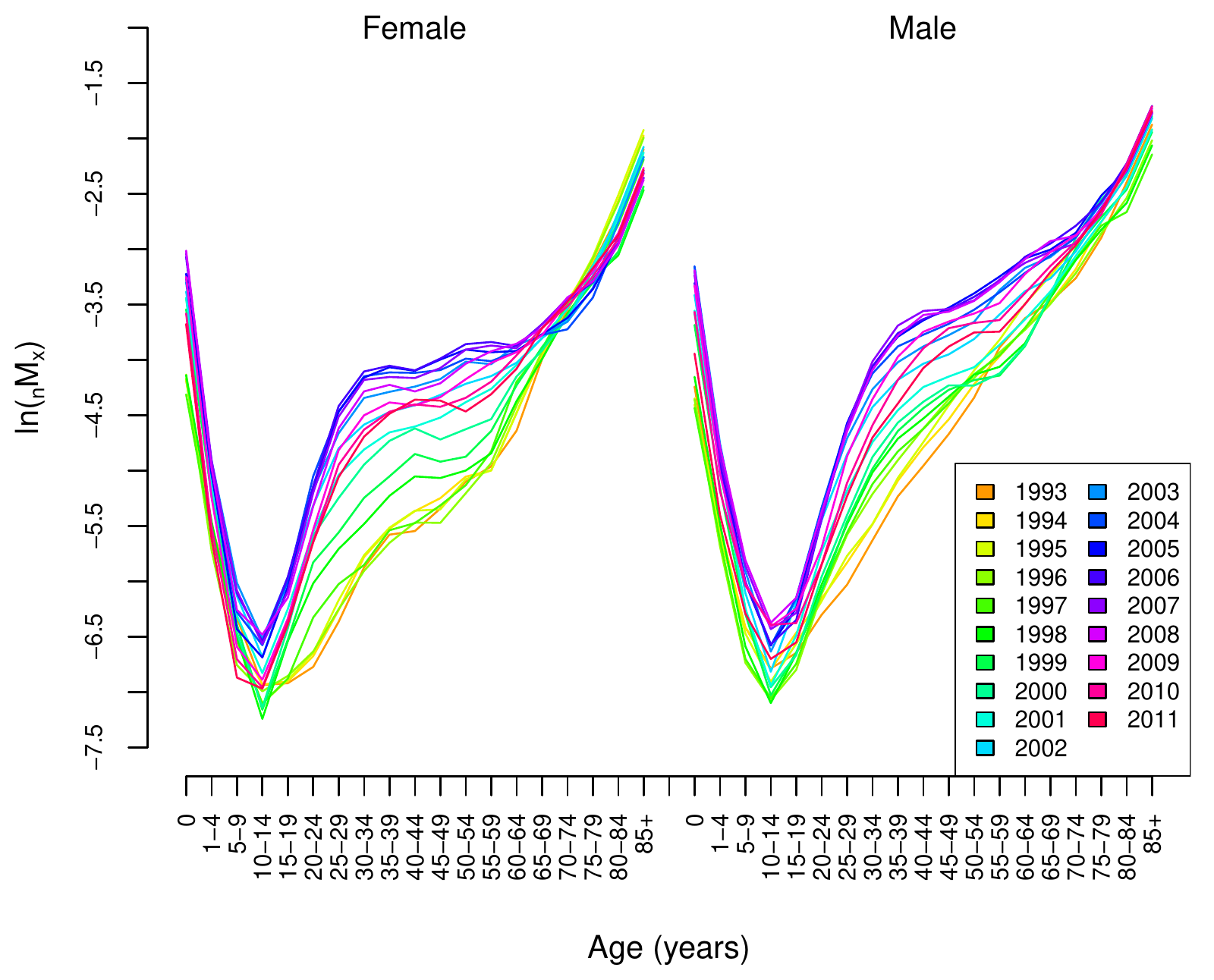} &
\includegraphics[width=0.48\textwidth]{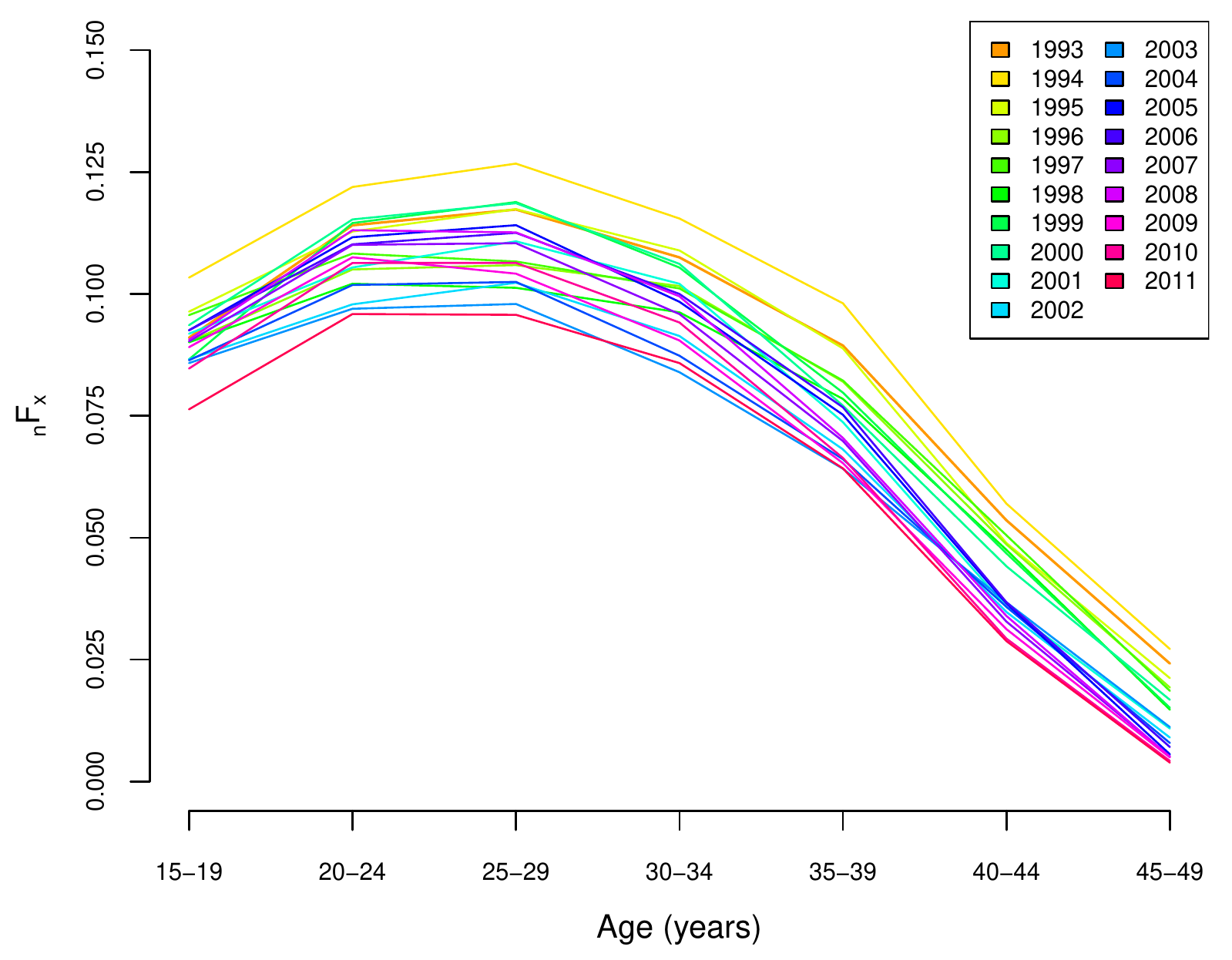}
\end{tabular}
\captionsetup{format=plain,font=small,margin=0cm,justification=justified}
\caption{{\bf Example Data from Agincourt.} \textbf{(A)}: Single-year, age-specific log mortality from 1993--2011.  These data have been smoothed to make demonstration of the method easier to follow and understand.  The counts of births and female person years used to calculate these rates were smoothed in both age and time using a kernel smoother. \textbf{(B)}: Single-year, age-specific fertility from 1993--2011.  These data have been smoothed to make demonstration of the method easier to follow and understand.  The counts of births and person-years used to calculate these rates were smoothed in both age and time using a kernel smoother.}
\label{fig:aginData}
\end{center}
\end{figure}

The models we develop below will be demonstrated using mortality and fertility data from the Agincourt HDSS in South Africa.  The Agincourt HDSS has monitored roughly 90,000 people for 22 years between 1992 and the present \cite{kahn2012agin}.  The entire population of the study area is included in the study, and each household is visited annually to update records on vital events, migrations and a variety of other topics. These records allow us to categorize aggregate observed person-time at risk of death and the counts of births and deaths by time, sex and age, and in the case of births, age of the mother as well.  This allows us to calculate age-specific event-exposure rates for mortality and fertility through time.  Tables \ref{app:exampleMx}.\ref{tab:aginMx3}, \ref{app:aginMx}.\ref{tab:aginMxAll1}, \ref{app:aginMx}.\ref{tab:aginMxAll2}, \ref{app:aginFx}.\ref{tab:aginFxAll1} and \ref{app:aginFx}.\ref{tab:aginFxAll2} in appendices \ref{app:exampleMx} -- \ref{app:aginFx} contain the data used in the examples, and Figure \ref{fig:aginData} displays the data.  The purpose of the examples is to illustrate and demonstrate the structure and various uses of the general model, \textit{not} to extract substantive conclusions about the age-profiles of mortality and fertility at Agincourt. In keeping with that aim, the data have been smoothed in both time and age in order to remove distracting stochastic features and to prevent any meaningful substantive interpretation. The time-sex-age-specific counts of deaths, births and all/female person years were smoothed in both age and time using a customized kernel smoother. The Agincourt data were chosen for this purpose because over the past fifteen years mortality has changed dramatically in response to the HIV epidemic in South Africa and this has produced a series of unusual and often difficult-to-model age patterns of mortality.  There is a publicly available version of the Agincourt mortality and fertility data available from the INDEPTH Network public data repositories (\url{http://www.indepth-ishare.org/} and \url{http://www.indepth-ishare.org/indepthstats/}).

\subsection{Component Model for Mortality}\label{sec:compModMx}

In this and following Sections \ref{sec:compModMx} -- \ref{sec:covsMx}  we define and demonstrate a component model of mortality.  We use the example mortality data listed in Tables \ref{app:aginMx}.\ref{tab:aginMxAll1} and \ref{app:aginMx}.\ref{tab:aginMxAll2} and displayed in Figure \ref{fig:aginData}.  The data consist of annual female and male age-specific mortality rates for ages 0, 1--4, 5--9, 10--14, $\ldots$, 85+ for years 1993--2011. We take the log of these mortality rates and concatenate the female and male age schedules for each year into a single 38-element vector.  This is done to ensure that time-specific features of mortality are coupled for females and males.  The data also contain covariates: HIV prevalence (\% of population), ART Coverage (\% of population), expectation of life at birth (years), adult mortality $\tensor[_{45}]{q}{_{15}}$, and child mortality $\tensor[_{5}]{q}{_{0}}$.  Using the component model for mortality, we are able to predict mortality age schedules using these covariates.

\begin{figure}[htbp]
\begin{center}
\begin{tabular}{cc}
A: Mortality & B: Fertility \\
\includegraphics[width=0.48\textwidth]{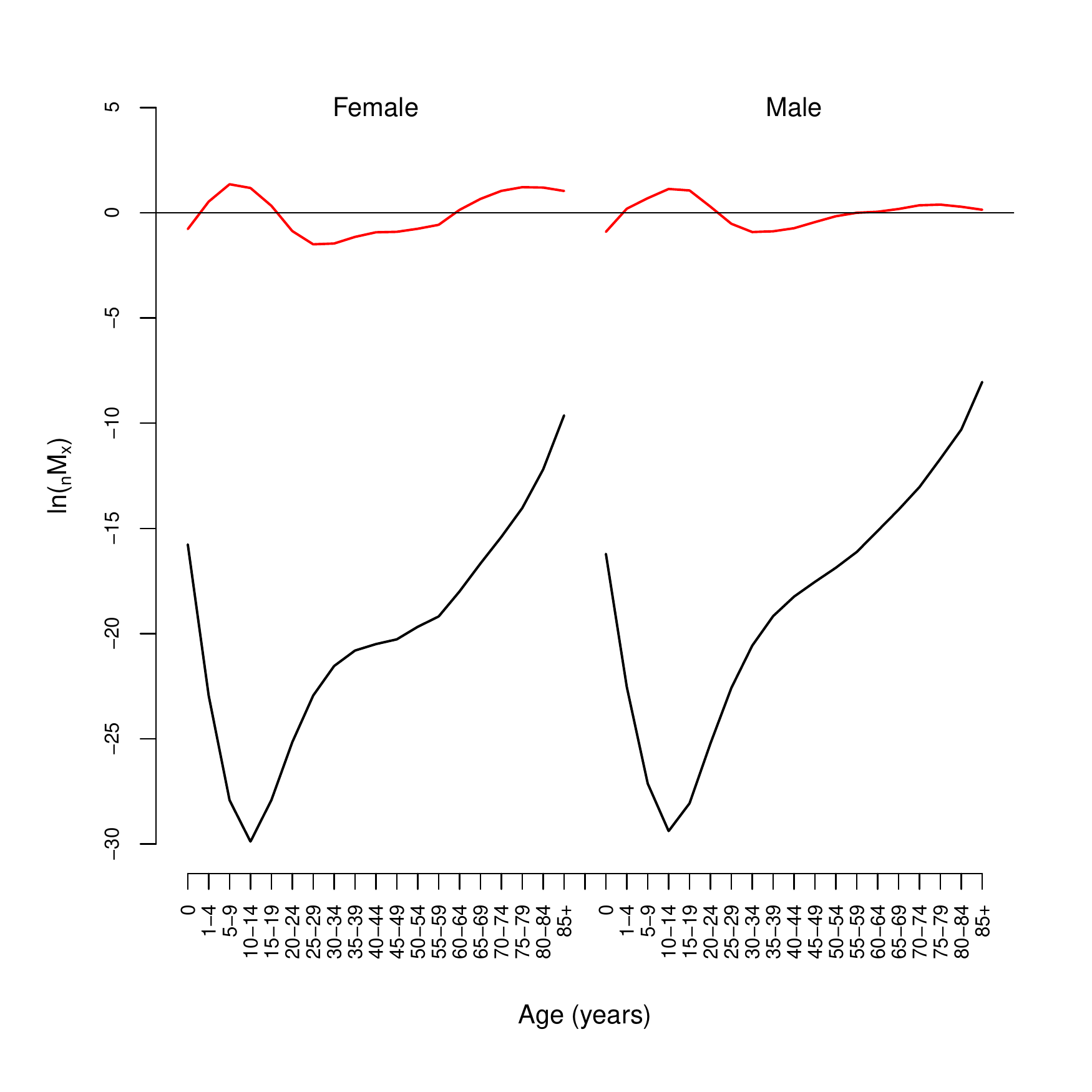} &
\includegraphics[width=0.48\textwidth]{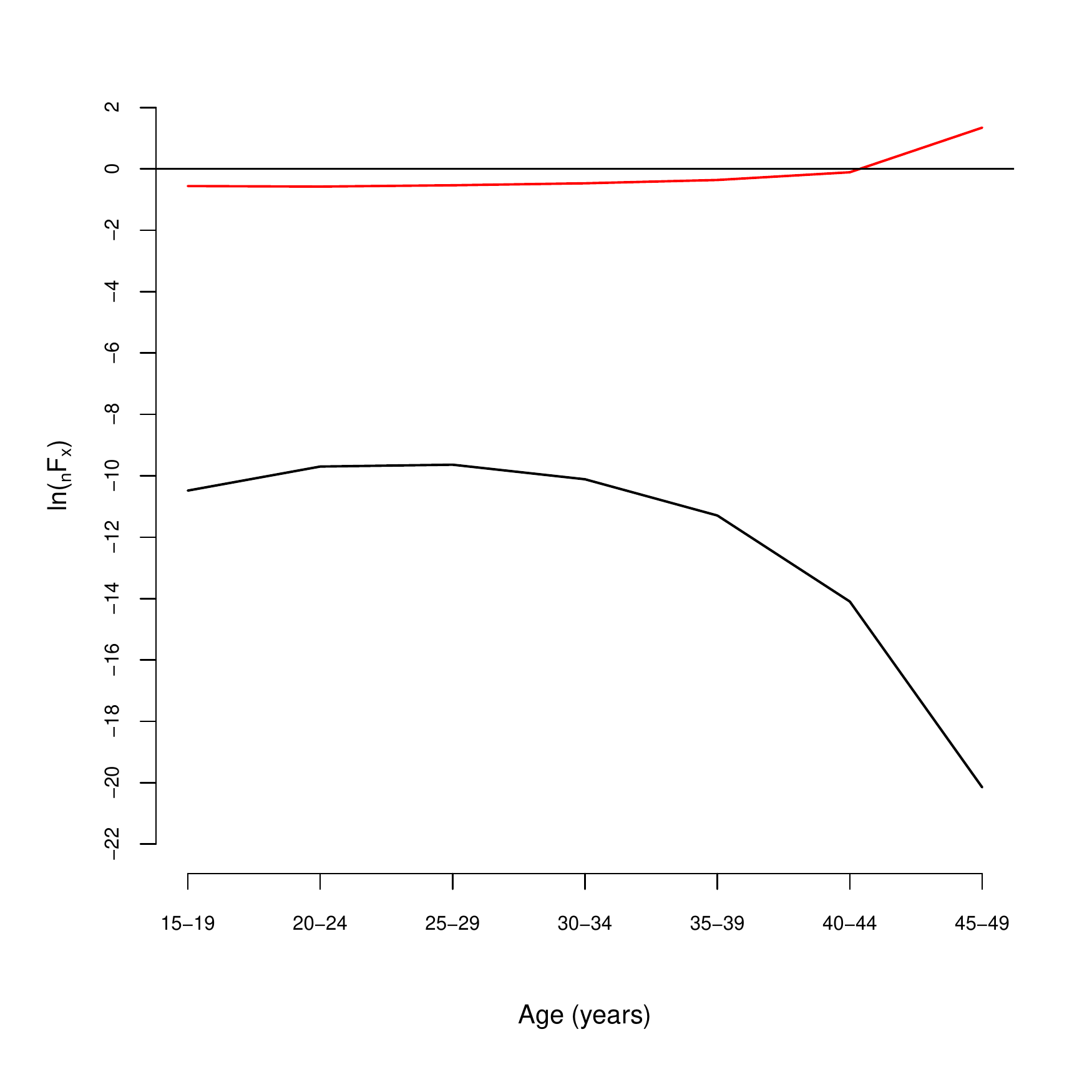}
\end{tabular}
\captionsetup{format=plain,font=small,margin=0cm,justification=justified}
\caption{{\bf Two Components of the Dimension-reduced Models of Mortality and Fertility at Agincourt.} \textsc{Panel} \textbf{(A)}: Mortality components. The black line is the first component -- the scaled first left singular vector $\tensor[_{\text{AM}}]{\mbf{\Lambda}}{_1}$ from the SVD decomposition of Agincourt log mortality rates.  The red line is the second component -- the scaled second left singular vector $\tensor[_{\text{AM}}]{\mbf{\Lambda}}{_2}$.  Notice that the first component is well below the x-axis, reflecting the fact that it `locates' the cloud of mortality points.  The second component crosses the x-axis because it fine-tunes the location of the mortality points within the cloud.  \textsc{Panel} \textbf{(B)}: Fertility components. Same as  \textsc{Panel} \textbf{(A)} for log fertility rates.}
\label{fig:Us}
\end{center}
\end{figure}

The component model of mortality follows the general SVD-based component model in Equation \ref{eq:genModel}.  For these example data we calculate the SVD of the $38 \times 19$ matrix $\mbf{AM}$ of concatenated female-male log mortality rates (38 sex-age groups and 19 calendar years). Following the description of the general model in Section \ref{sec:theModel}, we use the left singular vectors and singular values to construct mortality components $\tensor[_{\text{AM}}]{\mbf{\Lambda}}{_i} =  \tensor[_{\text{AM}}]{s}{_i} \cdot \tensor[_{\text{AM}}]{\mbf{u}}{_i}$.  The first four singular values $\tensor[_{\text{AM}}]{s}{_i} \ , \ i \in \{1,2,3,4\}$ are 123.8, 5.1, 1.7, and 1.2, with the remaining singular values $< 1.1$.  Consequently the new dimensions associated with the first four right singular vectors account for 99.8\%, 0.2\%, 0.02\%, and 0.009\% respectively of the total sum of squared perpendicular distances to all of the 38 points in the data set (see Section \ref{sec:sumRank1Mats}).  This indicates that the first two new dimensions effectively account for all of the variation in the original data (the remaining variation is lost in rounding error when presenting the results with a readable number of significant figures).  Consequently we adopt the following dimension-reduced model with two components, 
\begin{align}
\tensor[_{\text{AM}}]{\widehat{\mbf{m}}}{_{t}} &= \sum_{i = 1}^{2} \tensor[_{\text{AM}}]{\beta}{_{i,t}} \cdot \tensor[_{\text{AM}}]{\mbf{\Lambda}}{_i} \ , \label{eq:compModMx} \\
\tensor[]{\mbf{m}}{_{t}} &= \tensor[_{\text{AM}}]{\widehat{\mbf{m}}}{_{t}}  + \tensor[_{\text{AM}}]{\mbf{r}}{_{t}} \ . \nonumber
\end{align}
where $\tensor[_{\text{AM}}]{\widehat{\mbf{m}}}{_{t}}$ is the predicted sex-age mortality schedule for year $t$; $\tensor[_{\text{AM}}]{\beta}{_{i,t}}$ are the weights applied to the first two components $\tensor[_{\text{AM}}]{\mbf{\Lambda}}{_i}$ (left singular vectors scaled by their corresponding singular values); and $\tensor[_{\text{AM}}]{\mbf{r}}{_{t}}$ is the difference between the predicted and `real' sex-age mortality schedule, possible to calculate when $t$ is one of the years included in the $\mbf{AM}$ but otherwise an unknown residual when the model is used to predict a mortality schedule not included in $\mbf{AM}$.  Equation \ref{eq:compModMx} is a two-parameter model for sex-age schedules of mortality covering the years 1993--2011.  The two components $\tensor[_{\text{AM}}]{\mbf{\Lambda}}{_i} \ , \ i \in \{1,2\}$ are plotted in Figure \ref{fig:Us}.

The result of predicting the annual mortality schedules using Equation \ref{eq:compModMx} and the first and second weights from the SVD of $\mbf{AM}$ (the right singular vector weights as in Equation \ref{eq:genModelVs}) are displayed in Figures \ref{app:predMx}.\ref{fig:aginPredsNoCovars1} -- \ref{app:predMx}.\ref{fig:aginPredsNoCovars4} in Appendix \ref{app:predMx}.  The fits are very close; the total mean absolute error (MAE) is 0.080 ($\vert \widehat{m}-m\vert$ across both sexes and all years), and the five-number summary\footnote{$1^\text{st}$, $25^\text{th}$, $50^\text{th}$, $75^\text{th}$ and $99^\text{th}$ quantile.} of the distribution of absolute errors is (0.0013, 0.0322, 0.0651, 0.1100, 0.3000), Table \ref{tab:5nums}. The predictions are compared to the real values in a scatterplot in Figure \ref{fig:predErrorScatters}, Panel A.  

Knowing what we do about the SVD applied to demographic quantities correlated by age, this is not surprising.  The real value in our model is the ability to interpolate and extrapolate, and we demonstrate that in the next section using various covariates.

\subsubsection{Mortality Age Schedule Covariates and Predictors}\label{sec:covsMx}

As described in Section \ref{sec:paramsCovars}, a useful feature of the component model is the ability to relate covariates to age schedules through the weights.  The basic idea is to model the weights in terms of the covariates and then use that relationship to predict the weights and hence the age schedules.  Here we demonstrate this capability with the covariates listed in Tables \ref{app:aginMx}.\ref{tab:aginMxAll1} and \ref{app:aginMx}.\ref{tab:aginMxAll2}.  Life expectancy $\tensor[^{0}]{e}{}$ is a measure of overall level of mortality, HIV prevalence is the fraction of the population infected, ART coverage is the fraction of the population receiving antiretroviral therapy, child mortality $\tensor[_{5}]{q}{_{0}}$ is the probability of dying between birth and age 5, and adult mortality $\tensor[_{45}]{q}{_{15}}$ is the probability of dying between ages 15 and 60, conditional on surviving to age 15.  The fraction of the population that is HIV${^\text{+}}$ but not on ART is the group of people who die as a result of HIV and is referred to as $\Delta = (\text{HIV prevalence} - \text{ART coverage})$ from now on. 

\begin{figure}[htbp]
\begin{center}
\begin{tabular}{cc}
\includegraphics[width=0.48\textwidth]{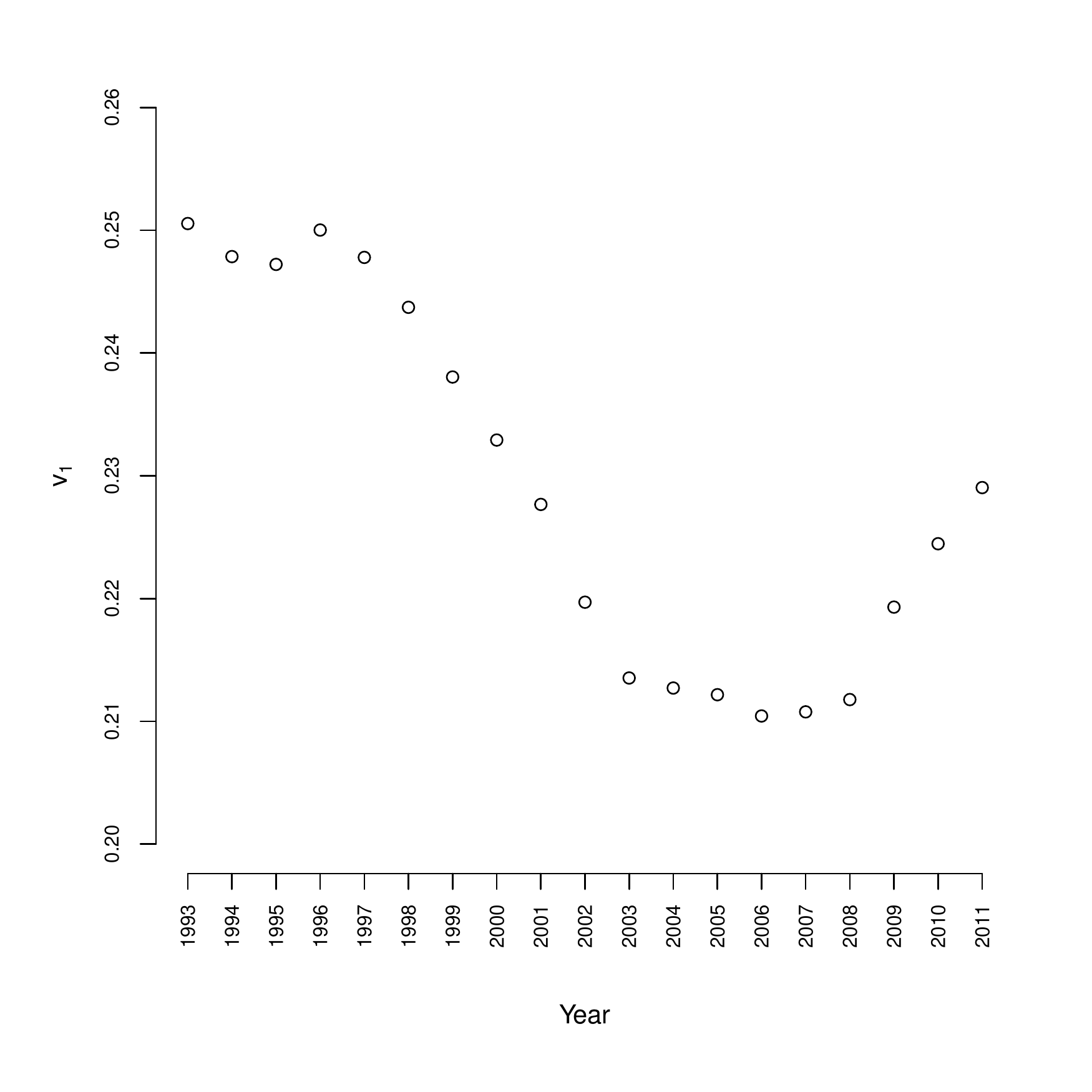} &
\includegraphics[width=0.48\textwidth]{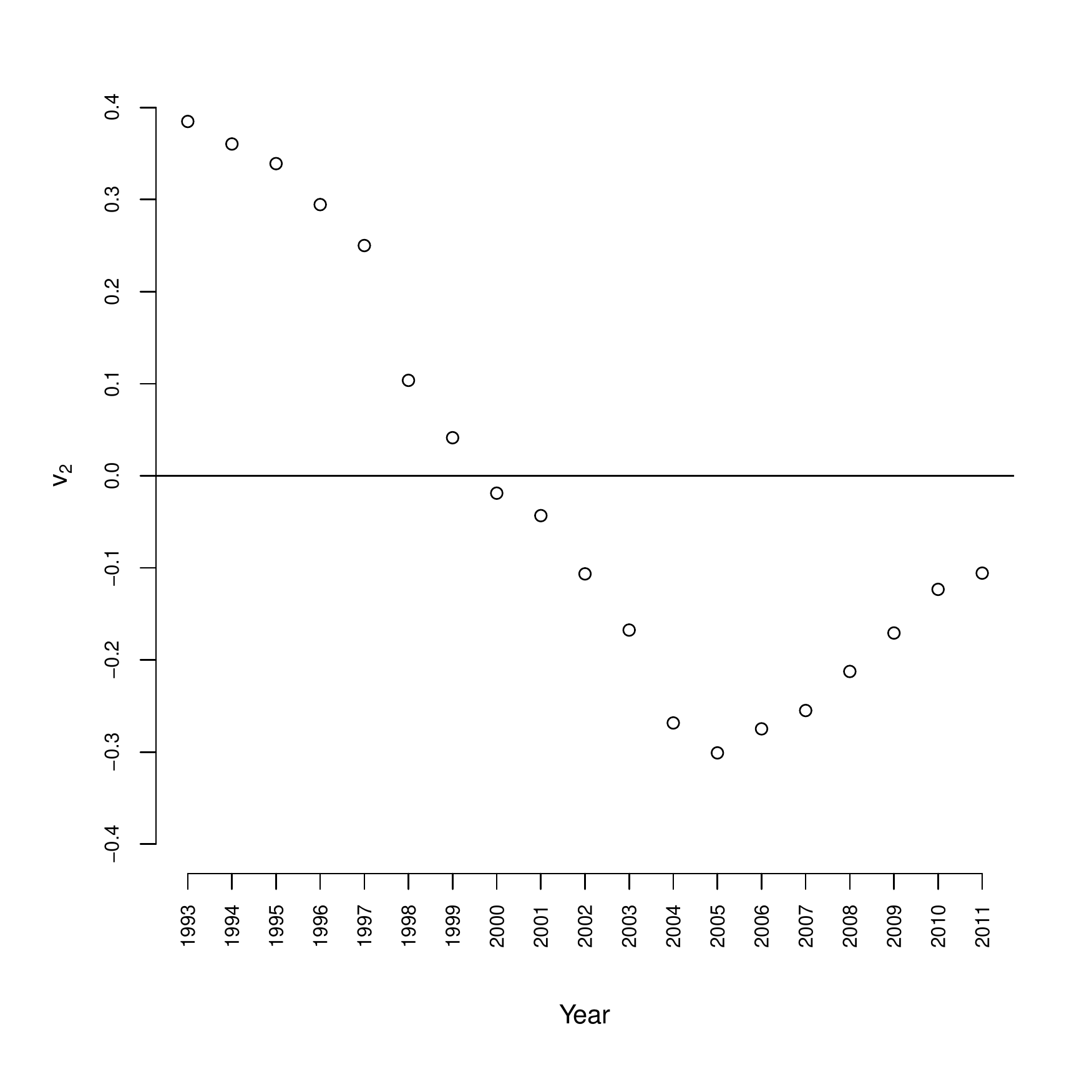}
\end{tabular}
\captionsetup{format=plain,font=small,margin=0.5cm,justification=justified}
\caption{{\bf Right Singular Vectors of Agincourt Log Mortality Rates by Year.} First and second right singular vectors $v_{1}$ and $v_{2}$.}
\label{fig:vsMx}
\end{center}
\end{figure}

\begin{figure}[htbp]
\begin{center}
\begin{tabular}{cc}
\includegraphics[width=0.48\textwidth]{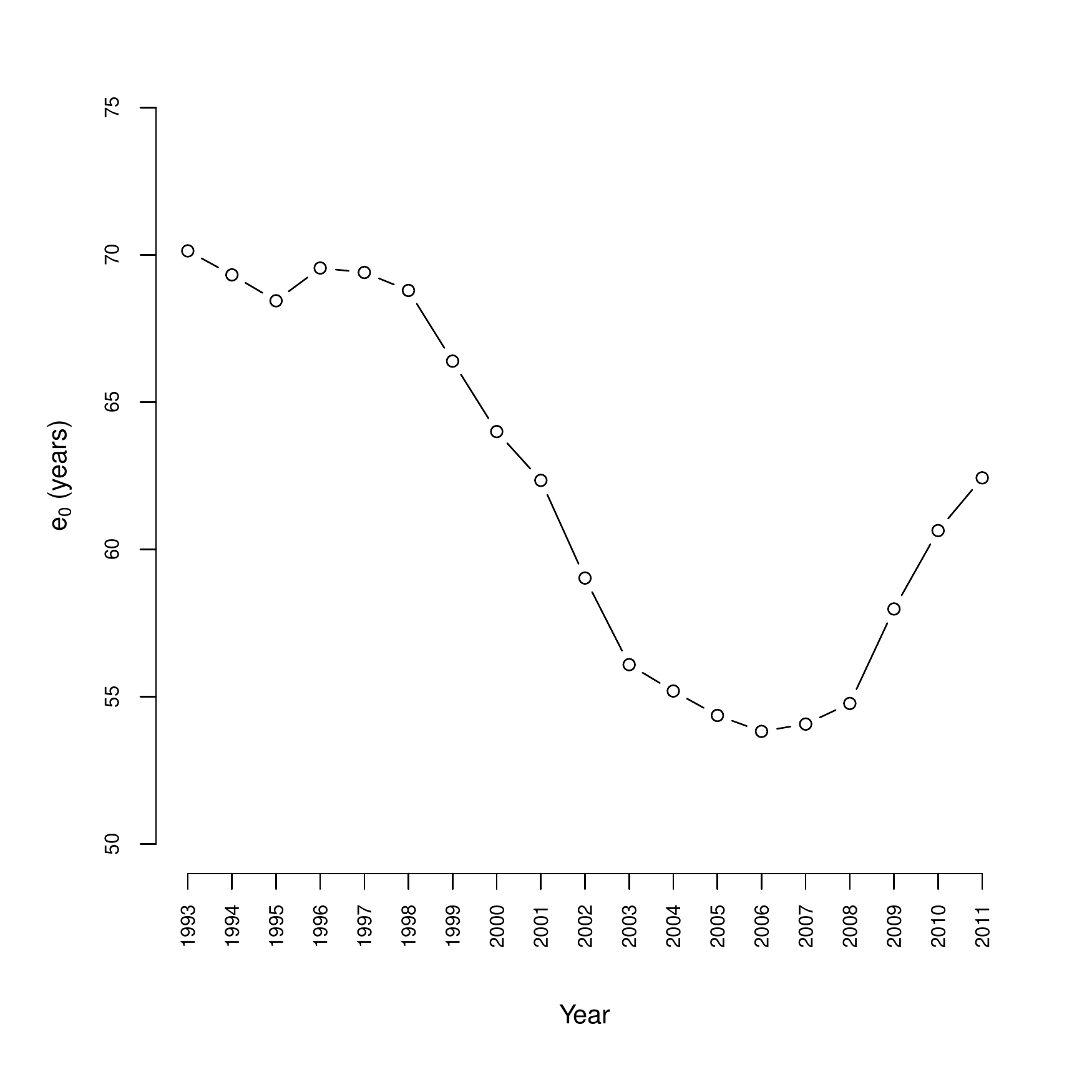} &
\includegraphics[width=0.48\textwidth]{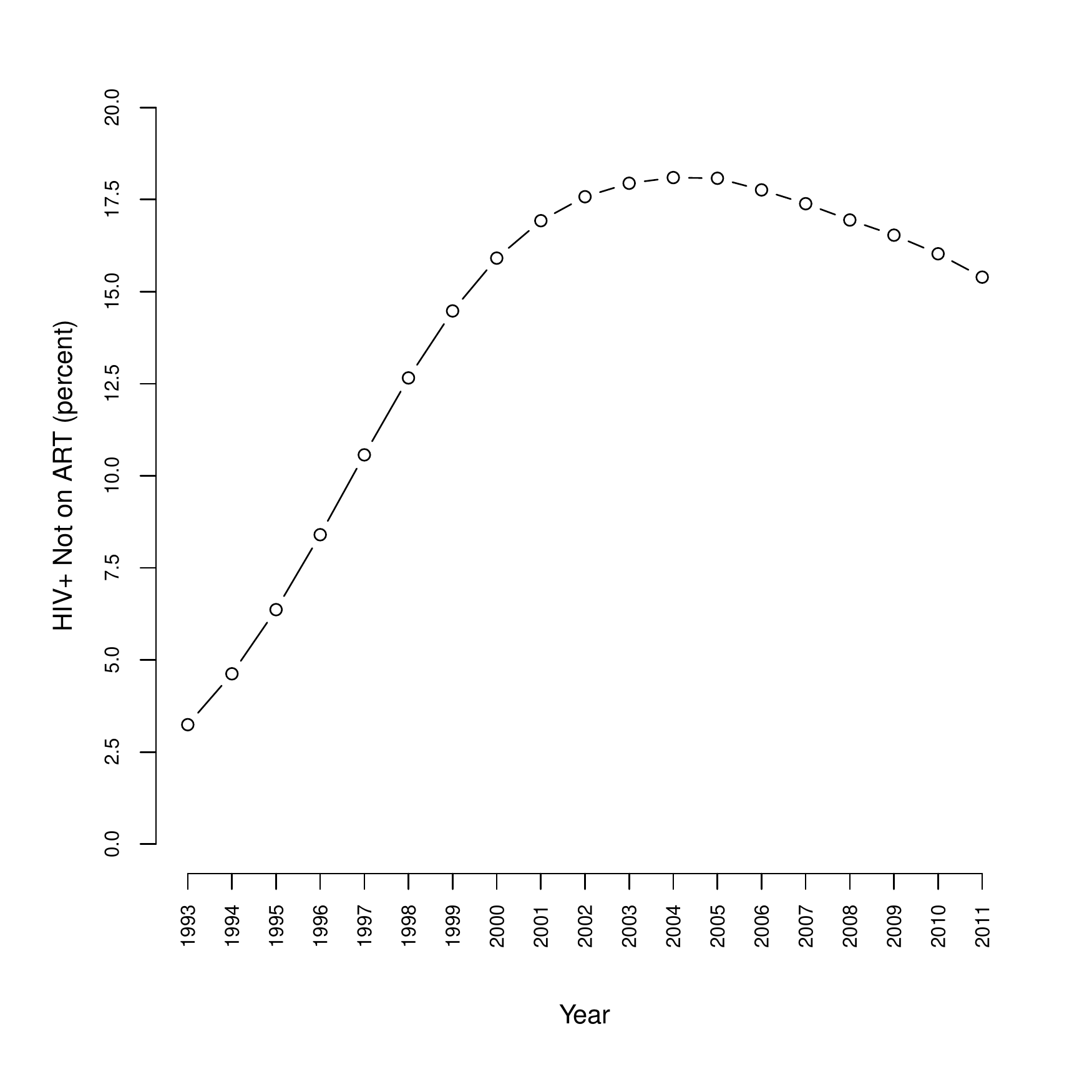}
\end{tabular}
\captionsetup{format=plain,font=small,margin=0.5cm,justification=justified}
\caption{{\bf  Agincourt Life Expectancy at Birth $\tensor[^{0}]{e}{}$ and Prevalence of Persons Who are HIV$^+$ and Not on ART $\Delta$ by Year.}}
\label{fig:e0noARTbyYear}
\end{center}
\end{figure}

\begin{figure}[htbp]
\begin{center}
\begin{tabular}{cc}
\includegraphics[width=0.48\textwidth]{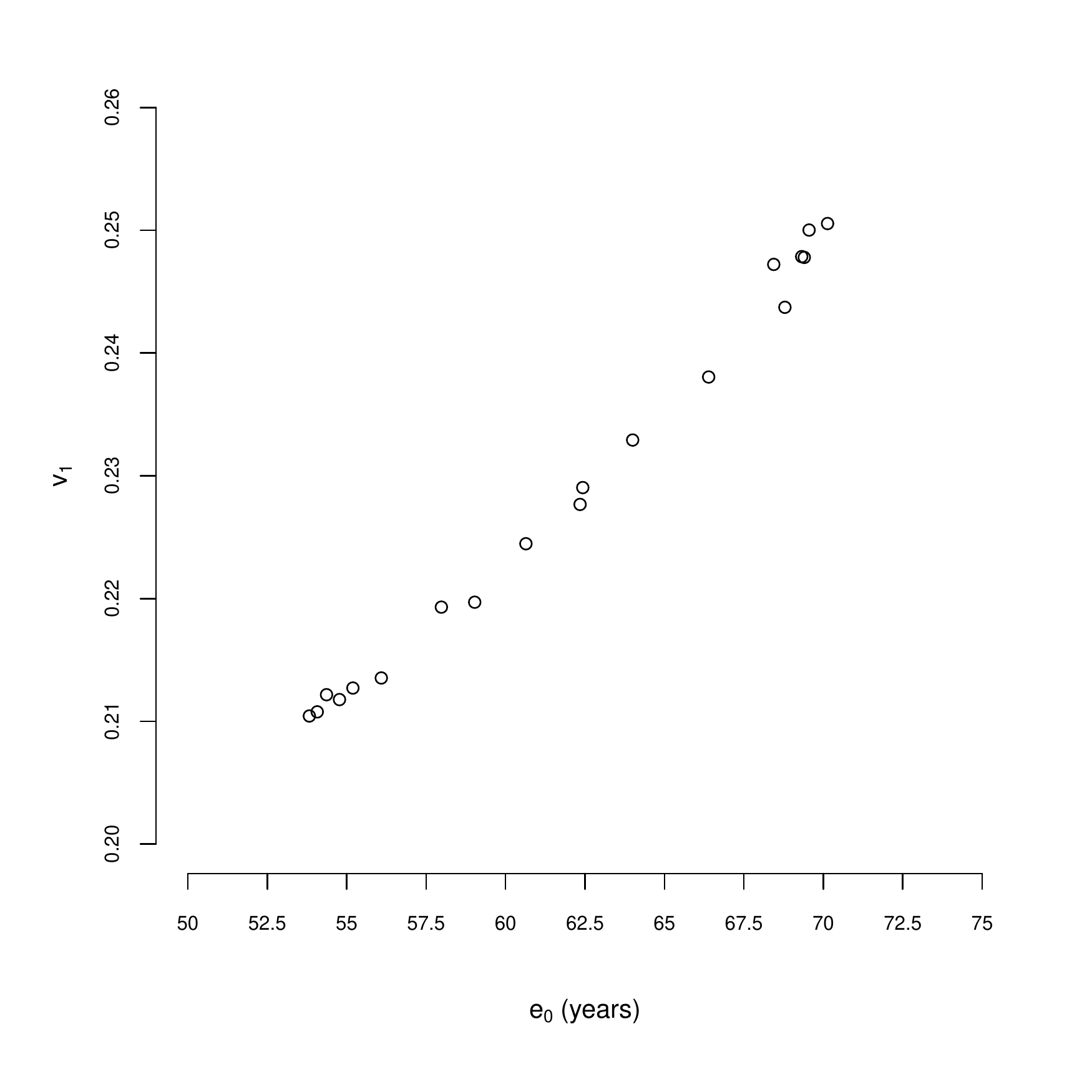} &
\includegraphics[width=0.48\textwidth]{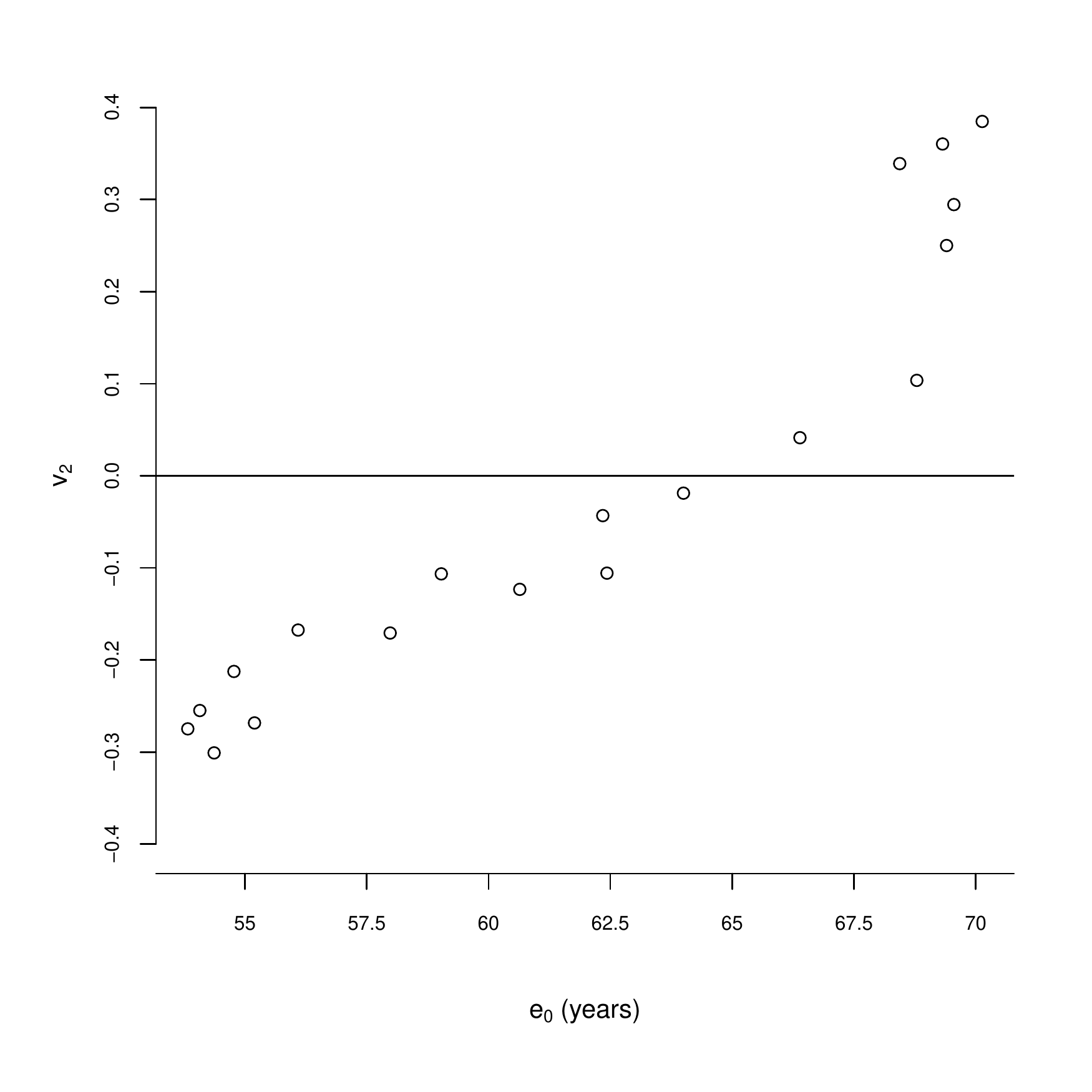}
\end{tabular}
\captionsetup{format=plain,font=small,margin=0.5cm,justification=justified}
\caption{{\bf Agincourt Age-specific Log Mortality Rate Right Singular Vectors $\mbf{v}_1$ and $\mbf{v}_2$ by Life Expectancy at Birth $\tensor[^{0}]{e}{}$.}}
\label{fig:mxVsByCovarsE0}
\end{center}
\end{figure}

\begin{figure}[htbp]
\begin{center}
\begin{tabular}{cc}
\includegraphics[width=0.48\textwidth]{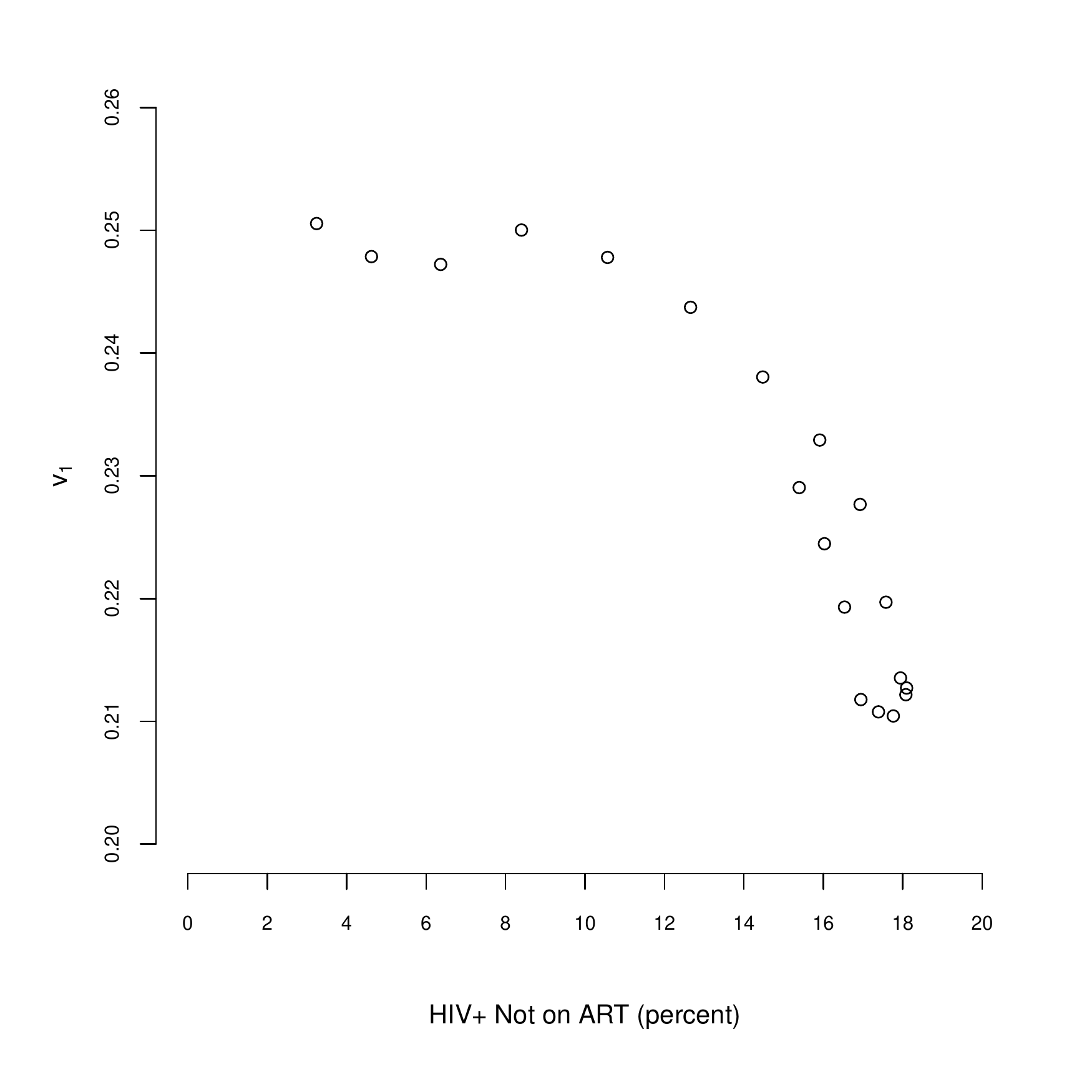} &
\includegraphics[width=0.48\textwidth]{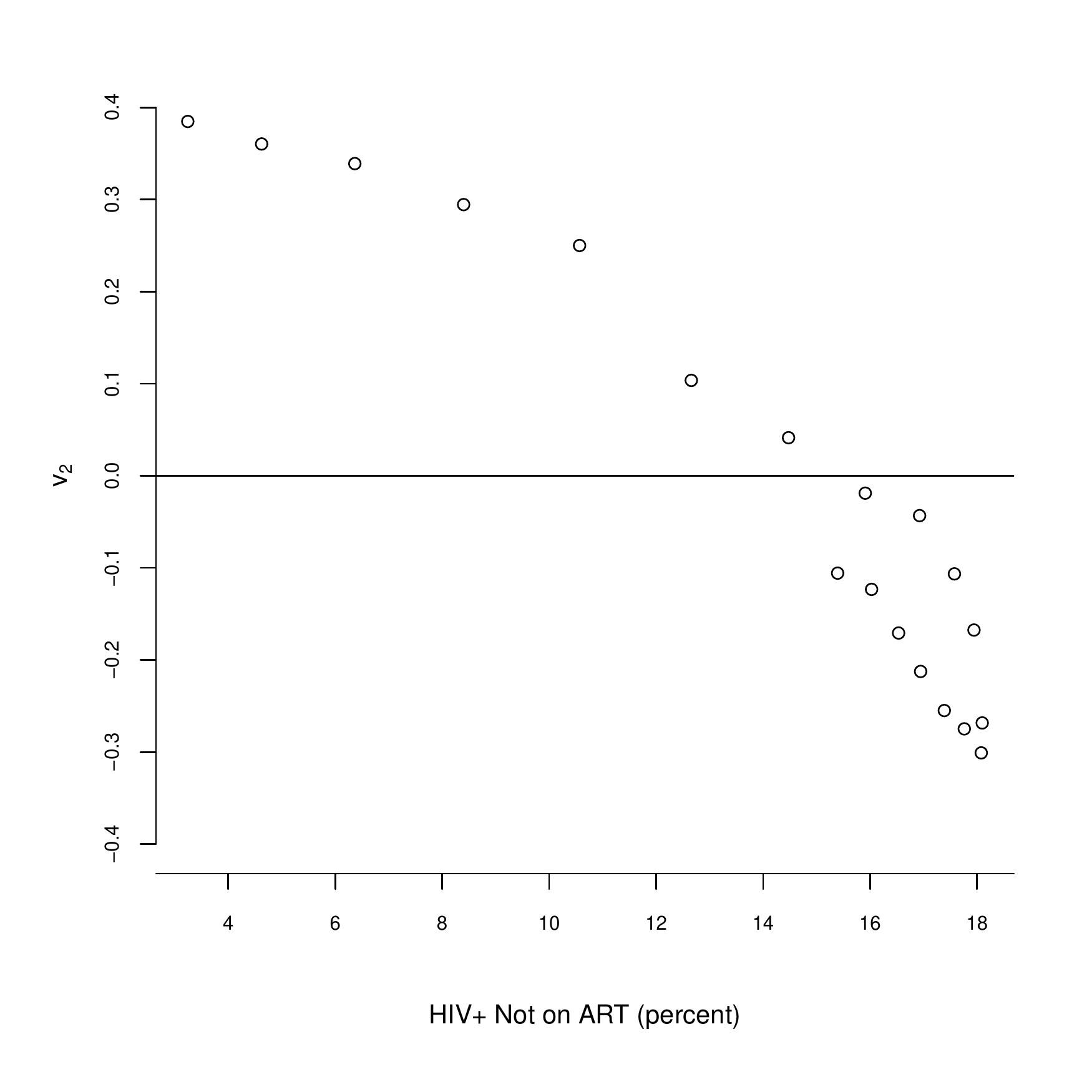}
\end{tabular}
\captionsetup{format=plain,font=small,margin=0.5cm,justification=justified}
\caption{{\bf Agincourt Age-specific Log Mortality Rate Right Singular Vectors $\mbf{v}_1$ and $\mbf{v}_2$ by Prevalence of Persons Who are HIV$^+$ and Not on ART $\Delta$.}}
\label{fig:mxVsByCovarsNoART}
\end{center}
\end{figure}

First we have a look at the weights associated with each age schedule that emerge from the SVD of the mortality schedules.  In our two-dimension model, these are the first two right singular vectors $\mbf{v}_i \ , \ i \in \{1,2\}$ ($\mbf{v}_1$ for the first component and $\mbf{v}_2$ for the second component), displayed by year in Figure \ref{fig:vsMx}.  Figure \ref{fig:e0noARTbyYear} displays the time trends in $\tensor[^{0}]{e}{}$ and $\Delta$.  It is clear that both $\mbf{v}_1$ and $\mbf{v}_2$ are strongly related $\tensor[^{0}]{e}{}$ in a positive, linear sense and related to $\Delta$ in a negative linear sense. These relationships are displayed in Figures \ref{fig:mxVsByCovarsE0} and \ref{fig:mxVsByCovarsNoART} which confirm that they are both approximately linear.  Based on this we used OLS regression to estimate two linear models that relate the $\mbf{v}$'s to the covariates through time $t$,
\begin{align}
\mbf{v}_{1,t} &= \tensor[_{v1}]{c}{} + \tensor[_{v1}]{\beta}{_{1}} \tensor[^{0}]{e}{}_{t} + \tensor[_{v1}]{\beta}{_{2}} \Delta_{t} + \tensor[_{v1}]{\epsilon}{_{t}} \ ,\phantom{.} \label{eq:v1lm} \\
\mbf{v}_{2,t} &= \tensor[_{v2}]{c}{} + \tensor[_{v2}]{\beta}{_{1}} \tensor[^{0}]{e}{}_{t} + \tensor[_{v2}]{\beta}{_{2}} \Delta_{t} + \tensor[_{v2}]{\epsilon}{_{t}} \ . \label{eq:v2lm}
\end{align}
The estimates are displayed in Tables \ref{tab:lmV1mx} and \ref{tab:lmV2mx}.

\begin{table}[ht]
\centering
\captionsetup{format=plain,font=small,margin=3cm,justification=justified}
\caption{Estimates for Equation \ref{eq:v1lm}: $R^2 = 0.9961$}
\begin{tabular}{lrrrr}
  \hline
 & Estimate & Std. Error & t value & Pr($>$$|$t$|$) \\ 
  \hline
$\tensor[_{v1}]{c}{}$ & 0.1024 & 0.0056 & 18.26 & 0.0000 \\ 
$\tensor[_{v1}]{\beta}{_{1}}$ & 0.0021 & 0.0001 & 29.73 & 0.0000 \\ 
$\tensor[_{v1}]{\beta}{_{2}}$ & -0.0005 & 0.0001 & -5.09 & 0.0001 \\ 
   \hline
\end{tabular}
\label{tab:lmV1mx}
\end{table}

\begin{table}[ht]
\centering
\captionsetup{format=plain,font=small,margin=3cm,justification=justified}
\caption{Estimates for Equation \ref{eq:v2lm}: $R^2 = 0.9779$}
\begin{tabular}{lrrrr}
  \hline
 & Estimate & Std. Error & t value & Pr($>$$|$t$|$) \\ 
  \hline
$\tensor[_{v2}]{c}{}$ & -0.8532 & 0.2046 & -4.17 & 0.0007 \\ 
$\tensor[_{v2}]{\beta}{_{1}}$ & 0.0192 & 0.0026 & 7.29 & 0.0000 \\ 
$\tensor[_{v2}]{\beta}{_{2}}$ & -0.0253 & 0.0034 & -7.53 & 0.0000 \\ 
   \hline
\end{tabular}
\label{tab:lmV2mx}
\end{table}

\begin{figure}[htbp]
\begin{center}
\begin{tabular}{ccc}
A: SVD & B: $\tensor[^{0}]{e}{}$ \& $\Delta$ & C: $\tensor[_{5}]{q}{_{0}}$ \\
\includegraphics[width=0.3\textwidth]{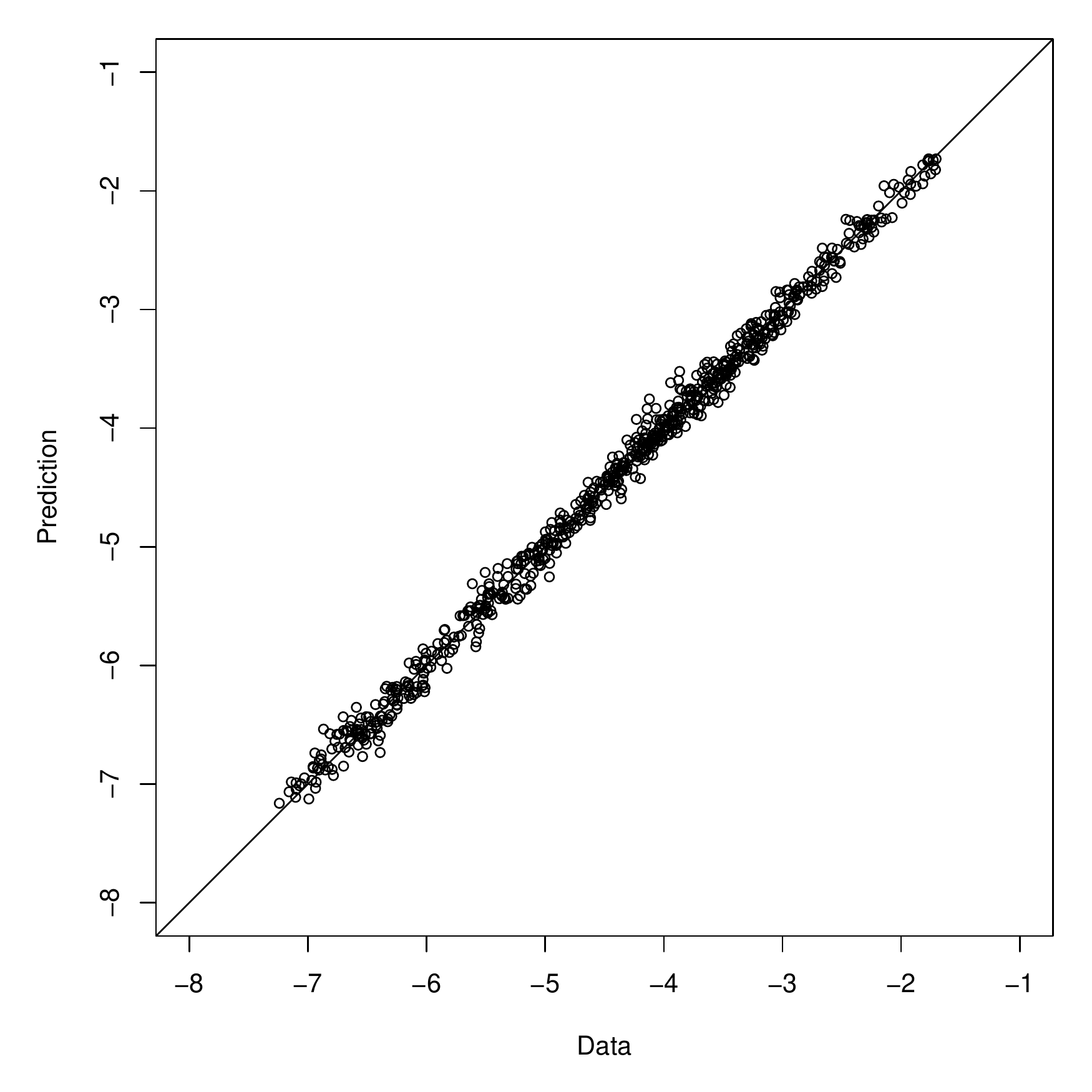} &
\includegraphics[width=0.3\textwidth]{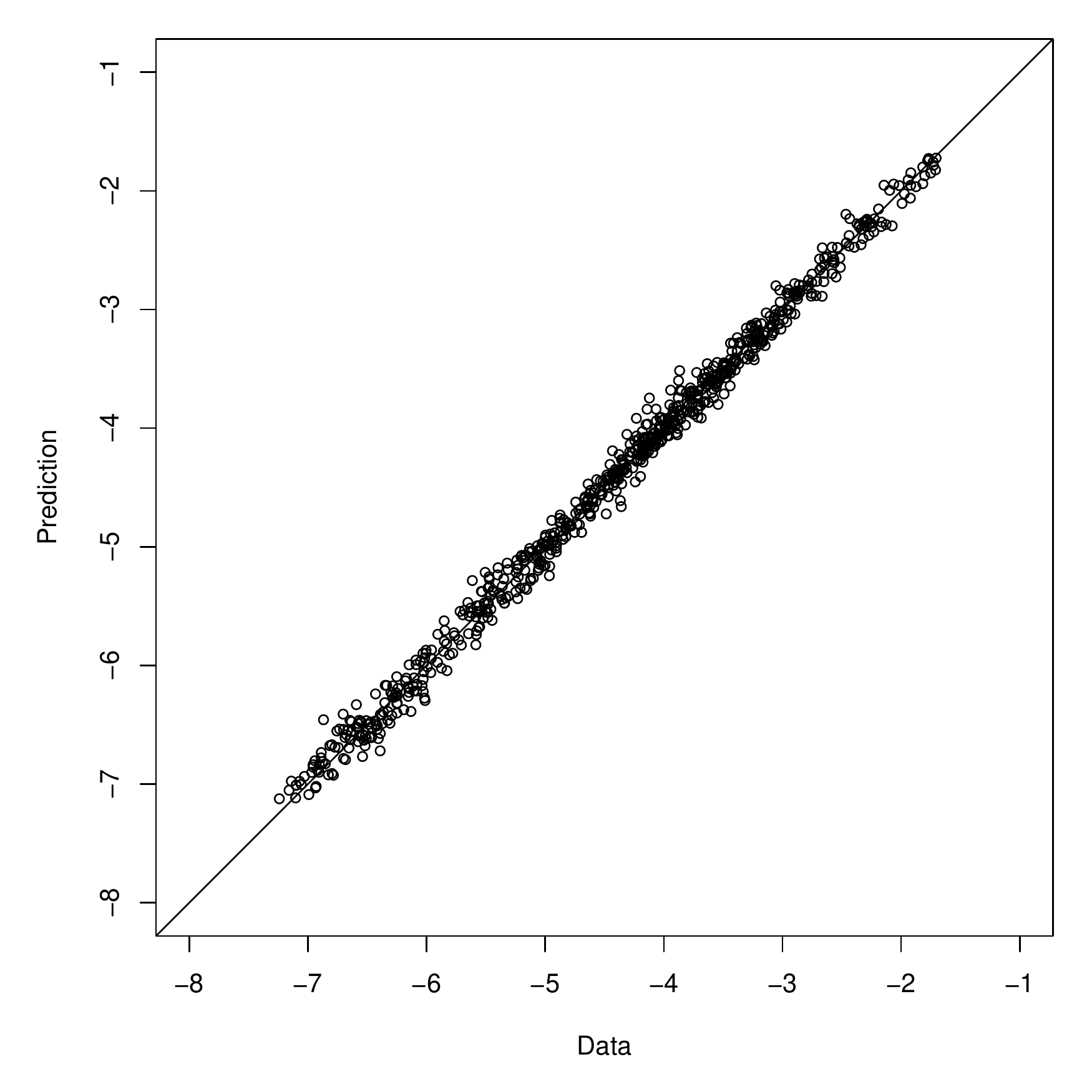} &
\includegraphics[width=0.3\textwidth]{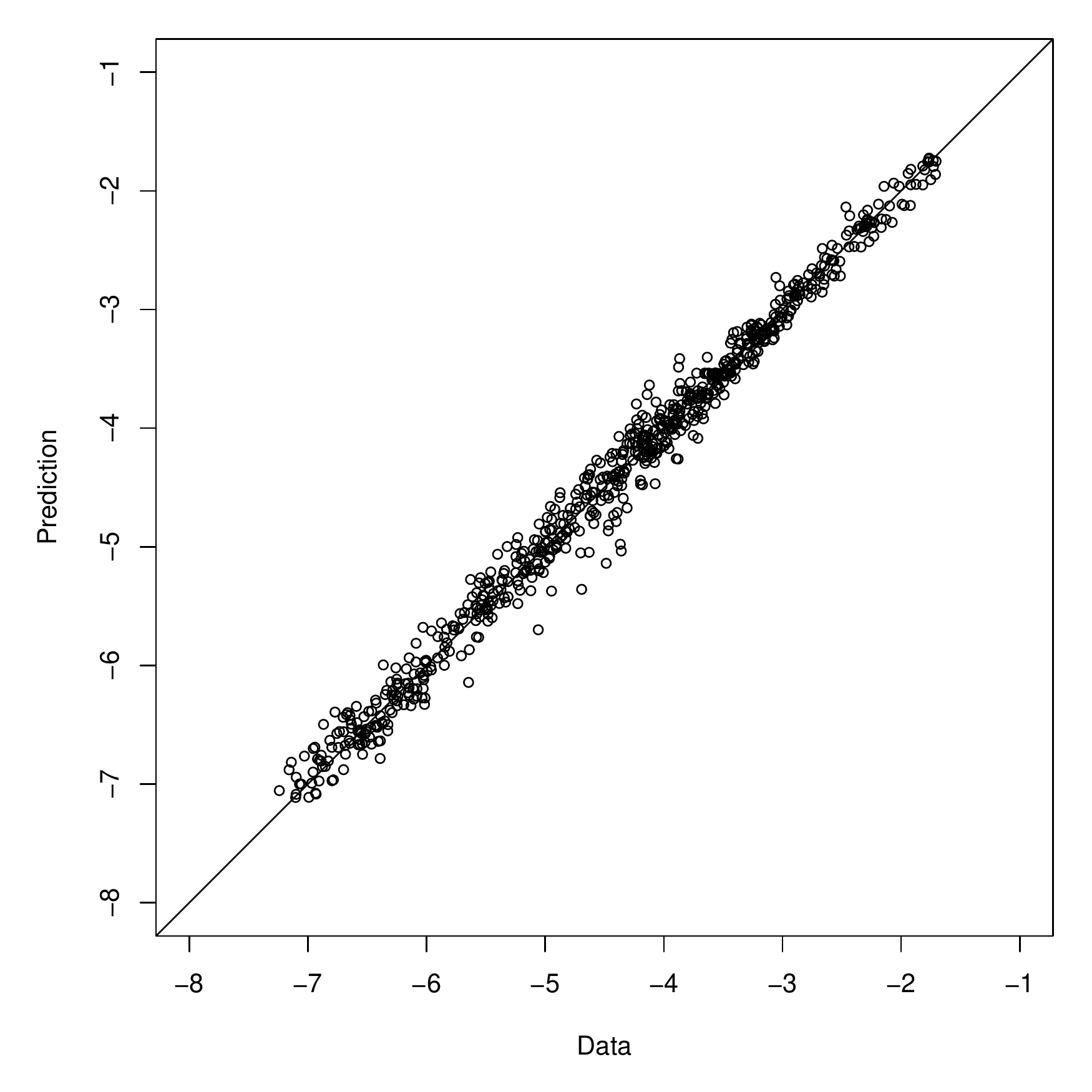} \\
D: $\tensor[_{45}]{q}{_{15}}$ & E: $\tensor[_{5}]{q}{_{0}}$ \& $\tensor[_{45}]{q}{_{15}}$ & F \\
\includegraphics[width=0.3\textwidth]{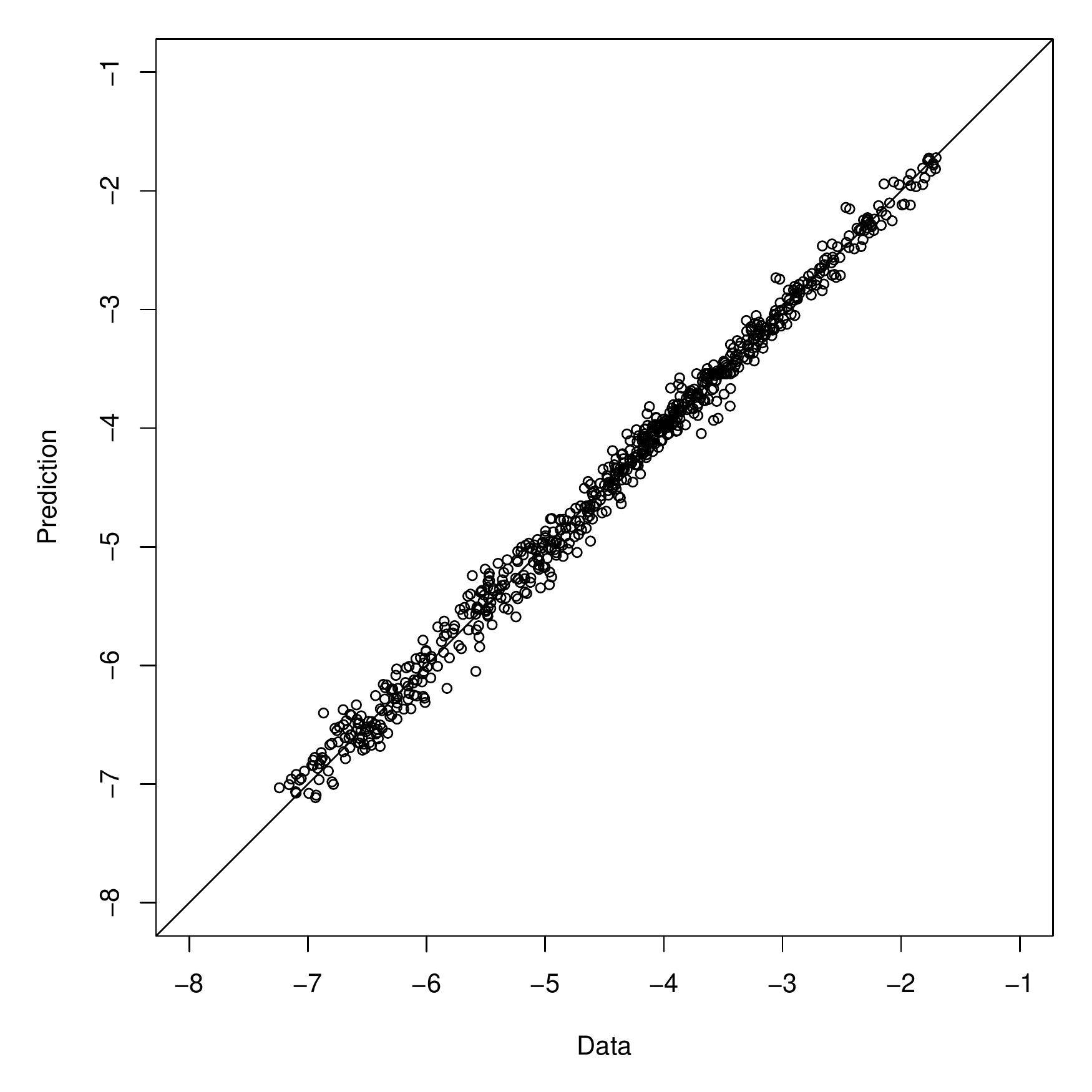} &
\includegraphics[width=0.3\textwidth]{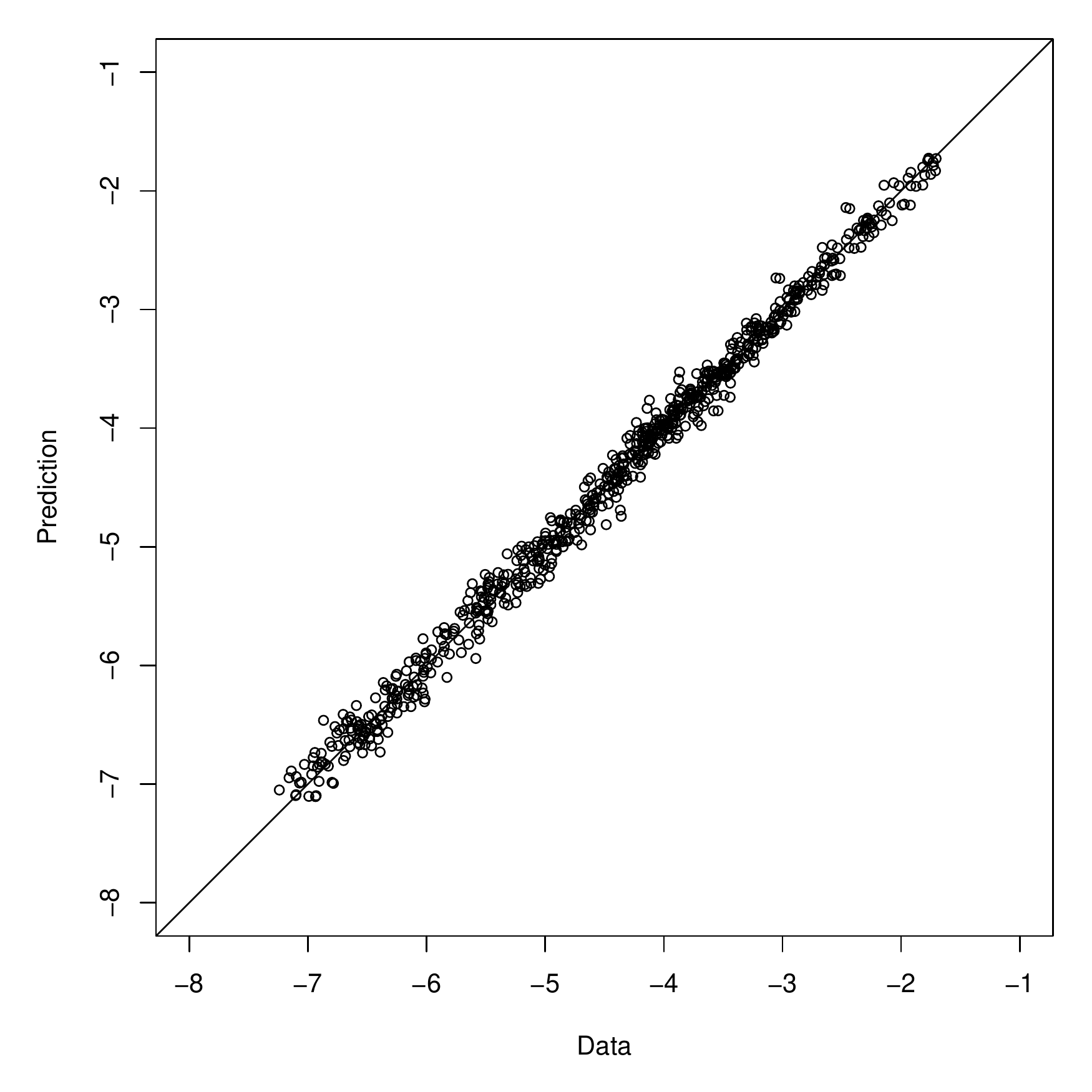} &
\includegraphics[width=0.3\textwidth]{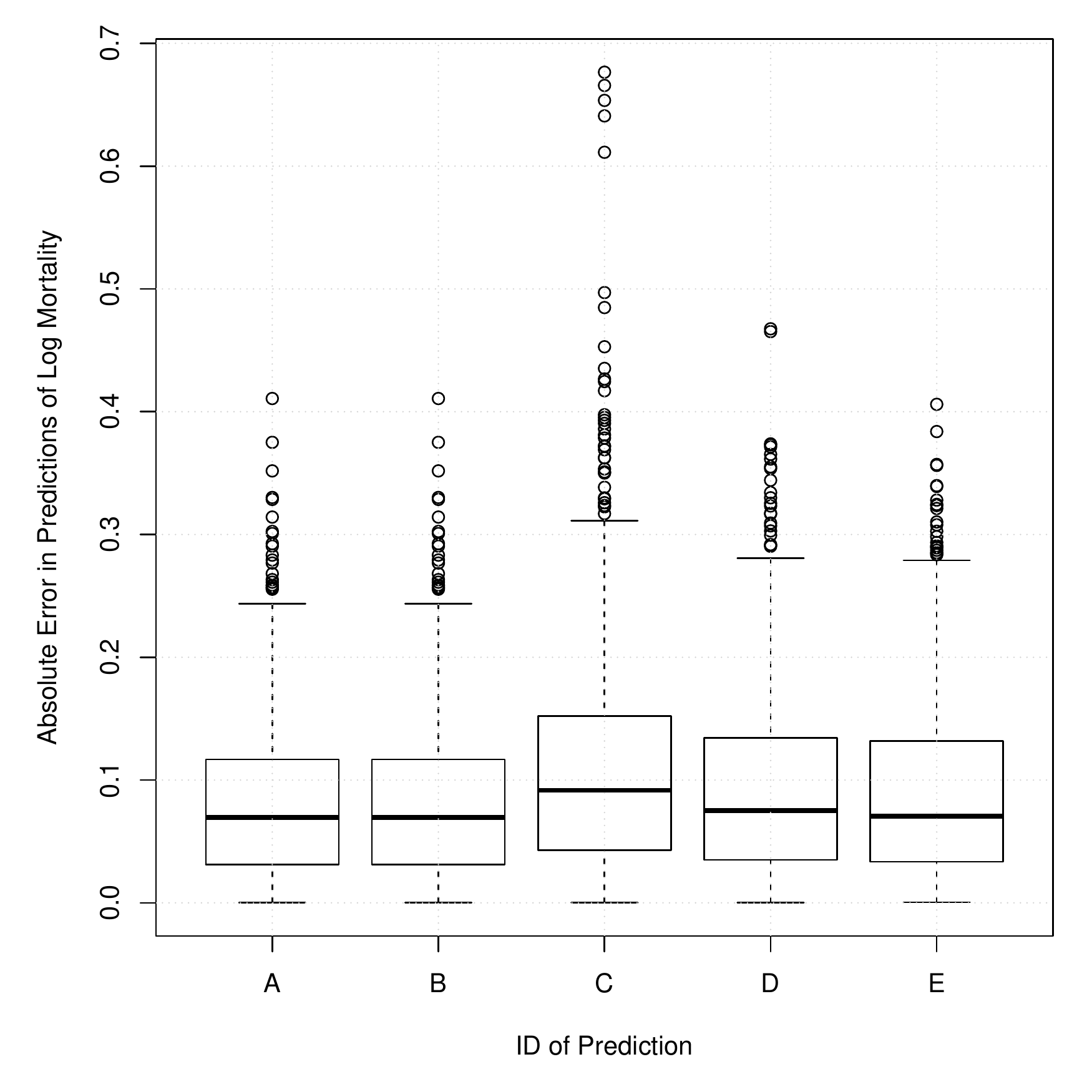}
\end{tabular}
\captionsetup{format=plain,font=small,margin=0.5cm,justification=justified}
\caption{{\bf Prediction Error in Agincourt Log Mortality Schedules: Scatterplots of Predictions vs. Data.} Each point in the scatterplots is a pair of time-sex-age-specific log mortality rates, 38 sex-age groups $\times$ 19 years = 722 points per plot. \textsc{Panel} \textbf{(A)}: Weights from the right singular vectors of the SVD of $\mbf{AM}$. \textsc{Panel} \textbf{(B)}: Weights predicted from $\tensor[^{0}]{e}{}$ and $\Delta$. \textsc{Panel} \textbf{(C)}: Weights predicted from child mortality $\tensor[_{5}]{q}{_{0}}$. \textsc{Panel} \textbf{(D)}: Weights predicted from adult mortality $\tensor[_{45}]{q}{_{15}}$. \textsc{Panel} \textbf{(E)}: Weights predicted from child $\tensor[_{5}]{q}{_{0}}$ and adult $\tensor[_{45}]{q}{_{15}}$ mortality.  \textsc{Panel} \textbf{(F)}: Boxplots of distributions of absolute error in predictions of Agincourt log Mortality. The letters on the horizontal axis correspond, in order, to the panels of this figure.}
\label{fig:predErrorScatters}
\end{center}
\end{figure}

These estimates can now be substituted into Equations \ref{eq:v1lm} and \ref{eq:v2lm} to create expressions that predict the $\mbf{v}$'s given values for $\tensor[^{0}]{e}{}$ and $\Delta$,
\begin{align}
\widehat{\mbf{v}}_{1,t} &= \phantom{-}0.1024 + 0.0021 \cdot \tensor[^{0}]{e}{}_{t} - 0.0005 \cdot \Delta_{t} \ , \label{eq:mxV1} \\
\widehat{\mbf{v}}_{2,t} &= -0.8532 + 0.0192 \cdot \tensor[^{0}]{e}{}_{t} - 0.0253 \cdot \Delta_{t} \ , \label{eq:mxV2} 
\end{align}
and these predicted $\mbf{v}$'s can in turn be substituted back into Equation \ref{eq:compModMx} to produce predicted sex-age schedules of mortality, net of the error $\mbf{r}$.  The results of doing this are plotted in Figures \ref{app:aginMx}.\ref{fig:aginPredsE01} through \ref{app:aginMx}.\ref{fig:aginPredsE04} in Appendix \ref{app:aginMx}.  As the figures make clear, the predicted values are very close to the real values.  The total MAE (calculated as before) for the predicted values is 0.083, and the five-number summary is (0.0022, 0.0312, 0.0697, 0.1169, 0.2990), Table \ref{tab:5nums} and  \textsc{Panel} \textbf{(F)} of Figure \ref{fig:predErrorScatters}.

\begin{table}[ht]
\captionsetup{format=plain,font=small,margin=2.5cm,justification=justified}
\centering
\caption{`Five Number' Quantiles of the Distributions of Absolute Prediction Errors for Log Sex-Age-Specific Agincourt Mortality.}
\begin{tabular}{lccccc}
  \hline
  Predictor(s) & 1\% & 25\% & 50\% & 75\% & 99\% \\ 
  \hline
  SVD & 0.001311 & 0.032202 & 0.065119 & 0.110022 & 0.299972 \\ 
  $\tensor[^{0}]{e}{}$ \& $\Delta$ & 0.002174 & 0.031208 & 0.069663 & 0.116879 & 0.299046 \\ 
  $\tensor[_{5}]{q}{_{0}}$ & 0.002252 & 0.043255 & 0.091623 & 0.152007 & 0.449184 \\ 
  $\tensor[_{45}]{q}{_{15}}$ & 0.001827 & 0.035096 & 0.075179 & 0.133854 & 0.354826 \\ 
  $\tensor[_{5}]{q}{_{0}}$ \& $\tensor[_{45}]{q}{_{15}}$ & 0.002496 & 0.033556 & 0.070406 & 0.131788 & 0.324306 \\ 
   \hline
\end{tabular}
\label{tab:5nums}
\end{table}



A similar exercise was conducted to predict sex-age schedules of mortality using child mortality $\tensor[_{5}]{q}{_{0}}$ alone, adult mortality $\tensor[_{45}]{q}{_{15}}$ alone and child and adult mortality together.  The predictions are displayed in Figures \ref{app:aginMx}.\ref{fig:aginPredsChild1} -- \ref{app:aginMx}.\ref{fig:aginPredsChildAdult4}, and the predictions and their errors are characterized in scatterplots in Figure \ref{fig:predErrorScatters}.  Similar to the predictions discussed already, in all cases the predictions are very close to the real values.  The distributions of absolute errors are described in Table \ref{tab:5nums} and  \textsc{Panel} \textbf{(F)} of Figure \ref{fig:predErrorScatters}.  The distributions are similar with medians around 0.065 and interquartile ranges around 0.85, except for child mortality which is only slightly worse.  This makes sense because child mortality covers only five years at one end of the age schedule and therefore does not contain as much information as the other predictors; nonetheless, it still does well.

\subsubsection{Identifying `Common' Age Schedules of Mortality}

The sequence or `pattern' of weights applied to the first few components in a SVD-based component model contain the information necessary to reconstruct the original data to within a level of precision related to the number of components used in the model.  Treating the components as fixed parameters, the weights themselves contain all the information.  As described in Section \ref{sec:clustering} and illustrated in Figure \ref{fig:3dimRecons}, it is possible to use the information contained in the weights to identify similar age schedules, and hence groups of similar age schedules.  

\begin{figure}[htbp]
\begin{center}
\includegraphics[width=1.0\textwidth]{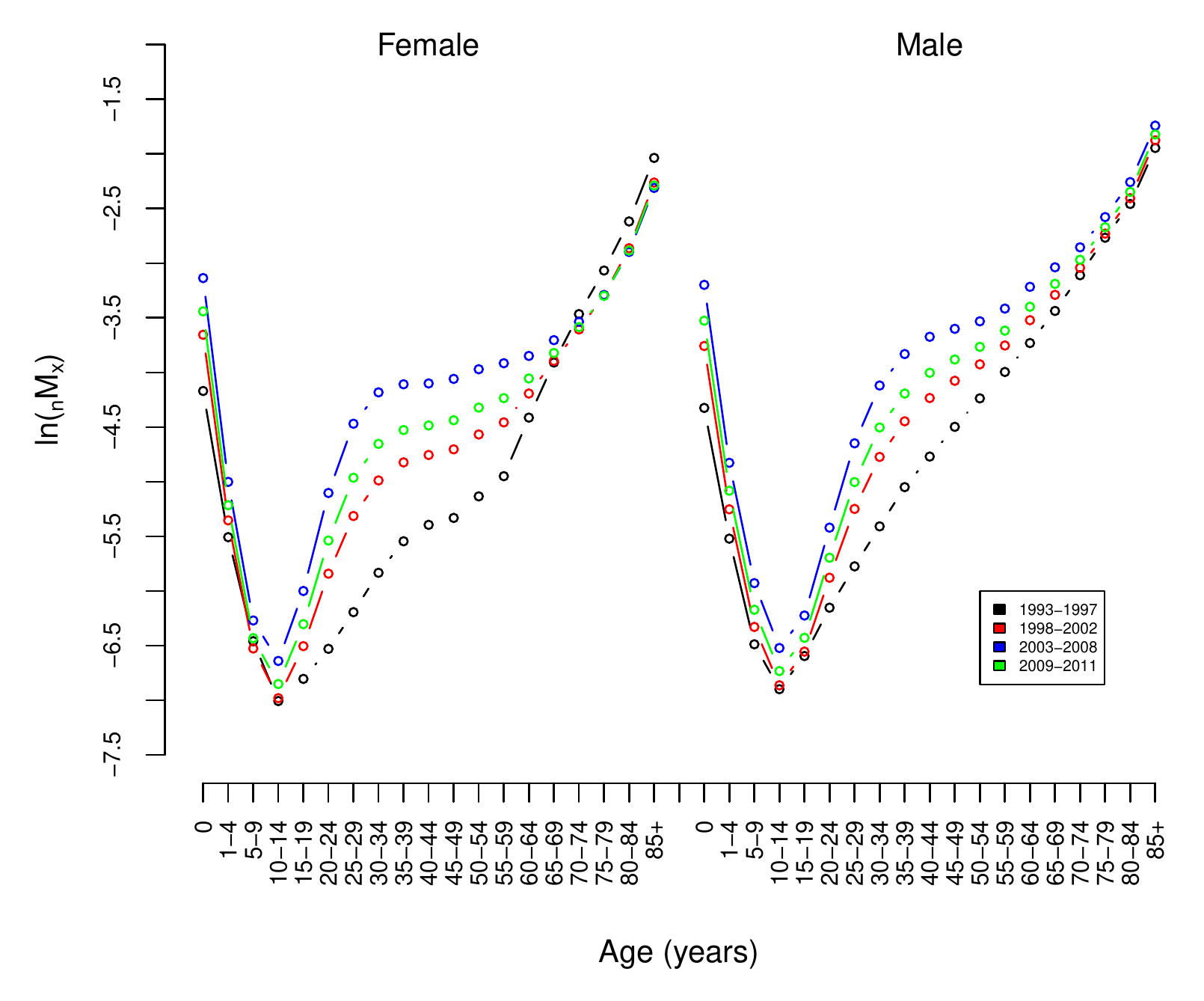}
\captionsetup{format=plain,font=small,margin=0cm,justification=justified}
\caption{{\bf Four Characteristic Age Patterns of Agincourt Log Mortality Rates.} The Mclust clustering method was used to identify four groups of similar weights on the two components retained in the reduced dimension model of Agincourt mortality. The model patterns generated using the median weights for each group are 1: 1993--1997, 2: 1998--2002, 3: 2003--2008, and 4: 2009--2011.}
\label{fig:clusterPatternsMx}
\end{center}
\end{figure}

We briefly demonstrate this potential using the log Agincourt mortality schedules discussed above in Sections \ref{sec:compModMx} and \ref{sec:covsMx}.  We examine the first two right singular vectors of the SVD of $\mbf{AM}$ which contain the weights used to reconstruct the rank-2 version of $\mbf{AM}$.  Treating these as a two-dimensional, compact representation of the age schedules contained in $\mbf{AM}$, we use the model-based clustering algorithm Mclust \citep{fraley2002,fraley2009mclust} to identify clusters of similar row vectors in the matrix formed by the first two right singular vectors, %
\begin{align}
\left[
\begin{matrix}
  | & | \\
  \mbf{v}_1 & \mbf{v}_2 \\
  | & | 
\end{matrix}
\right].
\end{align}
Mclust simultaneously identifies the optimal number of clusters and classifies each row of the dataset into one of the clusters.  Applied to the nineteen Agincourt mortality schedules, Mclust identified four clusters -- 1: 1993--1997, 2: 1998--2002, 3: 2003--2008, and 4: 2009-2011.  Within each of these clusters, we calculated the median values of $\mbf{v}_1$ and $\mbf{v}_2$ and reconstructed the log mortality age schedules for each cluster using those median values, displayed in Figure \ref{fig:clusterPatternsMx}.  These form the `characteristic' smoothed (reduced-dimension) age patterns for each of the periods identified by Mclust.  

Theses periods are substantively sensible -- 1: pre-HIV (no hump, generally low mortality), 2: developing HIV epidemic and no ART (hump developing and child mortality increasing), 3: height of HIV epidemic and beginning of sporadic roll-out of ART (very pronounced effects of HIV), and 4: ART available, coverage continuing to increase and mortality-reduction effects of ART beginning to be felt (attenuation of HIV-related mortality).

\subsection{Component Model for Fertility}\label{sec:compModFx}

In this and following Sections \ref{sec:compModFx} -- \ref{sec:covsFx} we define and demonstrate a component model of fertility, following the same general procedure as we employed for mortality in Sections  \ref{sec:compModMx} -- \ref{sec:covsMx}.  We use the example mortality data listed in Tables \ref{app:aginFx}.\ref{tab:aginFxAll1} and \ref{app:aginFx}.\ref{tab:aginFxAll2} and displayed in Figure \ref{fig:aginData}.  The data consist of annual age-specific fertility rates for ages 15--19, 20-24, $\ldots$, 45-49 for years 1993--2011. In what follows we work with the log fertility rates.  The data also contain one covariate, the total fertility rate (TFR).  Using the component model for mortality, we predict fertility age schedules using the TFR.

As with mortality, the component model of fertility follows the general SVD-based component model in Equation \ref{eq:genModel}.  We calculate the SVD of the $7 \times 19$ matrix $\mbf{AF}$ of log fertility rates. Following the description of the general model in Section \ref{sec:theModel}, we use the left singular vectors and singular values to construct fertility components $\tensor[_{\text{AF}}]{\mbf{\Lambda}}{_i} =  \tensor[_{\text{AF}}]{s}{_i} \cdot \tensor[_{\text{AF}}]{\mbf{u}}{_i}$.  The first four singular values $\tensor[_{\text{AF}}]{s}{_i} \ , \ i \in \{1,2,3,4\}$ are 33.64, 1.76, 0.22, and 0.18, with the remaining singular values $< 0.13$.  Consequently the new dimensions associated with the first four right singular vectors account for 99.7\%, 0.3\%, 0.004\%, and 0.003\% respectively of the total sum of squared perpendicular distances to all of the 7 points in the data set (see Section \ref{sec:sumRank1Mats}).  This indicates that the first two new dimensions effectively account for all of the variation in the original data (the remaining variation is lost in rounding error when presenting the results with a readable number of significant figures).  Consequently we adopt the following dimension-reduced model with two components, 
\begin{align}
\tensor[_{\text{AF}}]{\widehat{\mbf{f}}}{_{t}} &= \sum_{i = 1}^{2} \tensor[_{\text{AF}}]{\beta}{_{i,t}} \cdot \tensor[_{\text{AF}}]{\mbf{\Lambda}}{_i} \ , \label{eq:compModFx} \\
\tensor[]{\mbf{f}}{_{t}} &= \tensor[_{\text{AF}}]{\widehat{\mbf{f}}}{_{t}} + \tensor[_{\text{AF}}]{\mbf{r}}{_{t}} \ . \nonumber
\end{align}
where $\tensor[_{\text{AF}}]{\widehat{\mbf{f}}}{_{t}}$ is the predicted age-specific fertility schedule for year $t$; $\tensor[_{\text{AF}}]{\beta}{_{i,t}}$ are the weights applied to the first two components $\tensor[_{\text{AF}}]{\mbf{\Lambda}}{_i}$ (left singular vectors scaled by their corresponding singular values); and $\tensor[_{\text{AF}}]{\mbf{r}}{_{t}}$ is the difference between the predicted and `real' age-specific fertility schedule, possible to calculate when $t$ is one of the years included in the $\mbf{AF}$ but otherwise an unknown residual when the model is used to predict a mortality schedule not included in $\mbf{AF}$.  Equation \ref{eq:compModFx} is a two-parameter model for age-specific schedules of fertility covering the years 1993--2011.  The two components $\tensor[_{\text{AF}}]{\mbf{\Lambda}}{_i} \ , \ i \in \{1,2\}$ are plotted in Figure \ref{fig:Us}.

\subsubsection{Fertility Age Schedule Covariates and Predictors}\label{sec:covsFx}

\begin{figure}[htbp]
\begin{center}
\includegraphics[width=0.6\textwidth]{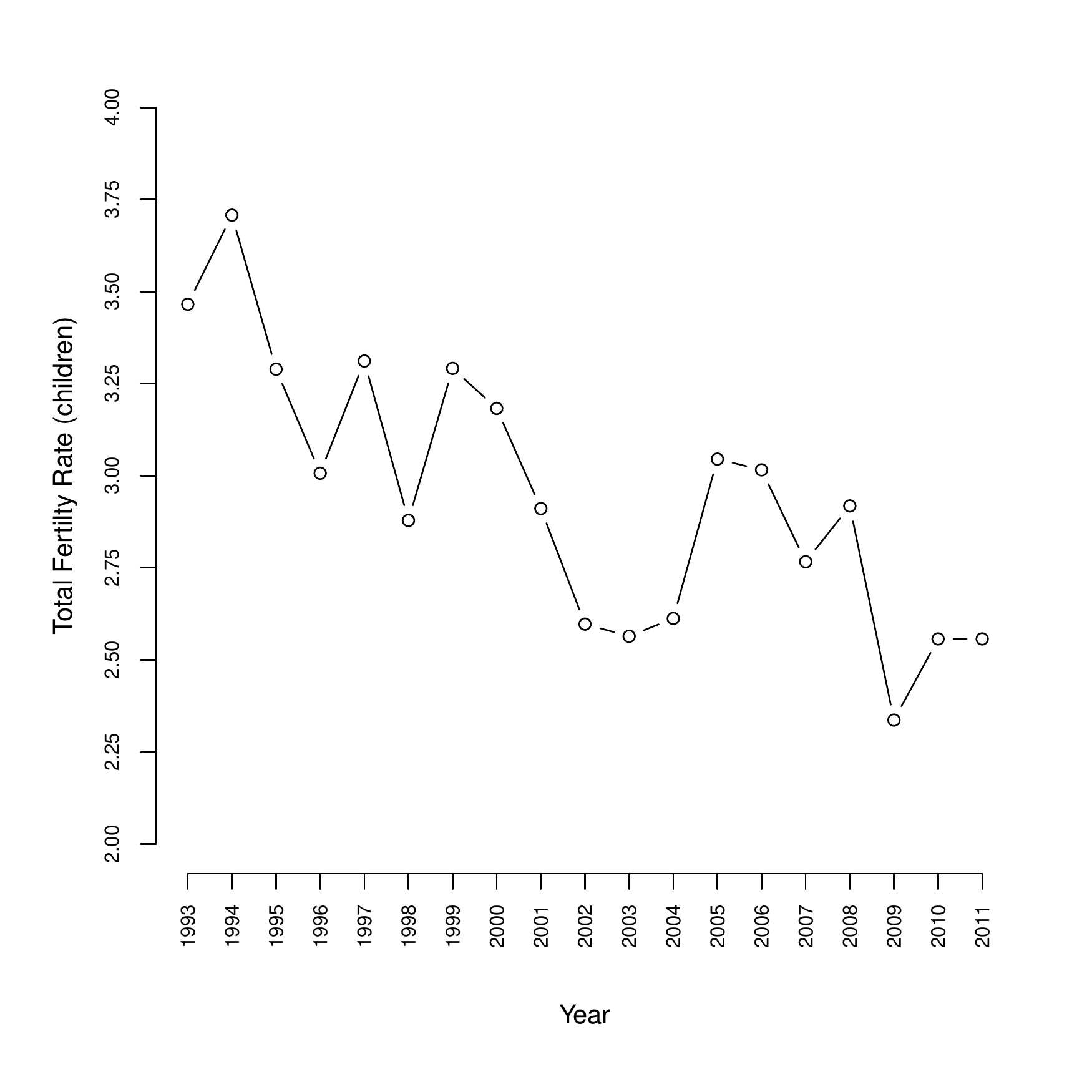}
\captionsetup{format=plain,font=small,margin=1cm,justification=justified}
\caption{{\bf Agincourt Total Fertility Rate (TFR) by Year.}}
\label{fig:TFR}
\end{center}
\end{figure}

\begin{figure}[htbp]
\begin{center}
\begin{tabular}{cc}
\includegraphics[width=0.48\textwidth]{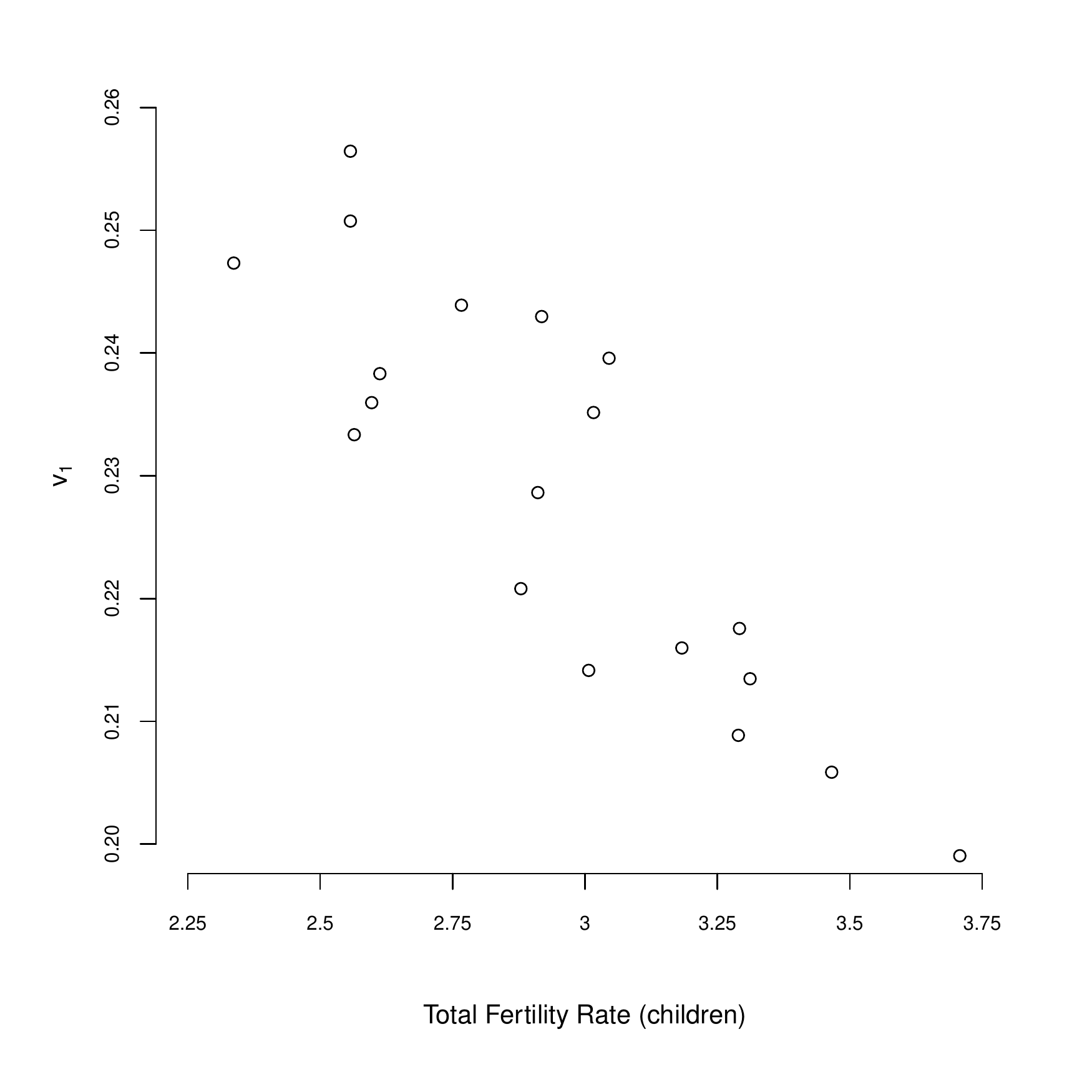} &
\includegraphics[width=0.48\textwidth]{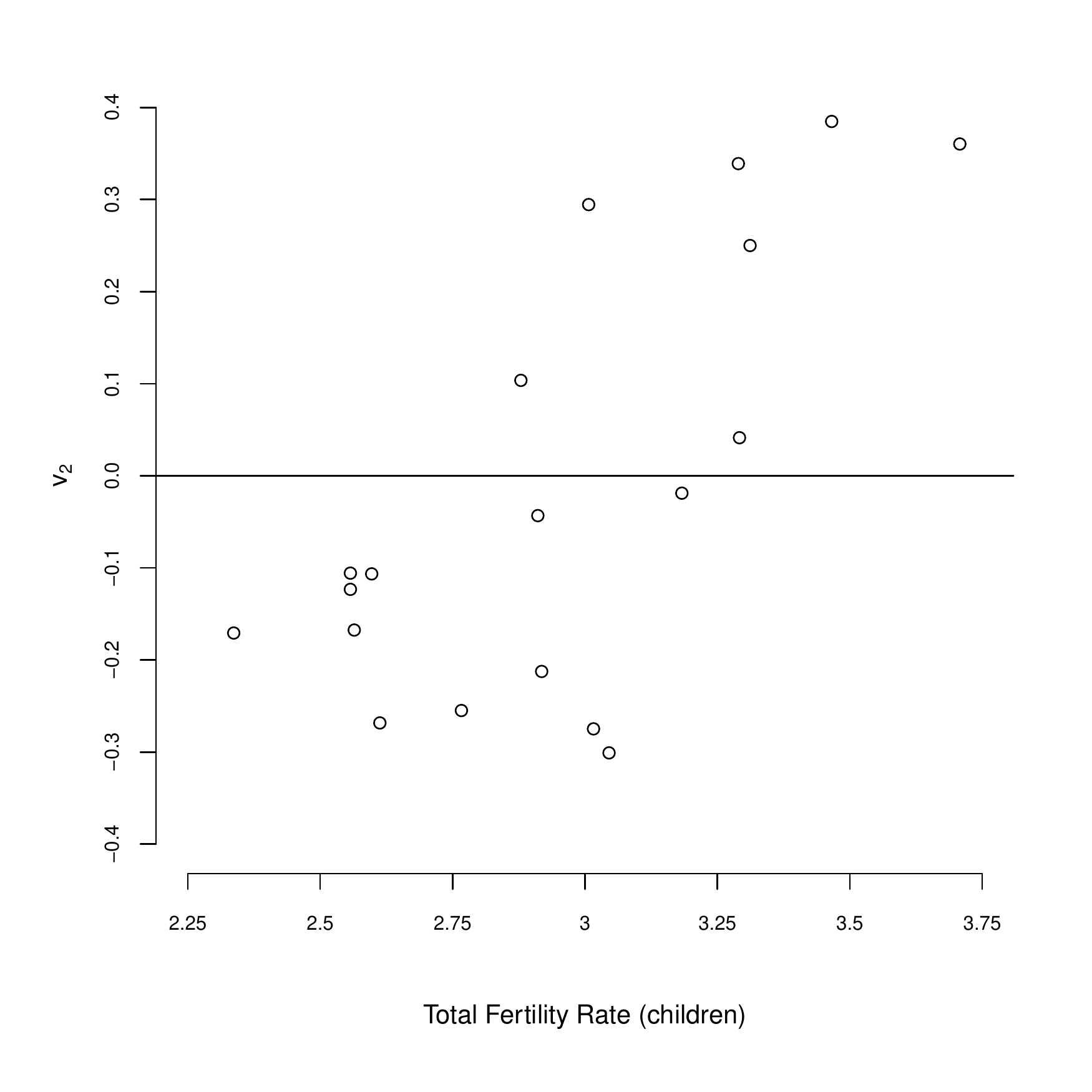}
\end{tabular}
\captionsetup{format=plain,font=small,margin=0cm,justification=justified}
\caption{{\bf Agincourt Age-specific Log Fertility Rate Right Singular Vectors $\mbf{v}_1$ and $\mbf{v}_2$ by Total Fertility Rate (TFR).}}
\label{fig:vsByTFR}
\end{center}
\end{figure}

As we did with mortality, we can construct a model of the weights that predict fertility in Equation \ref{eq:compModFx} using a covariate to predict the weights.  For fertility we will use the TFR or overall level as our predictor and see if we can accurately predict the age-pattern of fertility from TFR.  The time trend in the TFR is displayed in Figure \ref{fig:TFR}, and the relationship between the TFR and the first two right singular vectors of the SVD of the age-specific fertility rates $\mbf{AF}$ is displayed in Figure \ref{fig:vsByTFR}.  The relationships between the $\mbf{v}$'s and TFR is linear but less strong than the relationships between the $\mbf{v}$'s for mortality and their predictors.  Nonetheless, we model each with a simple linear equation and estimate the coefficients with OLS.  The results are displayed in Tables \ref{tab:fxV1} and \ref{tab:fxV2} with structures exactly analogous to Equations \ref{eq:v1lm} and \ref{eq:v2lm} used for the mortality model above in Section \ref{sec:covsMx}.

\begin{table}[ht]
\centering
\captionsetup{format=plain,font=small,margin=4cm,justification=justified}
\caption{OLS Estimates for Linear Model of Fertility $\mbf{v}_1$ by TFR: $R^2 = 0.746$}
\begin{tabular}{lrrrr}
  \hline
 & Estimate & Std. Error & t value & Pr($>$$|$t$|$) \\ 
  \hline
  $\tensor[_{v1}]{c}{}$ & 0.344799 & 0.016529 & 20.860334 & 0.000000 \\ 
  $\tensor[_{v1}]{\beta}{_{1}}$ & -0.039329 & 0.005566 & -7.065603 & 0.000002 \\ 
   \hline
\end{tabular}
\label{tab:fxV1}
\end{table}

\begin{table}[ht]
\centering
\captionsetup{format=plain,font=small,margin=4cm,justification=justified}
\caption{OLS Estimates for Linear Model of Fertility $\mbf{v}_2$ by TFR: $R^2 = 0.431$}
\begin{tabular}{lrrrr}
  \hline
 & Estimate & Std. Error & t value & Pr($>$$|$t$|$) \\ 
  \hline
  $\tensor[_{v2}]{c}{}$ & -1.236473 & 0.350591 & -3.526825 & 0.002589 \\ 
  $\tensor[_{v2}]{\beta}{_{1}}$ & 0.424483 & 0.118063 & 3.595388 & 0.002231 \\ 
   \hline
\end{tabular}
\label{tab:fxV2}
\end{table}

Using these relationships to predict the weights in Equation \ref{eq:compModFx} we predict the age-specific fertility schedules displayed on their natural scale in Figures \ref{app:predFx}.\ref{fig:aginFertPredsNatScale1} to \ref{app:predFx}.\ref{fig:aginFertPredsNatScale4} in Appendix \ref{app:predFx}.  The MAE for the prediction errors across age groups and years on the log scale is 0.0834, and the five-number summary is 0.0021, 0.0167, 0.0395, 0.0795, 0.6440.  As with mortality the predictions are surprisingly close to the data given the single predictor; this can be verified visually by looking at the Figures in Appendix \ref{app:predFx}.

\section{Discussion}

The SVD is a classic matrix decomposition that factorizes an arbitrary matrix into three new matrices with useful properties.  The right singular vectors $\mbf{V}$ identify a new orthonormal basis for the row vectors or points in the original matrix; the singular values correspond to how much variation among the points in the original matrix is captured by each of the new dimensions; and for each point, the left singular vectors $\mbf{U}$ stretch or shrink the new dimensions, scaled by their singular values, into a set of vectors whose sum locates the original points.

The SVD can be reorganized into a different form -- the Eckart-Young-Mirsky formula -- that expresses the original matrix as a sum or rank-1 matrices, and it has been shown that truncated sums of these rank-1 matrices represent the best reduced-rank approximations of the origin matrix in a perpendicular least squares sense.  An intermediate step on the way to deriving the EYM formula expresses each column vector in the original matrix as a weighted sum of the left singular vectors scaled by their singular values, with the weights being the right singular vectors.  Because they are equivalent to the EYM form, these weighted sums have the property that the first few terms contain the vast majority of the information necessary to represent the columns in the original matrix. This allows each column in the original matrix to be closely approximated by a weighted sum with potentially very few terms.  Not including the remaining terms also eliminates `noise', i.e. small magnitude variation that is not age-specific, and this provides a principled means by which to smooth the columns in the original matrix.

The SVD can be applied directly to age schedules of quantities that are correlated by age because data with that property are always arranged in a multidimensional cloud whose primary axis or dimension is very close to a line that intersects the origin.  This ensures that the first right singular vector derived from an SVD of these data is approximately lined up with the primary dimension of the cloud, and this in turn provides a standard interpretation of the first new dimension $\mbf{v}_1$ and more importantly the first left singular vector $\mbf{u}_1$.  $\mbf{u}_1$ for a set of points correlated by age is the principal shape of the age schedule with age, and the remaining $\mbf{u}$'s are age-specific deviations on that main shape, and typically only the first few represent systematic age-specific deviations, with the rest effectively being noise.

Combining our understanding of the SVD and the fact that the SVD of demographic quantitates correlated by age behaves in a predictable and interpretable way, we propose a general `component model' of demographic quantitates correlated by age.  This model given in Equation \ref{eq:genModel} represents an arbitrary age schedule as the weighted sum of components derived from the SVD of a matrix of similar age schedules.  The weights can come from the SVD that is used to define the components (if one wants to reconstruct or smooth the original data matrix) or by estimating an OLS linear regression model of an arbitrary age schedule as a function of the components, with no intercept. The model can be used for a variety of purposes including:
\begin{itemize}\addtolength{\itemsep}{-0.4\baselineskip}
\item to smooth age schedules;
\item to (dramatically) reduce the amount of information necessary to represent a large number of age schedules;
\item to represent a single age schedule with a small number of weights, typically 2--3, by treating the components as fixed parameters;
\item to predict new age schedules by supplying values for the weights, and
\item to cluster age schedules by applying a clustering algorithm to the weights.
\end{itemize}
A particularly useful application of the component model is to use it to represent age schedules as a function of arbitrary predictors or covariates. This is done by modeling the right singular vectors of the SVD used to create the components as functions of the predictors.  These models can be used to predict the weights using values supplied for the predictors/covariates, and the predicted weights can then be used to predict the age schedules.  This is a simple and general way to relate covariates to whole age structures and likely has many applications.

We demonstrate all of these properties and uses of the component model using example data from the Agincourt HDSS site in rural South Africa.  The component model is able to accurately reconstruct age-specific mortality and fertility rates starting with the original SVD-derived weights and/or weights predicted using simple models of the observed relationships between weights and covariates.  


\newpage
\bibliographystyle{chicago}
\bibliography{compModRef.bib}

\pagebreak
\appendix
\appendixpage
\addappheadtotoc

\section{Example Data}\label{app:exampleMx}

\vspace*{\fill}
\begin{table}[ht]
\captionsetup{format=plain,font=small,margin=4.8cm,justification=justified}
\caption{Example Mortality Rates: Agincourt Smoothed and Aggregated into Three Periods \citep{kahn2012agin}.} 
\centering
{\tiny
\begin{tabular}{ccccc}
  \hline
Sex & Age Group & 1992--1997 & 1998--2006 & 2007--2012 \\ 
  \hline \hline \\
Female & 0 & 0.01514 & 0.03575 & 0.02730 \\ 
  Female & 1-4 & 0.00398 & 0.00598 & 0.00227 \\ 
  Female & 5-9 & 0.00156 & 0.00214 & 0.00102 \\ 
  Female & 10-14 & 0.00082 & 0.00108 & 0.00084 \\ 
  Female & 15-19 & 0.00105 & 0.00217 & 0.00162 \\ 
  Female & 20-24 & 0.00124 & 0.00497 & 0.00370 \\ 
  Female & 25-29 & 0.00180 & 0.00869 & 0.00667 \\ 
  Female & 30-34 & 0.00271 & 0.01140 & 0.00909 \\ 
  Female & 35-39 & 0.00339 & 0.01246 & 0.01117 \\ 
  Female & 40-44 & 0.00359 & 0.01264 & 0.01249 \\ 
  Female & 45-49 & 0.00414 & 0.01364 & 0.01244 \\ 
  Female & 50-54 & 0.00508 & 0.01536 & 0.01104 \\ 
  Female & 55-59 & 0.00658 & 0.01619 & 0.01206 \\ 
  Female & 60-64 & 0.01161 & 0.01755 & 0.01784 \\ 
  Female & 65-69 & 0.02032 & 0.02174 & 0.02320 \\ 
  Female & 70-74 & 0.03176 & 0.02853 & 0.02896 \\ 
  Female & 75-79 & 0.05130 & 0.03989 & 0.03894 \\ 
  Female & 80-84 & 0.08233 & 0.05865 & 0.05500 \\ 
  Female & 85+ & 0.11596 & 0.09346 & 0.09200 \\ 
  \\ \hline \\
  Male & 0 & 0.01447 & 0.03821 & 0.02170 \\ 
  Male & 1-4 & 0.00390 & 0.00678 & 0.00443 \\ 
  Male & 5-9 & 0.00151 & 0.00259 & 0.00200 \\ 
  Male & 10-14 & 0.00088 & 0.00117 & 0.00124 \\ 
  Male & 15-19 & 0.00124 & 0.00181 & 0.00137 \\ 
  Male & 20-24 & 0.00202 & 0.00413 & 0.00259 \\ 
  Male & 25-29 & 0.00312 & 0.00842 & 0.00549 \\ 
  Male & 30-34 & 0.00451 & 0.01356 & 0.00908 \\ 
  Male & 35-39 & 0.00631 & 0.01777 & 0.01318 \\ 
  Male & 40-44 & 0.00827 & 0.02061 & 0.01735 \\ 
  Male & 45-49 & 0.01122 & 0.02250 & 0.02124 \\ 
  Male & 50-54 & 0.01543 & 0.02492 & 0.02273 \\ 
  Male & 55-59 & 0.02025 & 0.02966 & 0.02600 \\ 
  Male & 60-64 & 0.02570 & 0.03621 & 0.03248 \\ 
  Male & 65-69 & 0.03285 & 0.04325 & 0.04452 \\ 
  Male & 70-74 & 0.04600 & 0.05516 & 0.05463 \\ 
  Male & 75-79 & 0.06416 & 0.07414 & 0.07191 \\ 
  Male & 80-84 & 0.07973 & 0.09298 & 0.09995 \\ 
  Male & 85+ & 0.10870 & 0.14229 & 0.14114 \\ 
  \\ \hline \hline
\end{tabular}
}
\label{tab:aginMx3}
\end{table}
\vspace*{\fill}

\pagebreak
\section{Smoothed Agincourt Mortality Rates}\label{app:aginMx}

\vspace*{\fill}
\begin{table}[ht]
\centering
\begin{threeparttable}
\caption{Agincourt Mortality Rates: Smoothed 1993--2002 \citep{kahn2012agin}.} 
\label{tab:aginMxAll1}
{\tiny
\begin{tabular}{cccccccccccc}
  \hline
& Indicator \\
Sex & Age Group & 1993 & 1994 & 1995 & 1996 & 1997 & 1998 & 1999 & 2000 & 2001 & 2002 \\ 
  \hline \hline \\ 
  Both & HIV Prevalence\tnote{a} & 0.03243 & 0.04624 & 0.06366 & 0.08401 & 0.10568 & 0.12657 & 0.14476 & 0.15909 & 0.16938 & 0.17607 \\ 
  Both & ART Coverage\tnote{b} & 0.00000 & 0.00000 & 0.00000 & 0.00000 & 0.00000 & 0.00000 & 0.00000 & 0.00000 & 0.00013 & 0.00030 \\ 
  Both & $\tensor[^{0}]{e}{}$ & 70.13 & 69.32 & 68.44 & 69.55 & 69.40 & 68.79 & 66.39 & 64.00 & 62.34 & 59.03 \\ 
  Both & $\tensor[_{45}]{q}{_{15}}$ & 0.21467 & 0.22811 & 0.23932 & 0.23759 & 0.24640 & 0.26771 & 0.28508 & 0.31503 & 0.35818 & 0.42472 \\ 
  Both & $\tensor[_{5}]{q}{_{0}}$ & 0.03066 & 0.03003 & 0.02982 & 0.02760 & 0.02748 & 0.03262 & 0.04829 & 0.05393 & 0.05486 & 0.05997 \\ 
  \\ \hline \\
  Female & 0 & 0.01537 & 0.01529 & 0.01606 & 0.01548 & 0.01342 & 0.01598 & 0.02893 & 0.03203 & 0.03214 & 0.03398 \\ 
  Female & 1-4 & 0.00414 & 0.00409 & 0.00387 & 0.00329 & 0.00379 & 0.00426 & 0.00535 & 0.00568 & 0.00595 & 0.00674 \\ 
  Female & 5-9 & 0.00179 & 0.00149 & 0.00143 & 0.00116 & 0.00148 & 0.00149 & 0.00168 & 0.00156 & 0.00182 & 0.00217 \\ 
  Female & 10-14 & 0.00097 & 0.00098 & 0.00083 & 0.00092 & 0.00082 & 0.00072 & 0.00078 & 0.00079 & 0.00108 & 0.00125 \\ 
  Female & 15-19 & 0.00099 & 0.00100 & 0.00102 & 0.00106 & 0.00103 & 0.00145 & 0.00146 & 0.00162 & 0.00193 & 0.00241 \\ 
  Female & 20-24 & 0.00115 & 0.00126 & 0.00130 & 0.00132 & 0.00179 & 0.00244 & 0.00294 & 0.00376 & 0.00388 & 0.00493 \\ 
  Female & 25-29 & 0.00172 & 0.00191 & 0.00209 & 0.00193 & 0.00242 & 0.00332 & 0.00386 & 0.00527 & 0.00648 & 0.00825 \\ 
  Female & 30-34 & 0.00281 & 0.00309 & 0.00314 & 0.00273 & 0.00288 & 0.00416 & 0.00527 & 0.00712 & 0.00815 & 0.01021 \\ 
  Female & 35-39 & 0.00378 & 0.00399 & 0.00403 & 0.00351 & 0.00394 & 0.00537 & 0.00640 & 0.00882 & 0.00952 & 0.01146 \\ 
  Female & 40-44 & 0.00391 & 0.00468 & 0.00470 & 0.00422 & 0.00420 & 0.00640 & 0.00784 & 0.00987 & 0.01004 & 0.01217 \\ 
  Female & 45-49 & 0.00479 & 0.00526 & 0.00475 & 0.00421 & 0.00493 & 0.00631 & 0.00731 & 0.00893 & 0.01089 & 0.01336 \\ 
  Female & 50-54 & 0.00610 & 0.00639 & 0.00624 & 0.00550 & 0.00591 & 0.00675 & 0.00764 & 0.00980 & 0.01251 & 0.01477 \\ 
  Female & 55-59 & 0.00700 & 0.00675 & 0.00676 & 0.00722 & 0.00802 & 0.00788 & 0.00964 & 0.01073 & 0.01410 & 0.01583 \\ 
  Female & 60-64 & 0.00967 & 0.01138 & 0.01134 & 0.01213 & 0.01504 & 0.01264 & 0.01457 & 0.01578 & 0.01750 & 0.01786 \\ 
  Female & 65-69 & 0.01843 & 0.01899 & 0.01979 & 0.01975 & 0.02022 & 0.01845 & 0.02054 & 0.01993 & 0.02229 & 0.02231 \\ 
  Female & 70-74 & 0.02871 & 0.03111 & 0.03030 & 0.02979 & 0.03076 & 0.02781 & 0.02747 & 0.02676 & 0.02906 & 0.02639 \\ 
  Female & 75-79 & 0.04341 & 0.04483 & 0.04643 & 0.04582 & 0.04059 & 0.03822 & 0.03682 & 0.04130 & 0.04300 & 0.03919 \\ 
  Female & 80-84 & 0.06807 & 0.07707 & 0.08109 & 0.07562 & 0.06203 & 0.04711 & 0.04865 & 0.06367 & 0.06609 & 0.06958 \\ 
  Female & 85+ & 0.12280 & 0.13903 & 0.14629 & 0.13642 & 0.11190 & 0.08499 & 0.08777 & 0.11485 & 0.11923 & 0.12552 \\ 
  \\ \hline \\
  Male & 0 & 0.01440 & 0.01224 & 0.01295 & 0.01282 & 0.01189 & 0.01573 & 0.02512 & 0.02786 & 0.02860 & 0.03297 \\ 
  Male & 1-4 & 0.00375 & 0.00408 & 0.00380 & 0.00341 & 0.00359 & 0.00418 & 0.00577 & 0.00698 & 0.00700 & 0.00740 \\ 
  Male & 5-9 & 0.00164 & 0.00164 & 0.00155 & 0.00119 & 0.00123 & 0.00138 & 0.00185 & 0.00193 & 0.00194 & 0.00223 \\ 
  Male & 10-14 & 0.00113 & 0.00112 & 0.00100 & 0.00085 & 0.00086 & 0.00083 & 0.00089 & 0.00095 & 0.00097 & 0.00110 \\ 
  Male & 15-19 & 0.00130 & 0.00156 & 0.00141 & 0.00112 & 0.00118 & 0.00129 & 0.00131 & 0.00127 & 0.00152 & 0.00205 \\ 
  Male & 20-24 & 0.00183 & 0.00212 & 0.00208 & 0.00221 & 0.00229 & 0.00241 & 0.00227 & 0.00259 & 0.00325 & 0.00431 \\ 
  Male & 25-29 & 0.00241 & 0.00294 & 0.00312 & 0.00377 & 0.00385 & 0.00416 & 0.00426 & 0.00451 & 0.00563 & 0.00780 \\ 
  Male & 30-34 & 0.00359 & 0.00415 & 0.00416 & 0.00546 & 0.00597 & 0.00664 & 0.00686 & 0.00762 & 0.00872 & 0.01196 \\ 
  Male & 35-39 & 0.00535 & 0.00616 & 0.00632 & 0.00741 & 0.00802 & 0.00900 & 0.00964 & 0.01040 & 0.01164 & 0.01531 \\ 
  Male & 40-44 & 0.00704 & 0.00833 & 0.00872 & 0.00985 & 0.00984 & 0.01079 & 0.01185 & 0.01256 & 0.01434 & 0.01771 \\ 
  Male & 45-49 & 0.00938 & 0.01071 & 0.01215 & 0.01284 & 0.01239 & 0.01315 & 0.01399 & 0.01456 & 0.01577 & 0.01926 \\ 
  Male & 50-54 & 0.01304 & 0.01473 & 0.01668 & 0.01558 & 0.01610 & 0.01603 & 0.01527 & 0.01454 & 0.01715 & 0.02213 \\ 
  Male & 55-59 & 0.02051 & 0.01895 & 0.02197 & 0.01960 & 0.01908 & 0.01722 & 0.01591 & 0.01622 & 0.02100 & 0.02761 \\ 
  Male & 60-64 & 0.02670 & 0.02448 & 0.03044 & 0.02405 & 0.02443 & 0.02129 & 0.02074 & 0.02093 & 0.02641 & 0.03407 \\ 
  Male & 65-69 & 0.03098 & 0.03070 & 0.03925 & 0.03037 & 0.03356 & 0.03210 & 0.03251 & 0.03297 & 0.03400 & 0.03825 \\ 
  Male & 70-74 & 0.03834 & 0.04164 & 0.05165 & 0.04011 & 0.04548 & 0.04520 & 0.05016 & 0.05526 & 0.04691 & 0.04881 \\ 
  Male & 75-79 & 0.05518 & 0.06372 & 0.07056 & 0.05732 & 0.06157 & 0.05930 & 0.06644 & 0.07824 & 0.06387 & 0.06913 \\ 
  Male & 80-84 & 0.09120 & 0.08723 & 0.09655 & 0.07910 & 0.06965 & 0.07560 & 0.08534 & 0.09819 & 0.08729 & 0.09653 \\ 
  Male & 85+ & 0.15344 & 0.14676 & 0.16244 & 0.13309 & 0.11719 & 0.12719 & 0.14358 & 0.16520 & 0.14687 & 0.16242 \\ 
  \\
   \hline\hline
\end{tabular}
}
\begin{tablenotes}
{\small
\item[a] Source: \cite{unpd2011}
\item[b] Source: ART coverage is number on ART divided by population.  For Mpumalanga Province, numbers on ART through 2008 from \cite{mpumalangaART}, extrapolated through 2011 using observed growth rate in previous years.  Population of Mpumalanga Province assumed to be 4M. 
}
\end{tablenotes}
\end{threeparttable} 
\end{table}
\vspace*{\fill}

\begin{table}[ht]
\centering
\begin{threeparttable}
\caption{Agincourt Mortality Rates: Smoothed 2003--2011 \citep{kahn2012agin}.} 
\label{tab:aginMxAll2}
{\tiny
\begin{tabular}{ccccccccccc}
  \hline
& Indicator \\ 
Sex & Age Group & 2003 & 2004 & 2005 & 2006 & 2007 & 2008 & 2009 & 2010 & 2011 \\
  \hline \hline \\ 
  Both & HIV Prevalence\tnote{a} & 0.17993 & 0.18174 & 0.18218 & 0.18059 & 0.17962 & 0.17870 & 0.17782 & 0.17689 & 0.17586 \\ 
  Both & ART Coverage\tnote{b} & 0.00050 & 0.00077 & 0.00140 & 0.00298 & 0.00575 & 0.00925 & 0.01249 & 0.01661 & 0.02192 \\ 
  Both & $\tensor[^{0}]{e}{}$ & 56.09 & 55.20 & 54.37 & 53.83 & 54.07 & 54.77 & 57.98 & 60.64 & 62.43 \\ 
  Both & $\tensor[_{45}]{q}{_{15}}$ & 0.47853 & 0.51034 & 0.53773 & 0.54112 & 0.53425 & 0.51198 & 0.45785 & 0.41664 & 0.38816 \\ 
  Both & $\tensor[_{5}]{q}{_{0}}$ & 0.06905 & 0.06710 & 0.06649 & 0.06772 & 0.07102 & 0.07275 & 0.05923 & 0.04653 & 0.03822 \\ 
  \\ \hline \\
  Female & 0 & 0.03691 & 0.03815 & 0.03990 & 0.04646 & 0.04800 & 0.04914 & 0.03873 & 0.02788 & 0.02533 \\ 
  Female & 1-4 & 0.00740 & 0.00654 & 0.00660 & 0.00721 & 0.00740 & 0.00691 & 0.00531 & 0.00406 & 0.00365 \\ 
  Female & 5-9 & 0.00245 & 0.00188 & 0.00161 & 0.00223 & 0.00228 & 0.00192 & 0.00137 & 0.00123 & 0.00104 \\ 
  Female & 10-14 & 0.00148 & 0.00139 & 0.00125 & 0.00144 & 0.00141 & 0.00153 & 0.00102 & 0.00095 & 0.00094 \\ 
  Female & 15-19 & 0.00257 & 0.00250 & 0.00235 & 0.00258 & 0.00227 & 0.00214 & 0.00177 & 0.00174 & 0.00168 \\ 
  Female & 20-24 & 0.00568 & 0.00641 & 0.00572 & 0.00587 & 0.00553 & 0.00489 & 0.00396 & 0.00356 & 0.00353 \\ 
  Female & 25-29 & 0.00944 & 0.01147 & 0.01152 & 0.01207 & 0.01097 & 0.00988 & 0.00811 & 0.00711 & 0.00636 \\ 
  Female & 30-34 & 0.01300 & 0.01581 & 0.01562 & 0.01650 & 0.01530 & 0.01379 & 0.01111 & 0.00976 & 0.00916 \\ 
  Female & 35-39 & 0.01373 & 0.01637 & 0.01715 & 0.01738 & 0.01570 & 0.01462 & 0.01248 & 0.01147 & 0.01127 \\ 
  Female & 40-44 & 0.01440 & 0.01630 & 0.01661 & 0.01668 & 0.01555 & 0.01379 & 0.01216 & 0.01226 & 0.01280 \\ 
  Female & 45-49 & 0.01542 & 0.01671 & 0.01838 & 0.01855 & 0.01702 & 0.01483 & 0.01309 & 0.01200 & 0.01269 \\ 
  Female & 50-54 & 0.01803 & 0.01852 & 0.02010 & 0.02113 & 0.02011 & 0.01771 & 0.01539 & 0.01301 & 0.01150 \\ 
  Female & 55-59 & 0.01764 & 0.01812 & 0.01963 & 0.02153 & 0.02086 & 0.01981 & 0.01784 & 0.01506 & 0.01344 \\ 
  Female & 60-64 & 0.02098 & 0.01980 & 0.01987 & 0.02077 & 0.02037 & 0.02121 & 0.01965 & 0.01912 & 0.01693 \\ 
  Female & 65-69 & 0.02286 & 0.02294 & 0.02310 & 0.02528 & 0.02526 & 0.02440 & 0.02289 & 0.02451 & 0.02433 \\ 
  Female & 70-74 & 0.02585 & 0.02415 & 0.02677 & 0.03120 & 0.03198 & 0.03223 & 0.03101 & 0.02971 & 0.03098 \\ 
  Female & 75-79 & 0.03505 & 0.03214 & 0.03469 & 0.03699 & 0.03734 & 0.03722 & 0.04059 & 0.03889 & 0.04193 \\ 
  Female & 80-84 & 0.06344 & 0.05615 & 0.05265 & 0.05507 & 0.05267 & 0.05169 & 0.05751 & 0.05464 & 0.05652 \\ 
  Female & 85+ & 0.11444 & 0.10129 & 0.09497 & 0.09935 & 0.09502 & 0.09324 & 0.10375 & 0.09857 & 0.10197 \\ 
  \\ \hline \\
  Male & 0 & 0.04226 & 0.04270 & 0.04089 & 0.03673 & 0.03934 & 0.04113 & 0.03583 & 0.02827 & 0.01940 \\ 
  Male & 1-4 & 0.00863 & 0.00801 & 0.00763 & 0.00713 & 0.00767 & 0.00839 & 0.00657 & 0.00560 & 0.00453 \\ 
  Male & 5-9 & 0.00272 & 0.00244 & 0.00247 & 0.00247 & 0.00286 & 0.00301 & 0.00242 & 0.00239 & 0.00185 \\ 
  Male & 10-14 & 0.00132 & 0.00139 & 0.00139 & 0.00141 & 0.00161 & 0.00171 & 0.00165 & 0.00168 & 0.00123 \\ 
  Male & 15-19 & 0.00212 & 0.00214 & 0.00195 & 0.00175 & 0.00186 & 0.00214 & 0.00194 & 0.00170 & 0.00143 \\ 
  Male & 20-24 & 0.00477 & 0.00483 & 0.00458 & 0.00434 & 0.00436 & 0.00464 & 0.00337 & 0.00289 & 0.00288 \\ 
  Male & 25-29 & 0.00902 & 0.01015 & 0.01034 & 0.01000 & 0.01007 & 0.00960 & 0.00767 & 0.00603 & 0.00535 \\ 
  Male & 30-34 & 0.01405 & 0.01621 & 0.01728 & 0.01731 & 0.01806 & 0.01673 & 0.01285 & 0.01016 & 0.00909 \\ 
  Male & 35-39 & 0.01795 & 0.02070 & 0.02260 & 0.02332 & 0.02498 & 0.02285 & 0.01884 & 0.01530 & 0.01236 \\ 
  Male & 40-44 & 0.02069 & 0.02300 & 0.02619 & 0.02650 & 0.02851 & 0.02749 & 0.02362 & 0.02034 & 0.01702 \\ 
  Male & 45-49 & 0.02292 & 0.02537 & 0.02935 & 0.02879 & 0.02901 & 0.02831 & 0.02591 & 0.02440 & 0.02067 \\ 
  Male & 50-54 & 0.02584 & 0.02882 & 0.03341 & 0.03238 & 0.03138 & 0.03123 & 0.02785 & 0.02559 & 0.02348 \\ 
  Male & 55-59 & 0.03443 & 0.03373 & 0.03889 & 0.03746 & 0.03778 & 0.03698 & 0.03059 & 0.02627 & 0.02365 \\ 
  Male & 60-64 & 0.04206 & 0.04020 & 0.04585 & 0.04645 & 0.04400 & 0.04552 & 0.03978 & 0.03347 & 0.03030 \\ 
  Male & 65-69 & 0.04638 & 0.04703 & 0.04959 & 0.05239 & 0.04873 & 0.05360 & 0.04851 & 0.04280 & 0.04052 \\ 
  Male & 70-74 & 0.05563 & 0.05532 & 0.05774 & 0.06153 & 0.05235 & 0.05629 & 0.05651 & 0.05320 & 0.05163 \\ 
  Male & 75-79 & 0.07527 & 0.07682 & 0.08076 & 0.07604 & 0.07241 & 0.07094 & 0.07097 & 0.06811 & 0.07070 \\ 
  Male & 80-84 & 0.10541 & 0.10065 & 0.10176 & 0.10799 & 0.10145 & 0.10184 & 0.10565 & 0.10748 & 0.10329 \\ 
  Male & 85+ & 0.17734 & 0.16934 & 0.17121 & 0.18170 & 0.17069 & 0.17134 & 0.17776 & 0.18084 & 0.17378 \\ 
  \\ \hline \hline
\end{tabular}
}
\begin{tablenotes}
{\small
\item[a] Source: \cite{unpd2011}
\item[b] Source: ART coverage is number on ART divided by population.  For Mpumalanga Province, numbers on ART through 2008 from \cite{mpumalangaART}, extrapolated through 2011 using observed growth rate in previous years.  Population of Mpumalanga Province assumed to be 4M. 
}
\end{tablenotes}
\end{threeparttable} 
\end{table}
\vspace*{\fill}

\clearpage
\pagebreak
\section{Smoothed Agincourt Fertility Rates}\label{app:aginFx}

\vspace*{\fill}
\begin{table}[ht]
\caption{Agincourt Fertility Rates: Smoothed 1993--2002 \citep{kahn2012agin}} 
\centering
{\tiny
\begin{tabular}{ccccccccccc}
  \hline
Age Group &1993 &1994 &1995 &1996 &1997 &1998 &1999 &2000 &2001 &2002 \\ 
  \hline \hline \\
TFR & 3.47 & 3.71 & 3.29 & 3.00 & 3.31 & 2.88 & 3.29 & 3.18 & 2.91 & 2.60 \\ 
  \\ \hline \\
  15-19 & 0.09103 & 0.10335 & 0.09637 & 0.09019 & 0.09566 & 0.09008 & 0.08658 & 0.09359 & 0.09184 & 0.08659 \\ 
  20-24 & 0.11408 & 0.12194 & 0.11281 & 0.10503 & 0.10829 & 0.10209 & 0.11455 & 0.11528 & 0.10543 & 0.09787 \\ 
  25-29 & 0.11732 & 0.12674 & 0.11740 & 0.10592 & 0.10666 & 0.10126 & 0.11885 & 0.11861 & 0.11079 & 0.10230 \\ 
  30-34 & 0.10750 & 0.11548 & 0.10892 & 0.10162 & 0.10112 & 0.09623 & 0.10542 & 0.10616 & 0.10208 & 0.09137 \\ 
  35-39 & 0.08941 & 0.09807 & 0.08883 & 0.08200 & 0.08224 & 0.07847 & 0.07976 & 0.07708 & 0.07370 & 0.06812 \\ 
  40-44 & 0.05357 & 0.05699 & 0.04891 & 0.04863 & 0.05050 & 0.04751 & 0.04679 & 0.04414 & 0.03583 & 0.03477 \\ 
  45-49 & 0.02418 & 0.02716 & 0.02118 & 0.01926 & 0.01858 & 0.01471 & 0.01510 & 0.01674 & 0.01084 & 0.00900 \\ 
  \\ \hline \hline
\end{tabular}
}
\label{tab:aginFxAll1}
\end{table}

\begin{table}[ht]
\caption{Agincourt Fertility Rates: 2003--2011 \citep{kahn2012agin}.} 
\centering
{\tiny
\begin{tabular}{cccccccccc}
  \hline
Age Group &2003 &2004 &2005 &2006 &2007 &2008 &2009 &2010 &2011 \\ 
  \hline \hline \\
TFR & 2.56 & 2.61 & 3.05 & 3.02 & 2.77 & 2.92 & 2.34 & 2.56 & 2.56 \\ 
  \\ \hline \\
  15-19 & 0.08582 & 0.08639 & 0.09255 & 0.09255 & 0.09026 & 0.09064 & 0.08908 & 0.08470 & 0.07633 \\ 
  20-24 & 0.09696 & 0.10184 & 0.11163 & 0.11018 & 0.11008 & 0.11308 & 0.10753 & 0.10636 & 0.09588 \\ 
  25-29 & 0.09794 & 0.10247 & 0.11411 & 0.11254 & 0.11041 & 0.11263 & 0.10416 & 0.10638 & 0.09572 \\ 
  30-34 & 0.08394 & 0.08733 & 0.09839 & 0.10000 & 0.09587 & 0.09951 & 0.09046 & 0.09415 & 0.08582 \\ 
  35-39 & 0.06415 & 0.06612 & 0.07522 & 0.07675 & 0.06983 & 0.07046 & 0.06529 & 0.06636 & 0.06421 \\ 
  40-44 & 0.03675 & 0.03582 & 0.03649 & 0.03678 & 0.03279 & 0.03395 & 0.03124 & 0.02930 & 0.02879 \\ 
  45-49 & 0.01115 & 0.00788 & 0.00557 & 0.00706 & 0.00515 & 0.00505 & 0.00492 & 0.00422 & 0.00386 \\ 
  \\ \hline \hline
\end{tabular}
}
\label{tab:aginFxAll2}
\end{table}
\vspace*{\fill}

\clearpage
\pagebreak
\section{Predicted Age-Specific Mortality Plots}\label{app:predMx}

\noindent
\begin{minipage}{\linewidth}
\captionsetup{format=plain,font=small,margin=1cm,justification=justified}
\makebox[\linewidth]{
\includegraphics[width=1.0\textwidth]{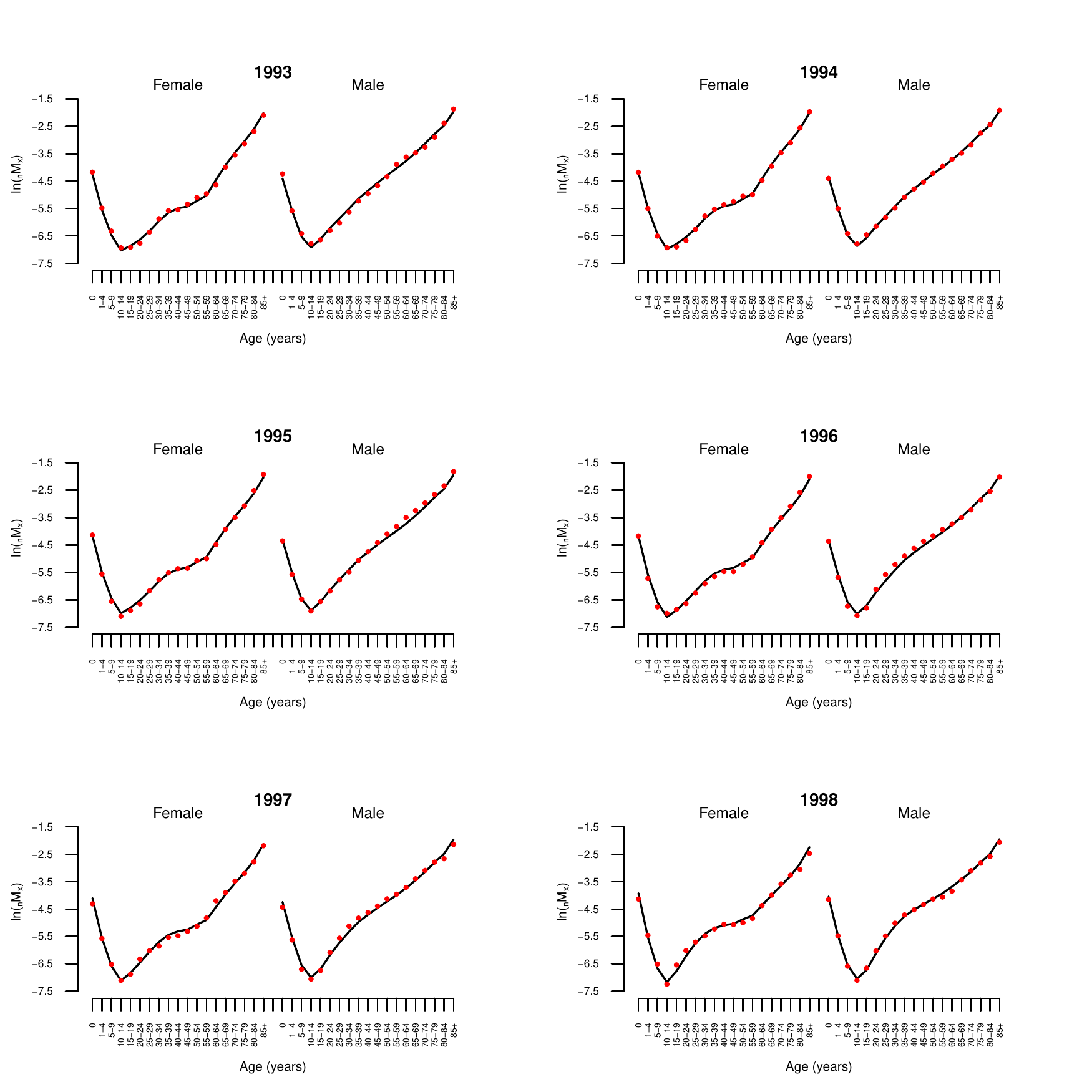}
}
\captionof{figure}{{\bf Age-specific Log Mortality Rates Predicted using Right Singular Vector Weights, 1993--1998.} The red dots are the data, and the solid black line indicates predicted values.}
\label{fig:aginPredsNoCovars1}
\end{minipage}

\noindent
\begin{minipage}{\linewidth}
\captionsetup{format=plain,font=small,margin=1cm,justification=justified}
\makebox[\linewidth]{
\includegraphics[width=1.0\textwidth]{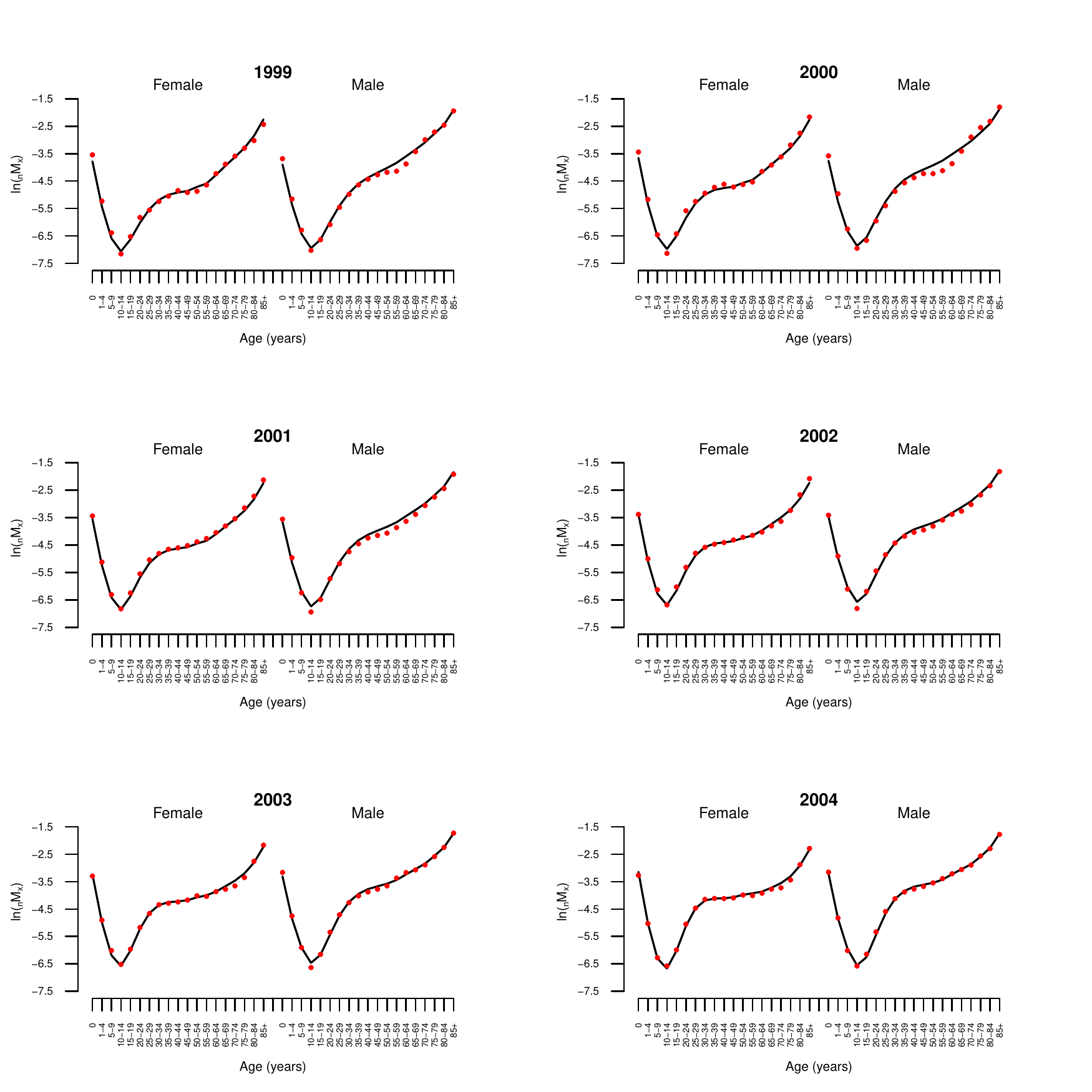}
}
\captionof{figure}{{\bf Age-specific Log Mortality Rates Predicted using Right Singular Vector Weights, 1999--2004.} The red dots are the data, and the solid black line indicates predicted values.}
\label{fig:aginPredsNoCovars2}
\end{minipage}

\noindent
\begin{minipage}{\linewidth}
\captionsetup{format=plain,font=small,margin=1cm,justification=justified}
\makebox[\linewidth]{
\includegraphics[width=1.0\textwidth]{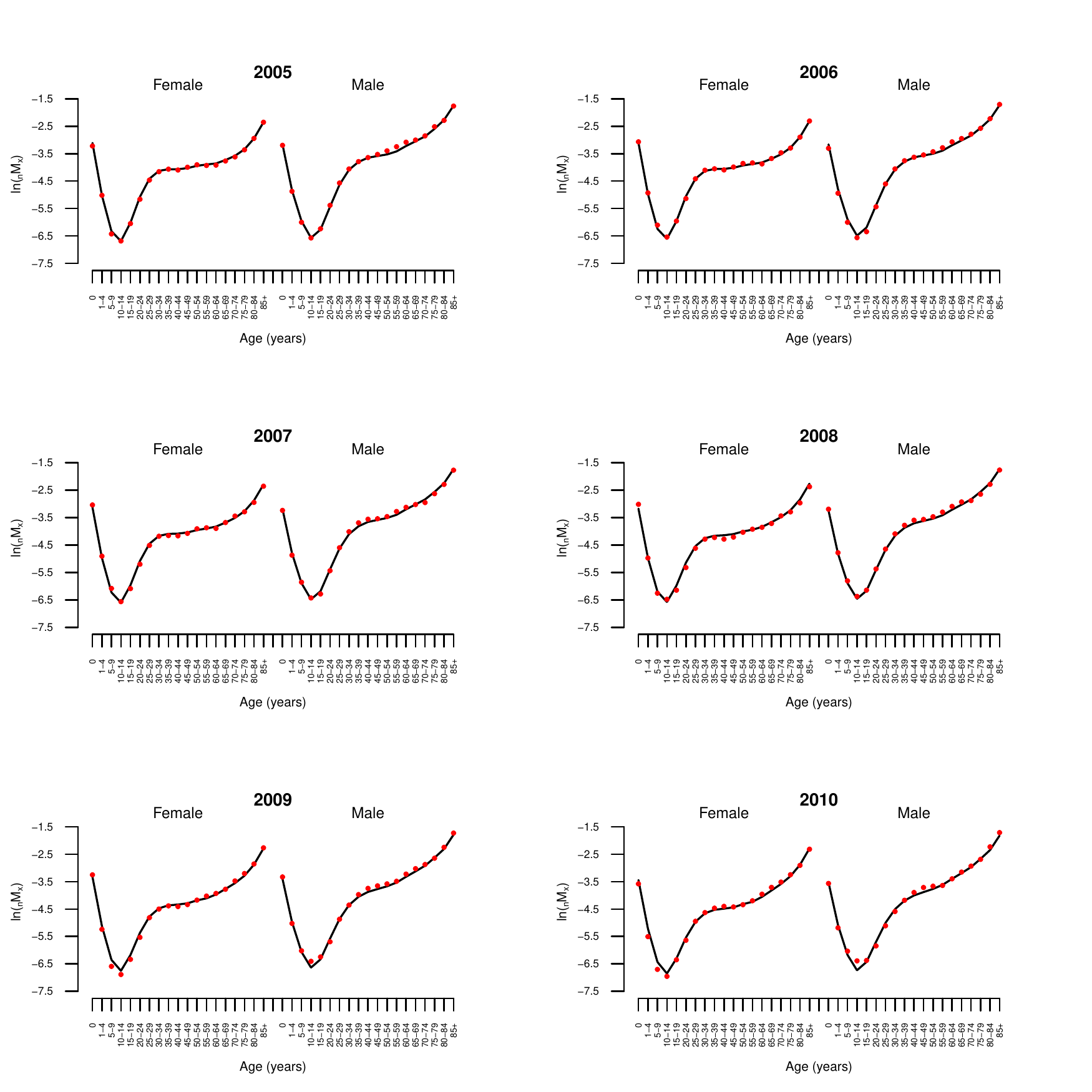}
}
\captionof{figure}{{\bf Age-specific Log Mortality Rates Predicted using Right Singular Vector Weights, 2005--2010.} The red dots are the data, and the solid black line indicates predicted values.}
\label{fig:aginPredsNoCovars3}
\end{minipage}

\noindent
\begin{minipage}{\linewidth}
\captionsetup{format=plain,font=small,margin=1cm,justification=justified}
\makebox[\linewidth]{
\includegraphics[width=1.0\textwidth]{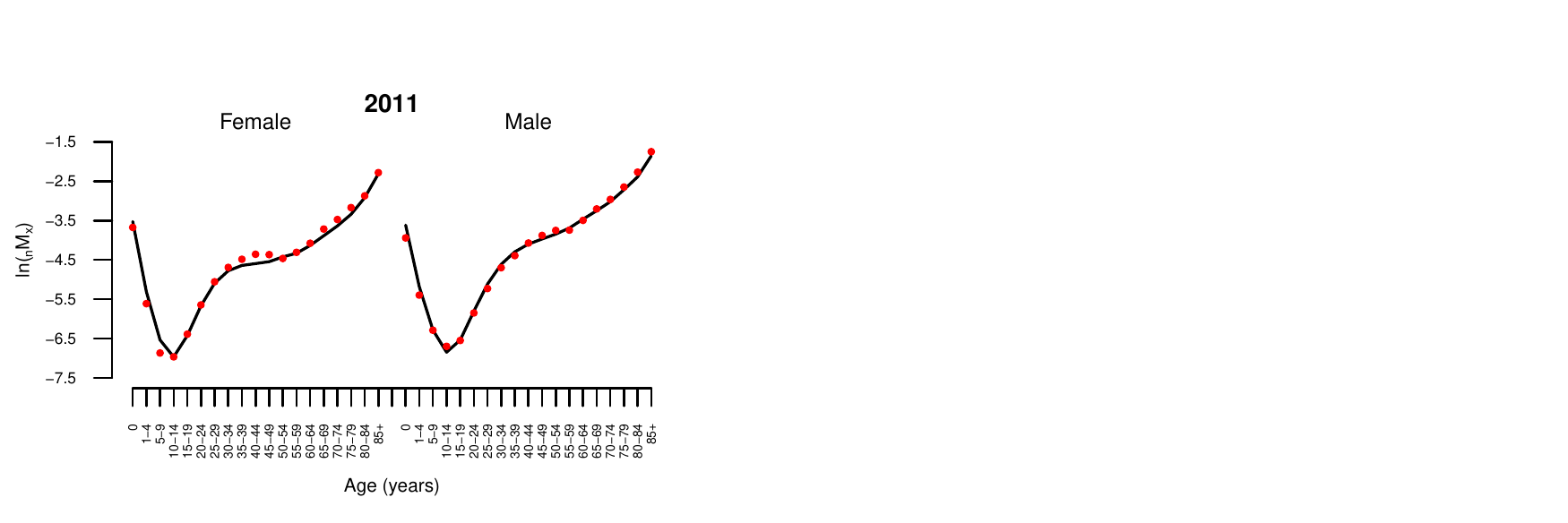}
}
\captionof{figure}{{\bf Age-specific Log Mortality Rates Predicted using Right Singular Vector Weights, 2011.} The red dots are the data, and the solid black line indicates predicted values.}
\label{fig:aginPredsNoCovars4}
\end{minipage}

\noindent
\begin{minipage}{\linewidth}
\captionsetup{format=plain,font=small,margin=1cm,justification=justified}
\makebox[\linewidth]{
\includegraphics[width=1.0\textwidth]{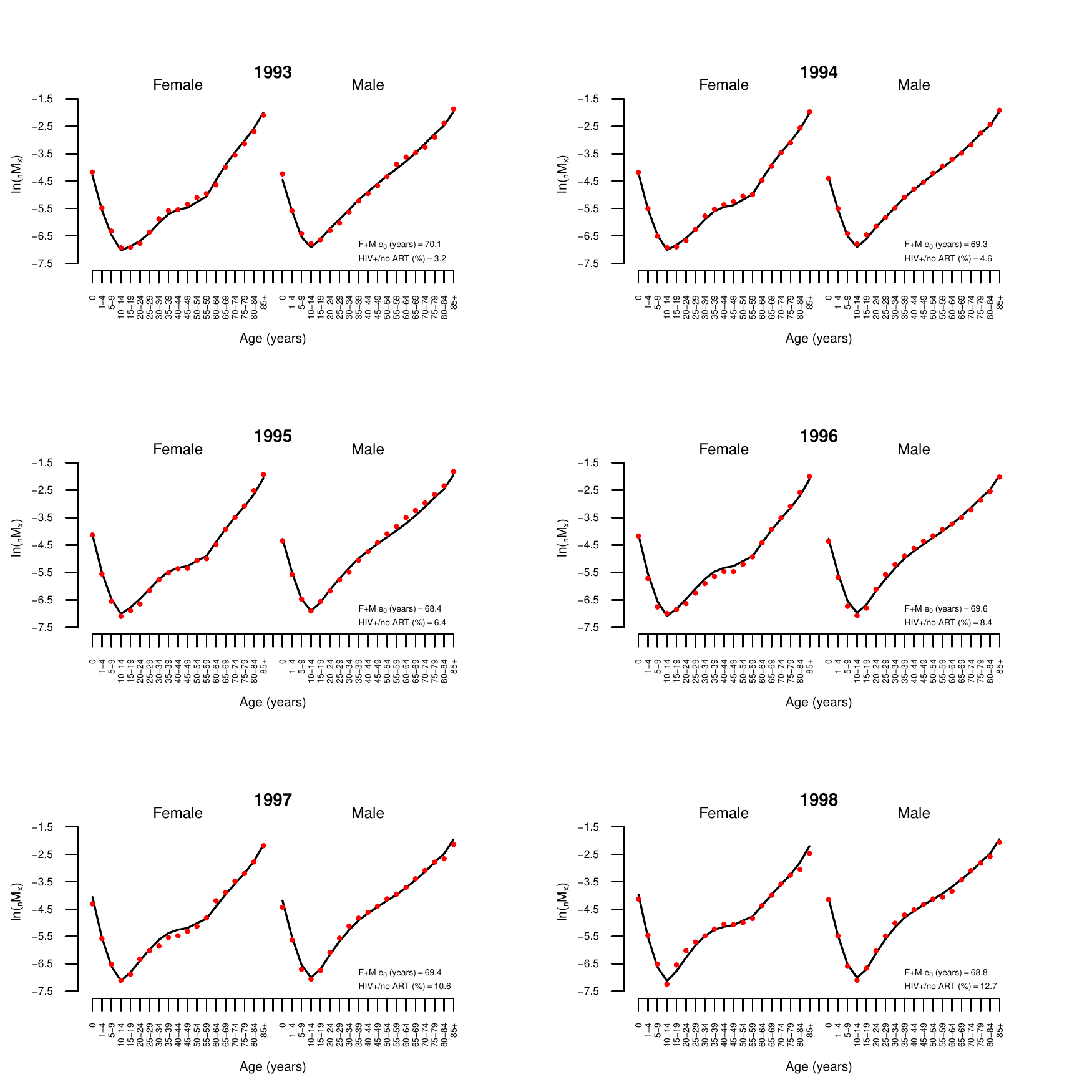}
}
\captionof{figure}{{\bf Age-specific Log Mortality Rates Predicted as a Function of \textit{Life Expectancy at Birth and Prevalence of Persons Who are HIV$^+$ and Not on ART}, 1993--1998.} The red dots are the data, and the solid black line indicates predicted values.}
\label{fig:aginPredsE01}
\end{minipage}

\noindent
\begin{minipage}{\linewidth}
\captionsetup{format=plain,font=small,margin=1cm,justification=justified}
\makebox[\linewidth]{
\includegraphics[width=1.0\textwidth]{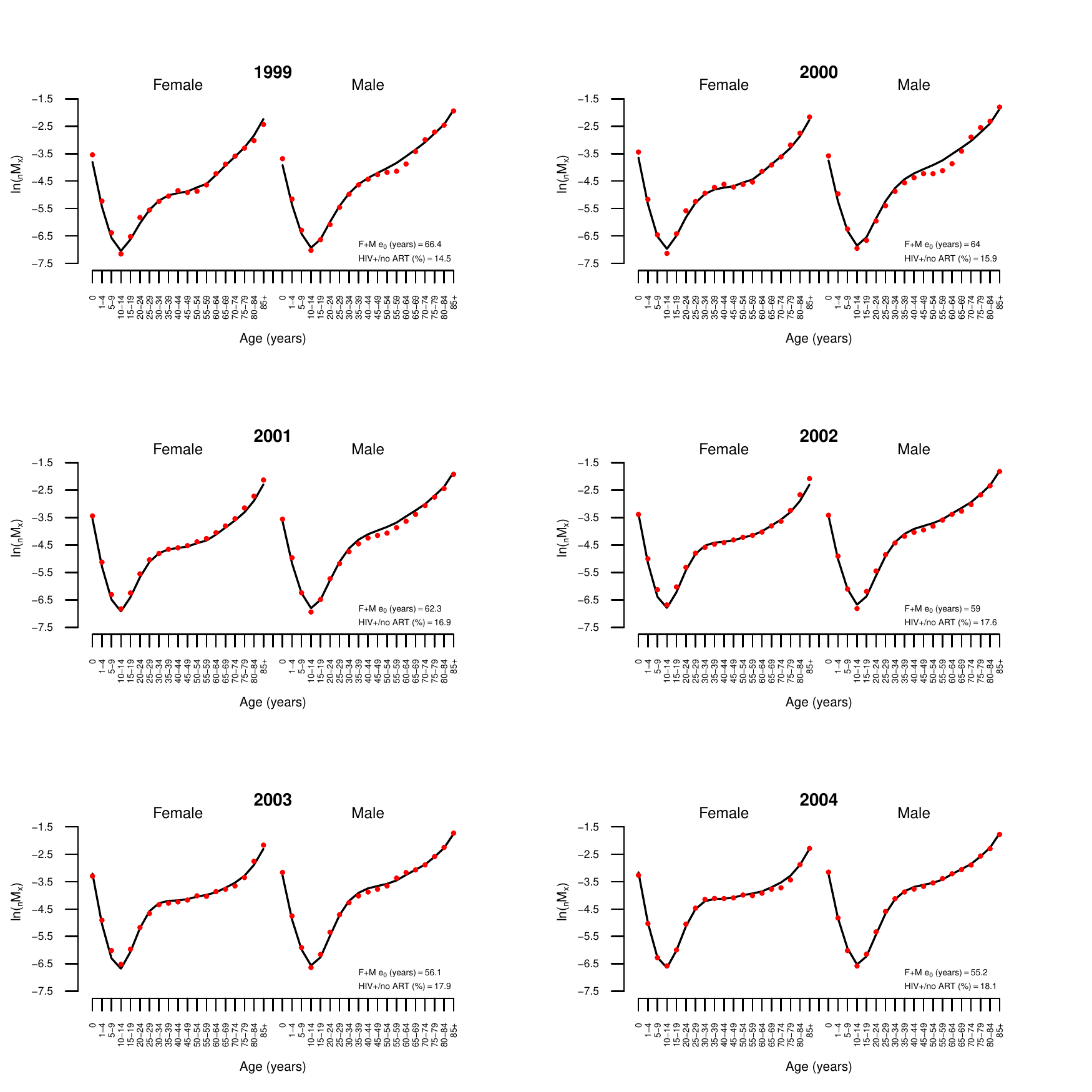}
}
\captionof{figure}{{\bf Age-specific Log Mortality Rates Predicted as a Function of \textit{Life Expectancy at Birth and Prevalence of Persons Who are HIV$^+$ and Not on ART}, 1999--2004.} The red dots are the data, and the solid black line indicates predicted values.}
\label{fig:aginPredsE02}
\end{minipage}

\noindent
\begin{minipage}{\linewidth}
\captionsetup{format=plain,font=small,margin=1cm,justification=justified}
\makebox[\linewidth]{
\includegraphics[width=1.0\textwidth]{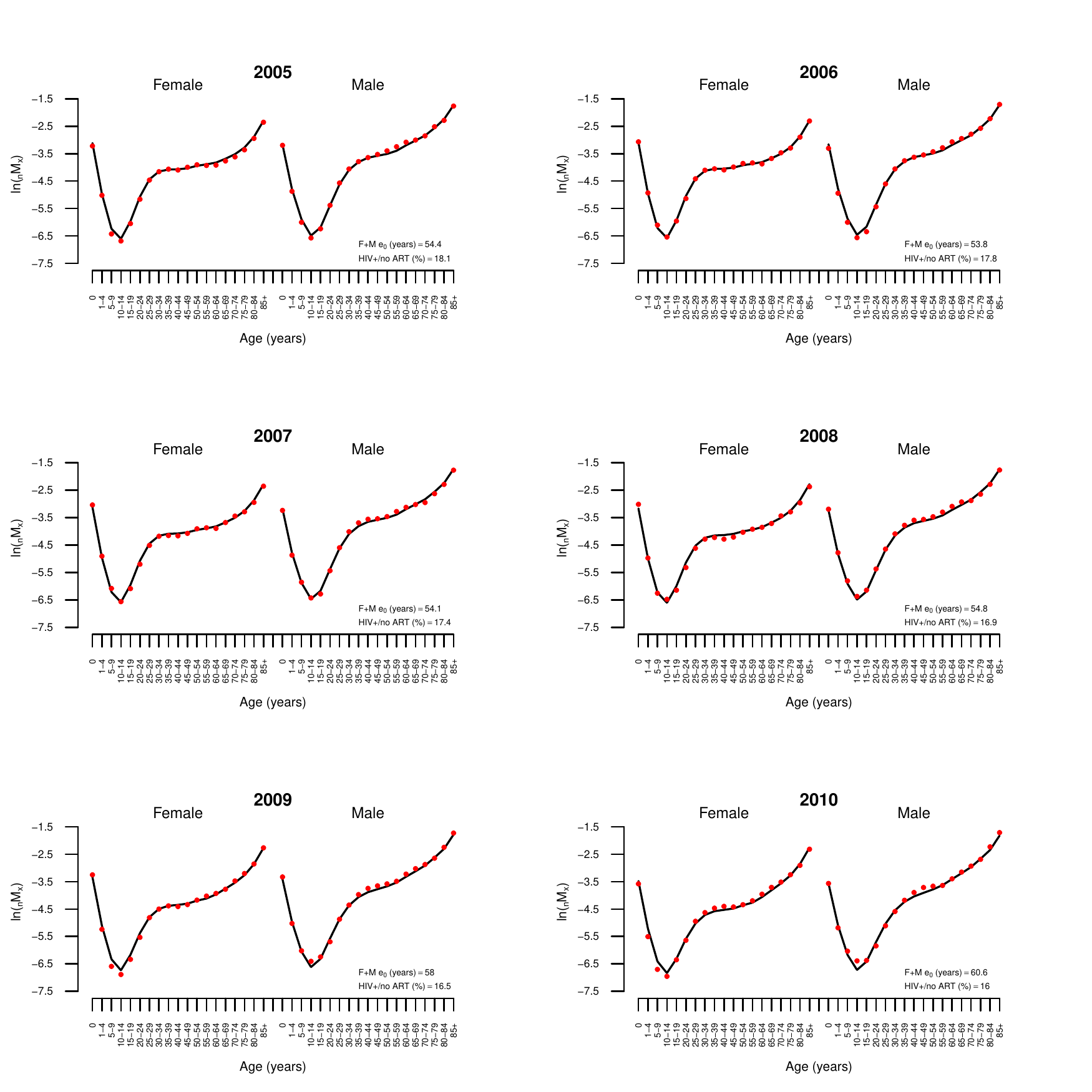}
}
\captionof{figure}{{\bf Age-specific Log Mortality Rates Predicted as a Function of\textit{Life Expectancy at Birth and Prevalence of Persons Who are HIV$^+$ and Not on ART}, 2005--2010.} The red dots are the data, and the solid black line indicates predicted values.}
\label{fig:aginPredsE03}
\end{minipage}

\noindent
\begin{minipage}{\linewidth}
\captionsetup{format=plain,font=small,margin=1cm,justification=justified}
\makebox[\linewidth]{
\includegraphics[width=1.0\textwidth]{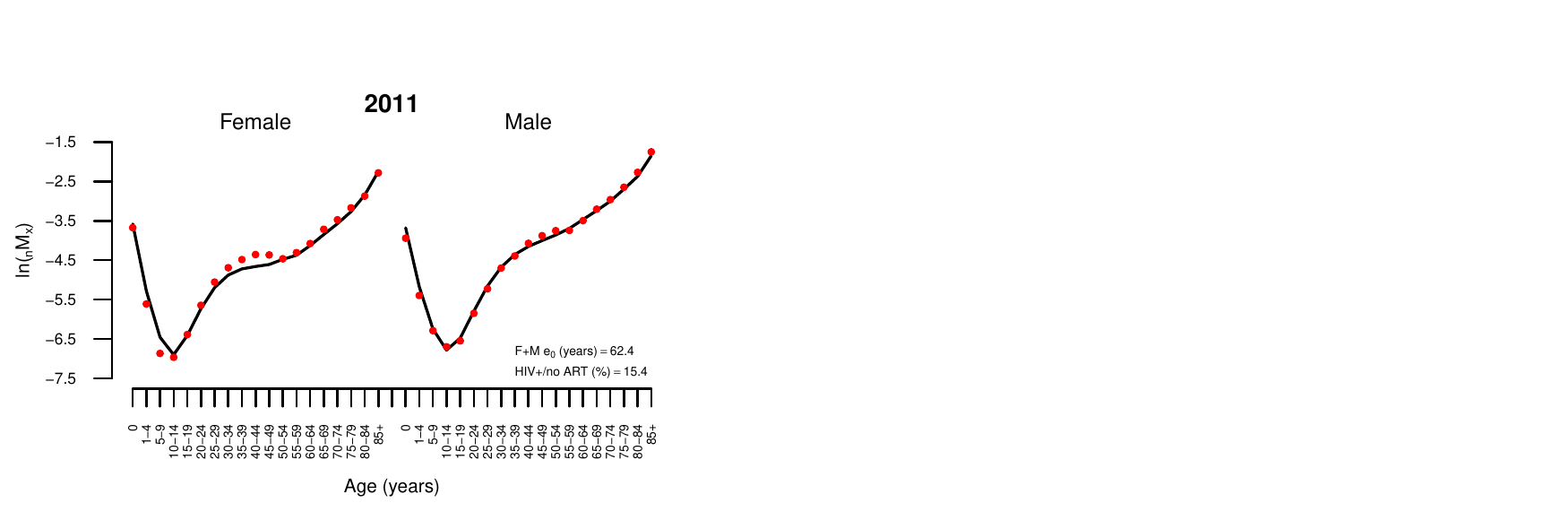}
}
\captionof{figure}{{\bf Age-specific Log Mortality Rates Predicted as a Function of \textit{Life Expectancy at Birth and Prevalence of Persons Who are HIV$^+$ and Not on ART}, 2011.} The red dots are the data, and the solid black line indicates predicted values.}
\label{fig:aginPredsE04}
\end{minipage}

\noindent
\begin{minipage}{\linewidth}
\captionsetup{format=plain,font=small,margin=1cm,justification=justified}
\makebox[\linewidth]{
\includegraphics[width=1.0\textwidth]{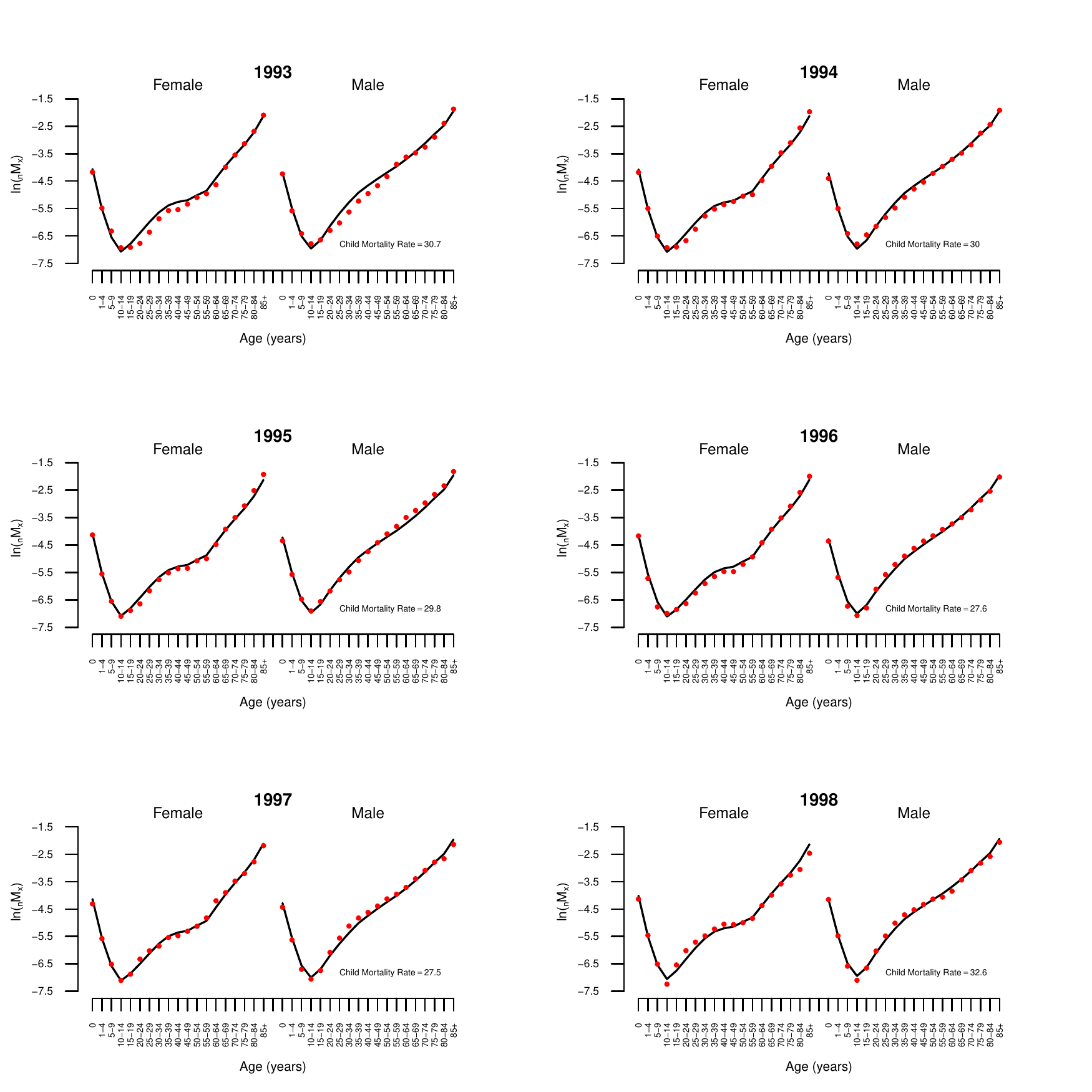}
}
\captionof{figure}{{\bf Age-specific Log Mortality Rates Predicted as a Function of \textit{Child} Mortality, 1993--1998.} The red dots are the data, and the solid black line indicates predicted values.}
\label{fig:aginPredsChild1}
\end{minipage}

\noindent
\begin{minipage}{\linewidth}
\captionsetup{format=plain,font=small,margin=1cm,justification=justified}
\makebox[\linewidth]{
\includegraphics[width=1.0\textwidth]{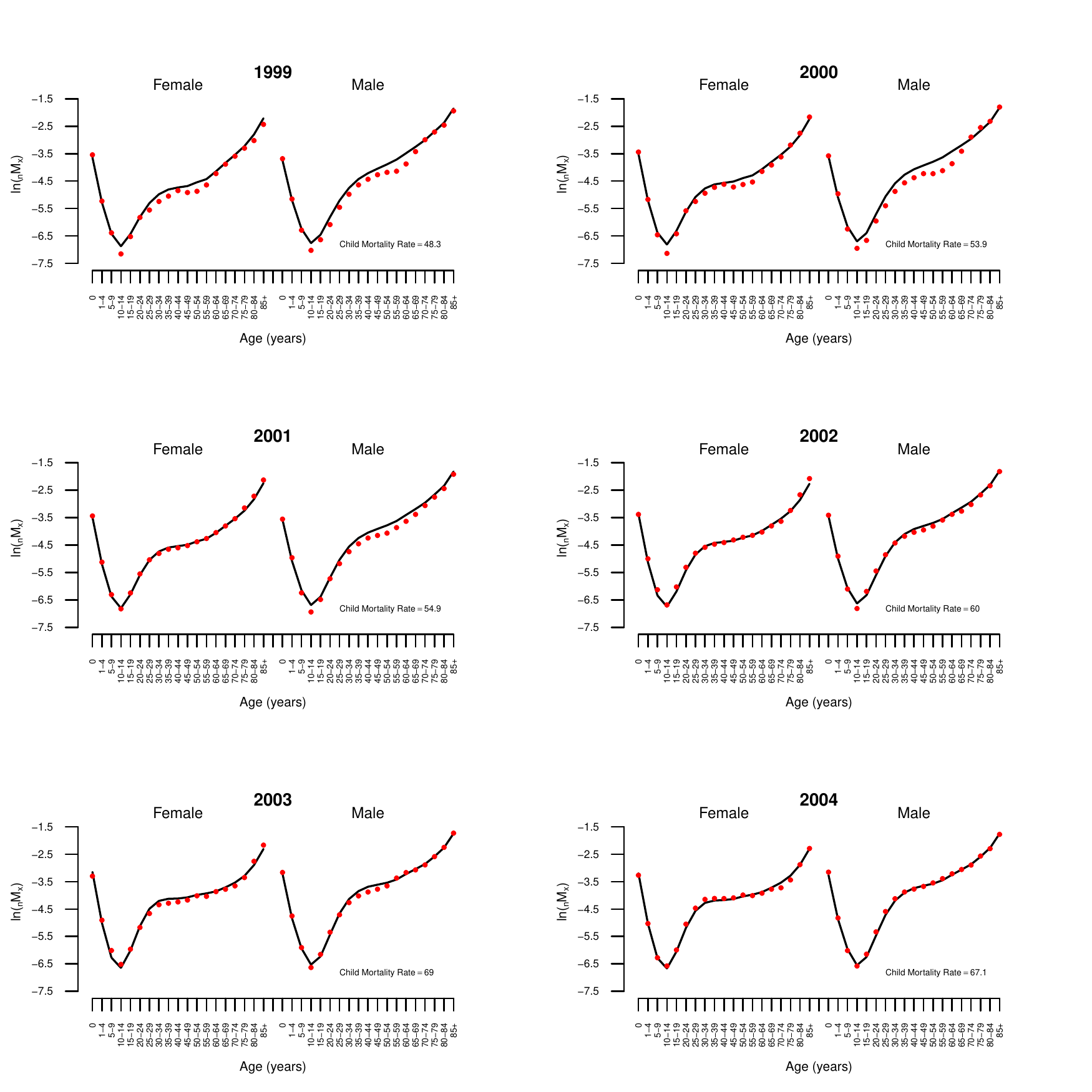}
}
\captionof{figure}{{\bf Age-specific Log Mortality Rates Predicted as a Function of \textit{Child} Mortality, 1999--2004.} The red dots are the data, and the solid black line indicates predicted values.}
\label{fig:aginPredsChild2}
\end{minipage}

\noindent
\begin{minipage}{\linewidth}
\captionsetup{format=plain,font=small,margin=1cm,justification=justified}
\makebox[\linewidth]{
\includegraphics[width=1.0\textwidth]{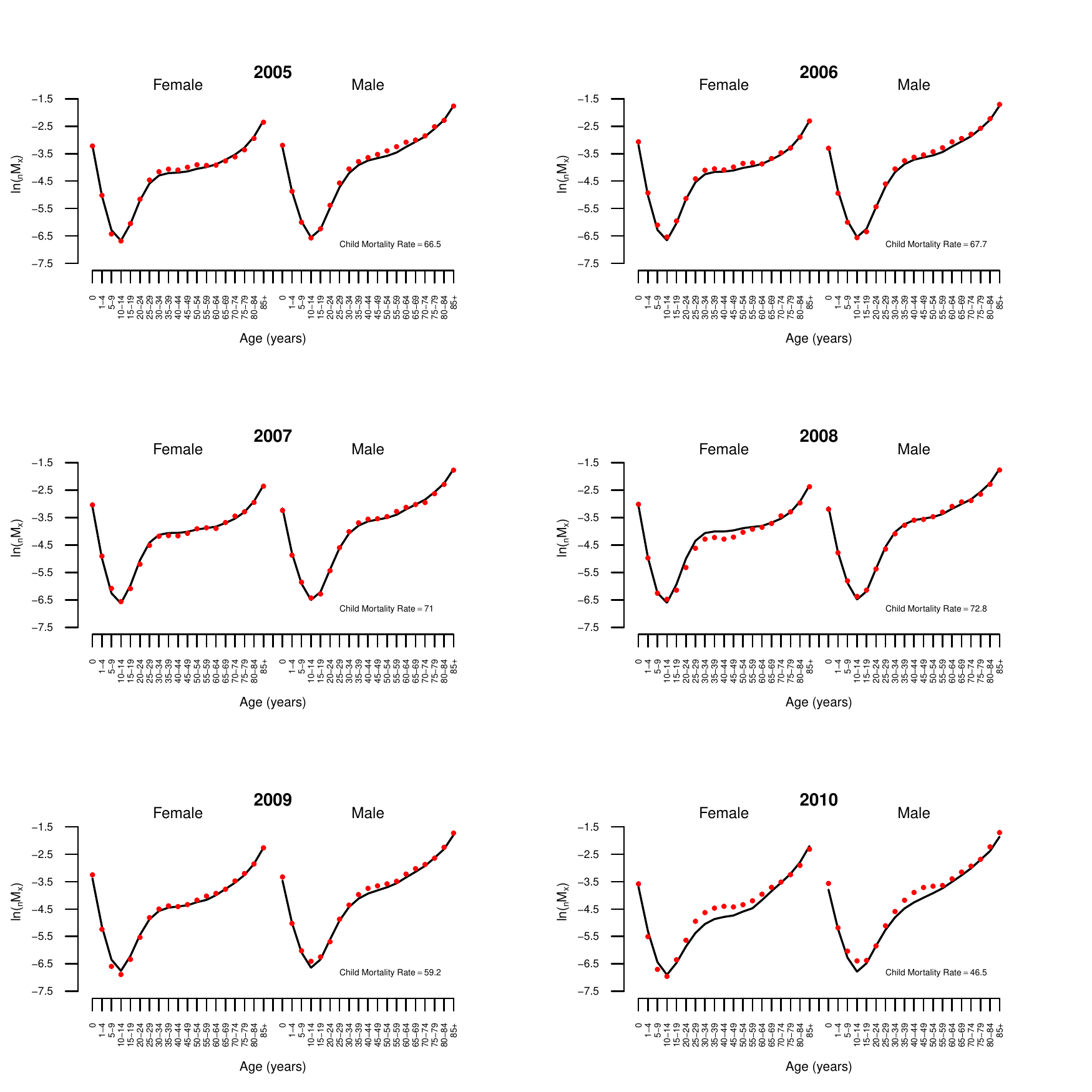}
}
\captionof{figure}{{\bf Age-specific Log Mortality Rates Predicted as a Function of \textit{Child} Mortality, 2005--2010.} The red dots are the data, and the solid black line indicates predicted values.}
\label{fig:aginPredsChild3}
\end{minipage}

\noindent
\begin{minipage}{\linewidth}
\captionsetup{format=plain,font=small,margin=1cm,justification=justified}
\makebox[\linewidth]{
\includegraphics[width=1.0\textwidth]{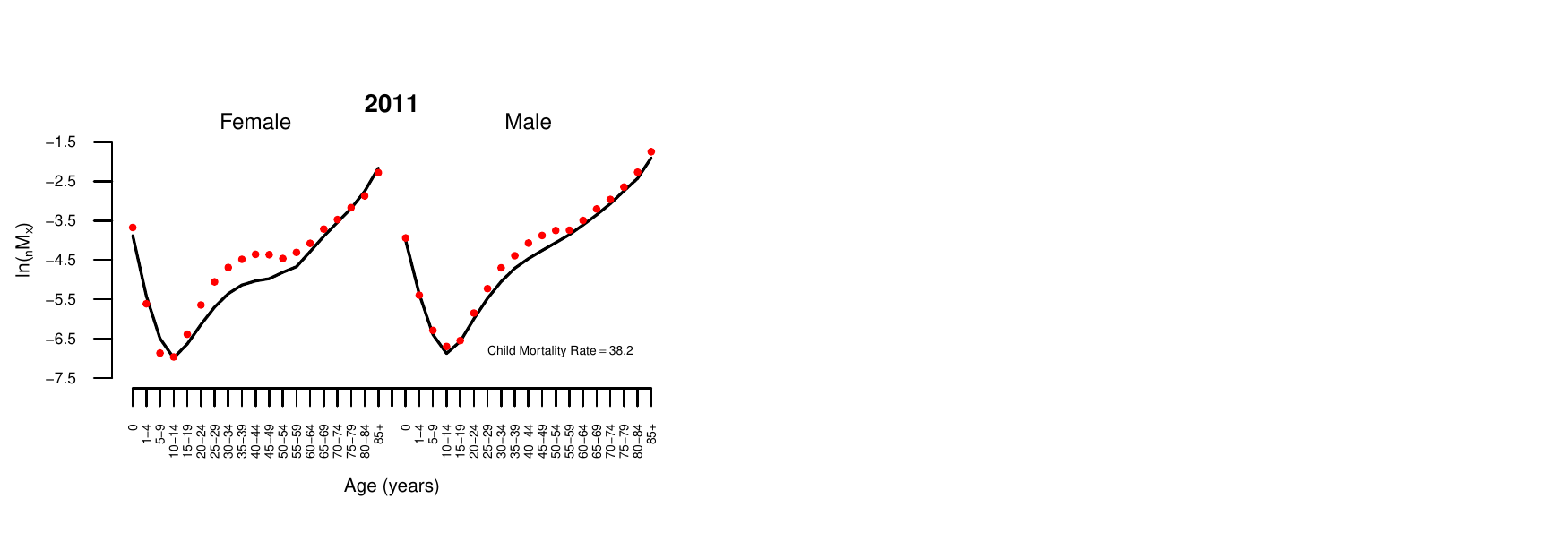}
}
\captionof{figure}{{\bf Age-specific Log Mortality Rates Predicted as a Function of \textit{Child} Mortality, 2011.} The red dots are the data, and the solid black line indicates predicted values.}
\label{fig:aginPredsChild4}
\end{minipage}

\noindent
\begin{minipage}{\linewidth}
\captionsetup{format=plain,font=small,margin=1cm,justification=justified}
\makebox[\linewidth]{
\includegraphics[width=1.0\textwidth]{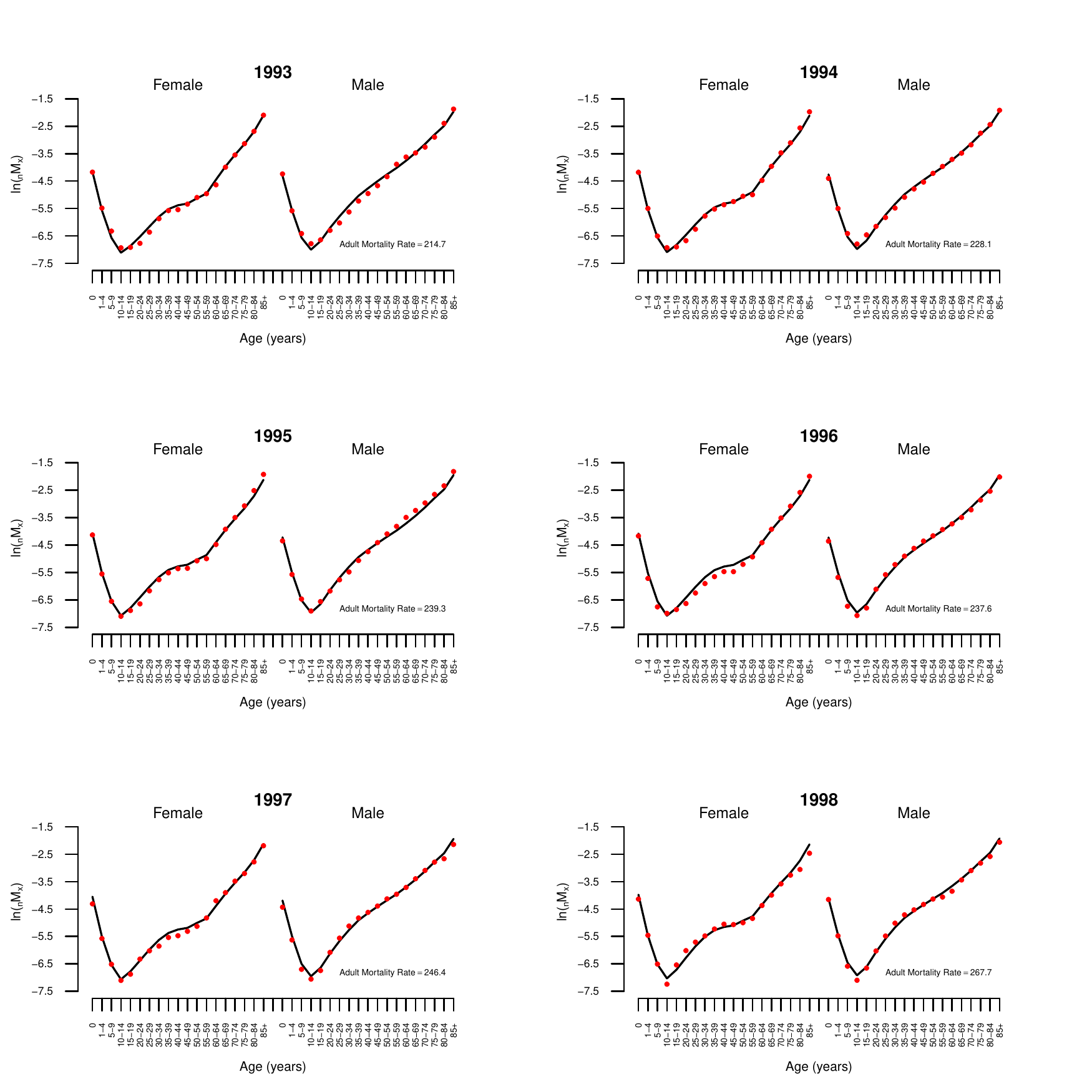}
}
\captionof{figure}{{\bf Age-specific Log Mortality Rates Predicted as a Function of \textit{Adult} Mortality, 1993--1998.} The red dots are the data, and the solid black line indicates predicted values.}
\label{fig:aginPredsAdult1}
\end{minipage}

\noindent
\begin{minipage}{\linewidth}
\captionsetup{format=plain,font=small,margin=1cm,justification=justified}
\makebox[\linewidth]{
\includegraphics[width=1.0\textwidth]{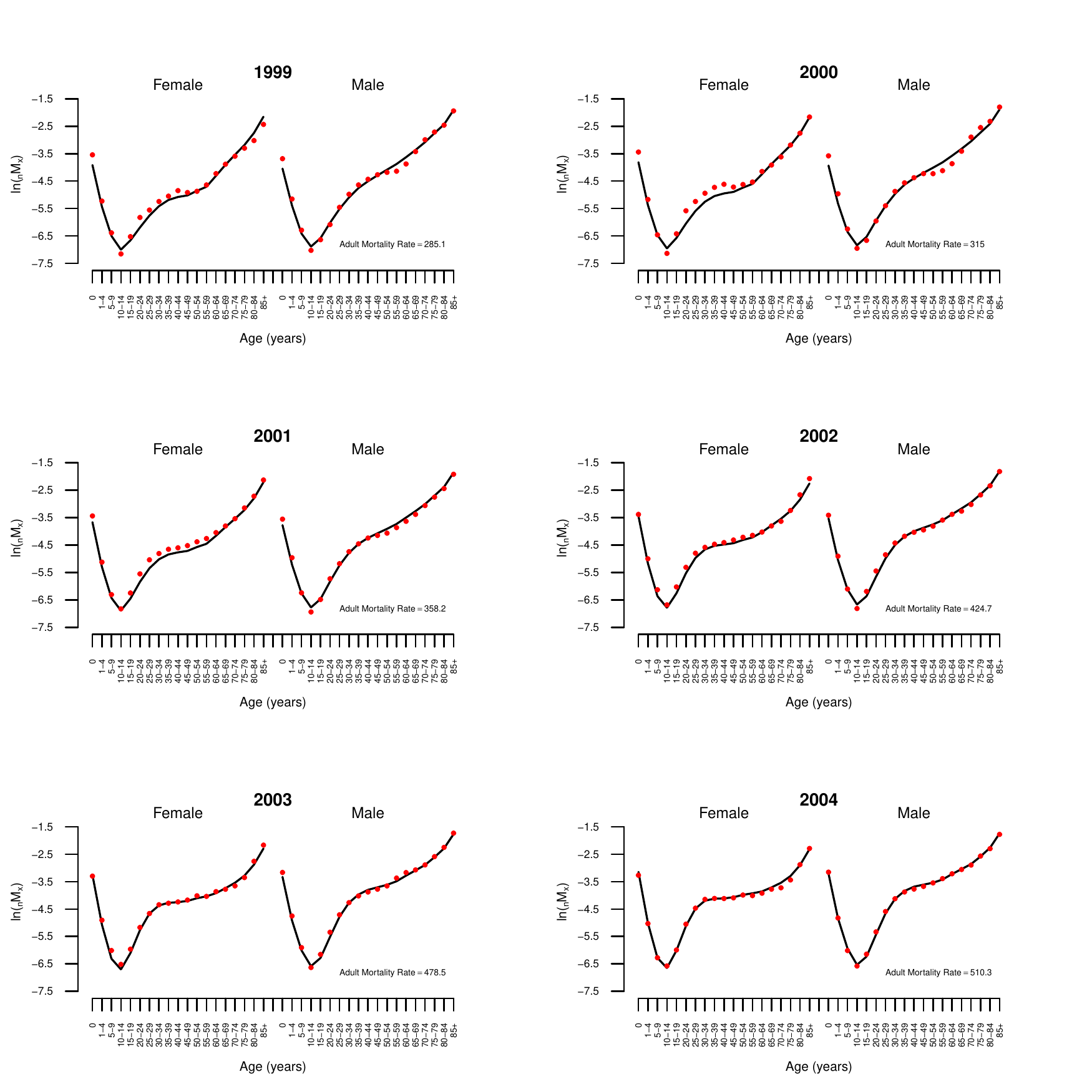}
}
\captionof{figure}{{\bf Age-specific Log Mortality Rates Predicted as a Function of \textit{Adult} Mortality, 1999--2004.} The red dots are the data, and the solid black line indicates predicted values.}
\label{fig:aginPredsAdult2}
\end{minipage}

\noindent
\begin{minipage}{\linewidth}
\captionsetup{format=plain,font=small,margin=1cm,justification=justified}
\makebox[\linewidth]{
\includegraphics[width=1.0\textwidth]{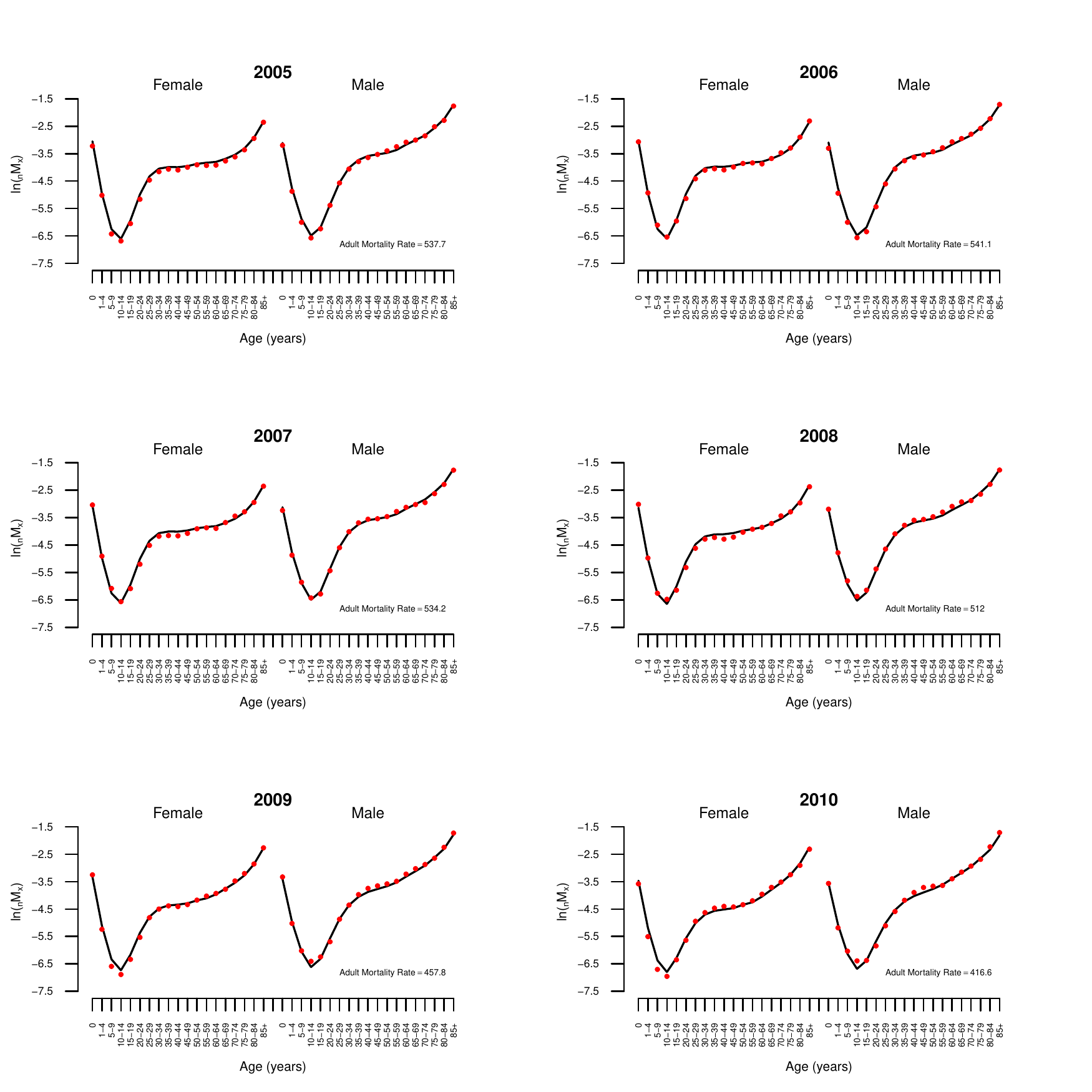}
}
\captionof{figure}{{\bf Age-specific Log Mortality Rates Predicted as a Function of \textit{Adult} Mortality, 2005--2010.} The red dots are the data, and the solid black line indicates predicted values.}
\label{fig:aginPredsAdult3}
\end{minipage}

\noindent
\begin{minipage}{\linewidth}
\captionsetup{format=plain,font=small,margin=1cm,justification=justified}
\makebox[\linewidth]{
\includegraphics[width=1.0\textwidth]{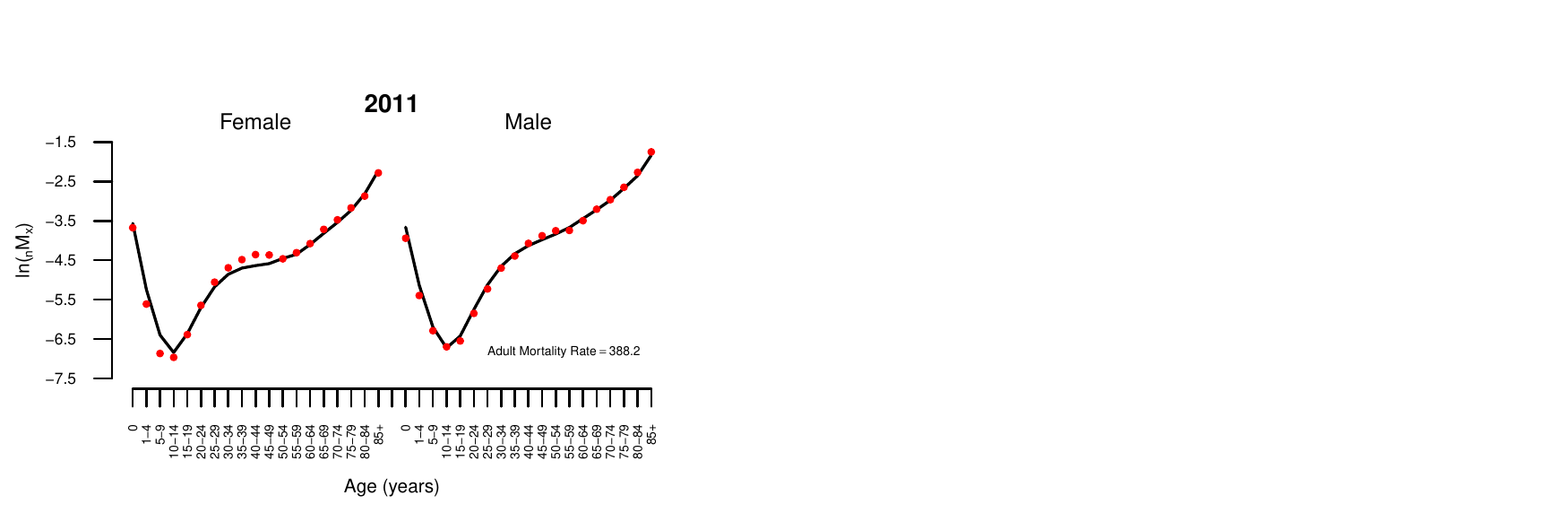}
}
\captionof{figure}{{\bf Age-specific Log Mortality Rates Predicted as a Function of \textit{Adult} Mortality, 2011.} The red dots are the data, and the solid black line indicates predicted values.}
\label{fig:aginPredsAdult4}
\end{minipage}

\noindent
\begin{minipage}{\linewidth}
\captionsetup{format=plain,font=small,margin=1cm,justification=justified}
\makebox[\linewidth]{
\includegraphics[width=1.0\textwidth]{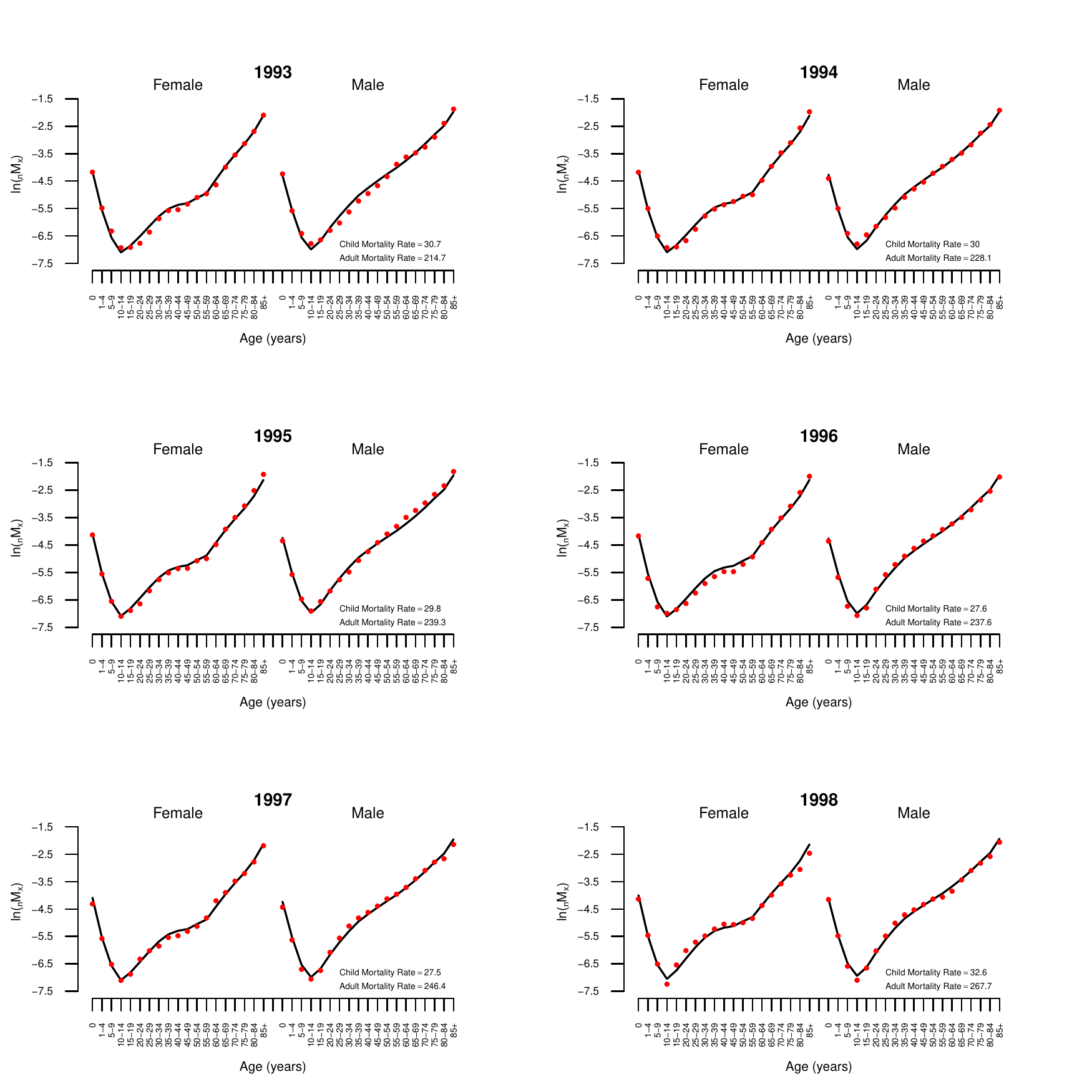}
}
\captionof{figure}{{\bf Age-specific Log Mortality Rates Predicted as a Function of \textit{Child and Adult} Mortality, 1993--1998.} The red dots are the data, and the solid black line indicates predicted values.}
\label{fig:aginPredsChildAdult1}
\end{minipage}

\noindent
\begin{minipage}{\linewidth}
\captionsetup{format=plain,font=small,margin=1cm,justification=justified}
\makebox[\linewidth]{
\includegraphics[width=1.0\textwidth]{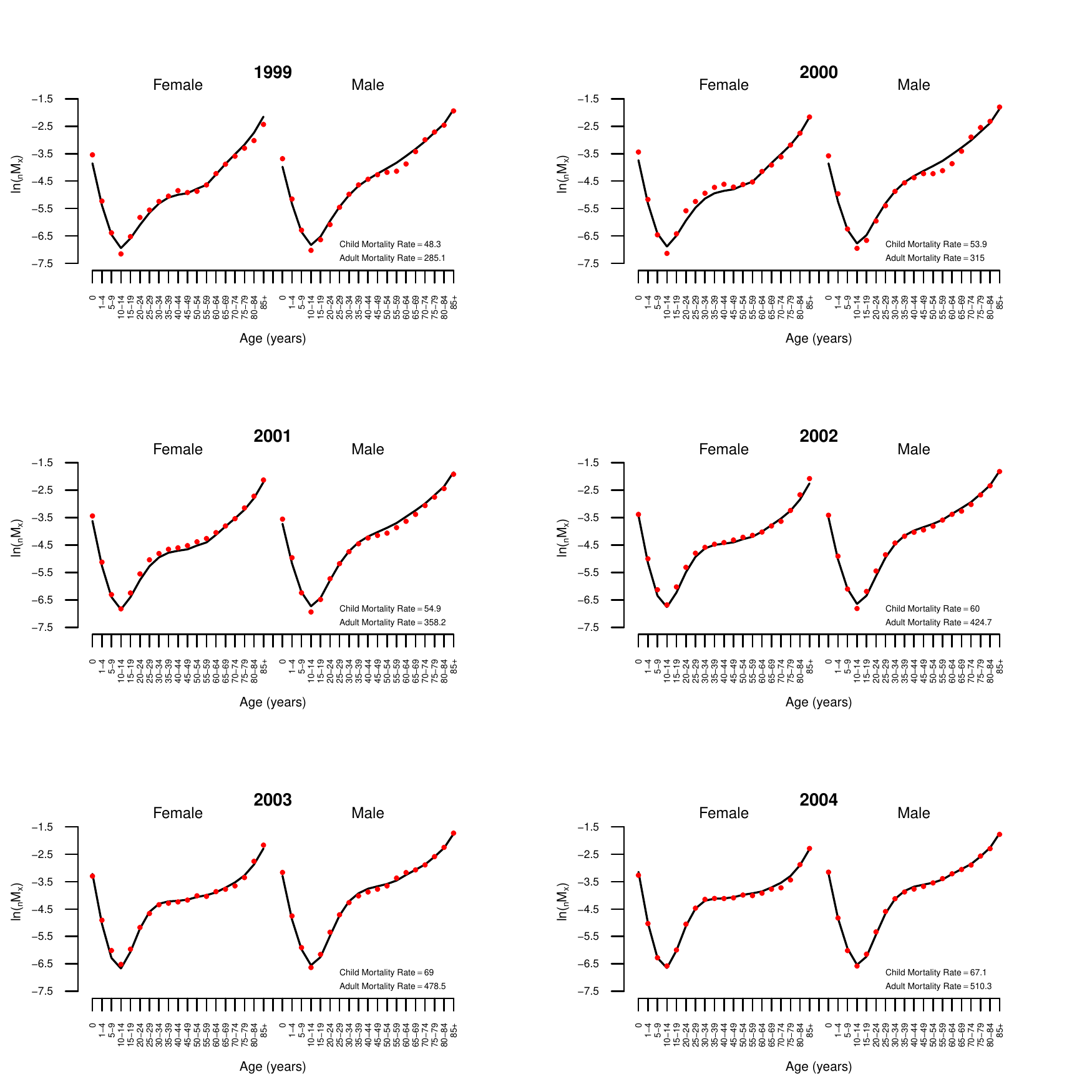}
}
\captionof{figure}{{\bf Age-specific Log Mortality Rates Predicted as a Function of \textit{Child and Adult} Mortality, 1999--2004.} The red dots are the data, and the solid black line indicates predicted values.}
\label{fig:aginPredsChildAdult2}
\end{minipage}

\noindent
\begin{minipage}{\linewidth}
\captionsetup{format=plain,font=small,margin=1cm,justification=justified}
\makebox[\linewidth]{
\includegraphics[width=1.0\textwidth]{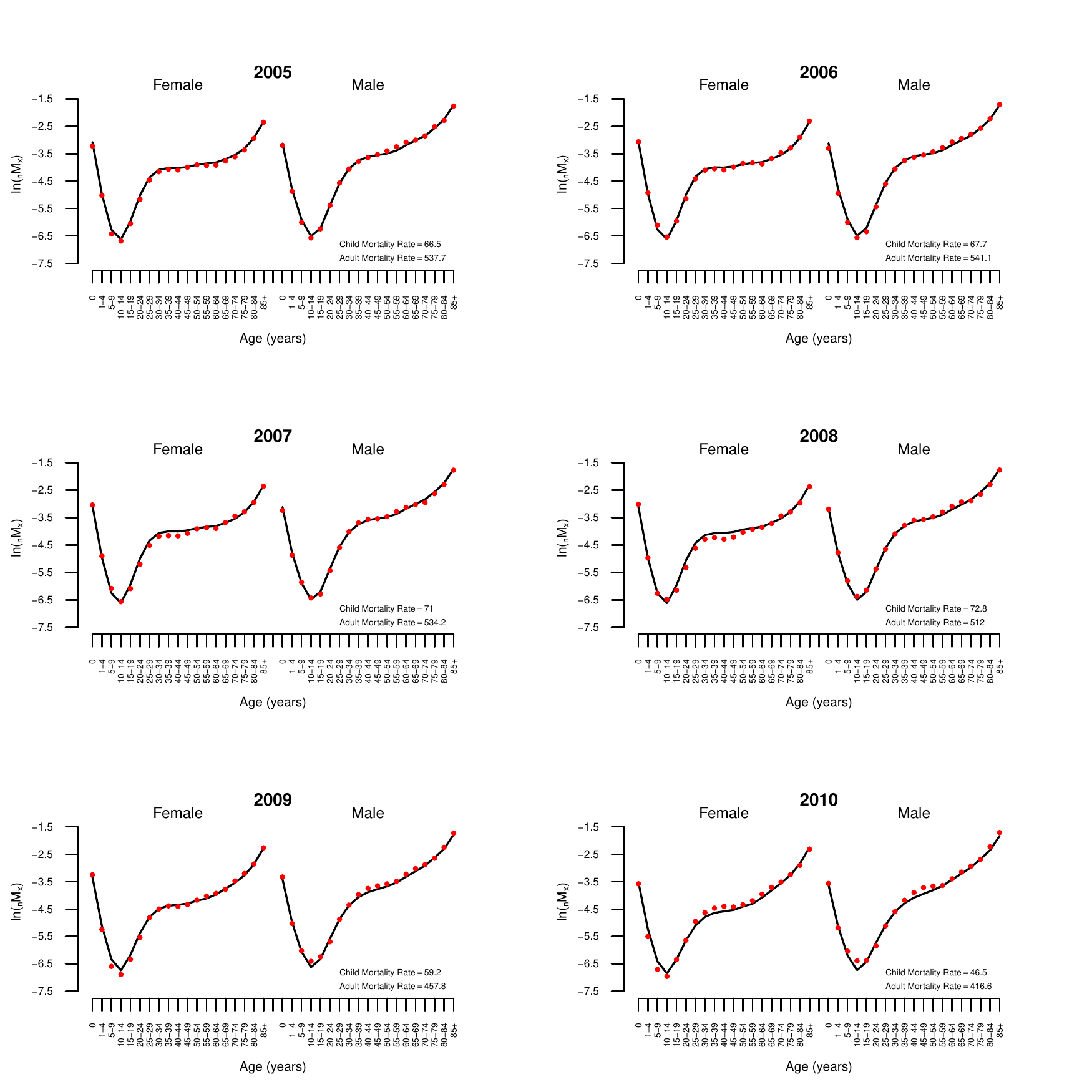}
}
\captionof{figure}{{\bf Age-specific Log Mortality Rates Predicted as a Function of \textit{Child and Adult} Mortality, 2005--2010.} The red dots are the data, and the solid black line indicates predicted values.}
\label{fig:aginPredsChildAdult3}
\end{minipage}

\noindent
\begin{minipage}{\linewidth}
\captionsetup{format=plain,font=small,margin=1cm,justification=justified}
\makebox[\linewidth]{
\includegraphics[width=1.0\textwidth]{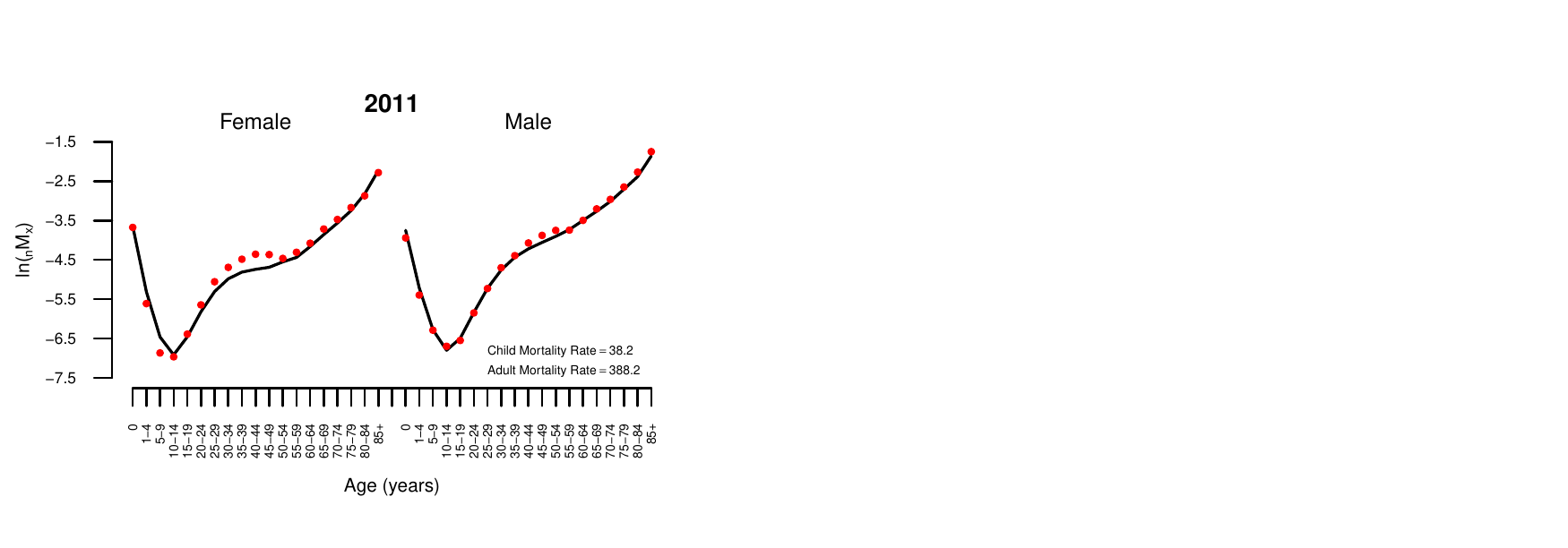}
}
\captionof{figure}{{\bf Age-specific Log Mortality Rates Predicted as a Function of \textit{Child and Adult} Mortality, 2011.} The red dots are the data, and the solid black line indicates predicted values.}
\label{fig:aginPredsChildAdult4}
\end{minipage}

\clearpage
\pagebreak
\section{Predicted Age-Specific Fertility Plots}\label{app:predFx}

\noindent
\begin{minipage}{\linewidth}
\captionsetup{format=plain,font=small,margin=1cm,justification=justified}
\makebox[\linewidth]{
\includegraphics[width=1.0\textwidth]{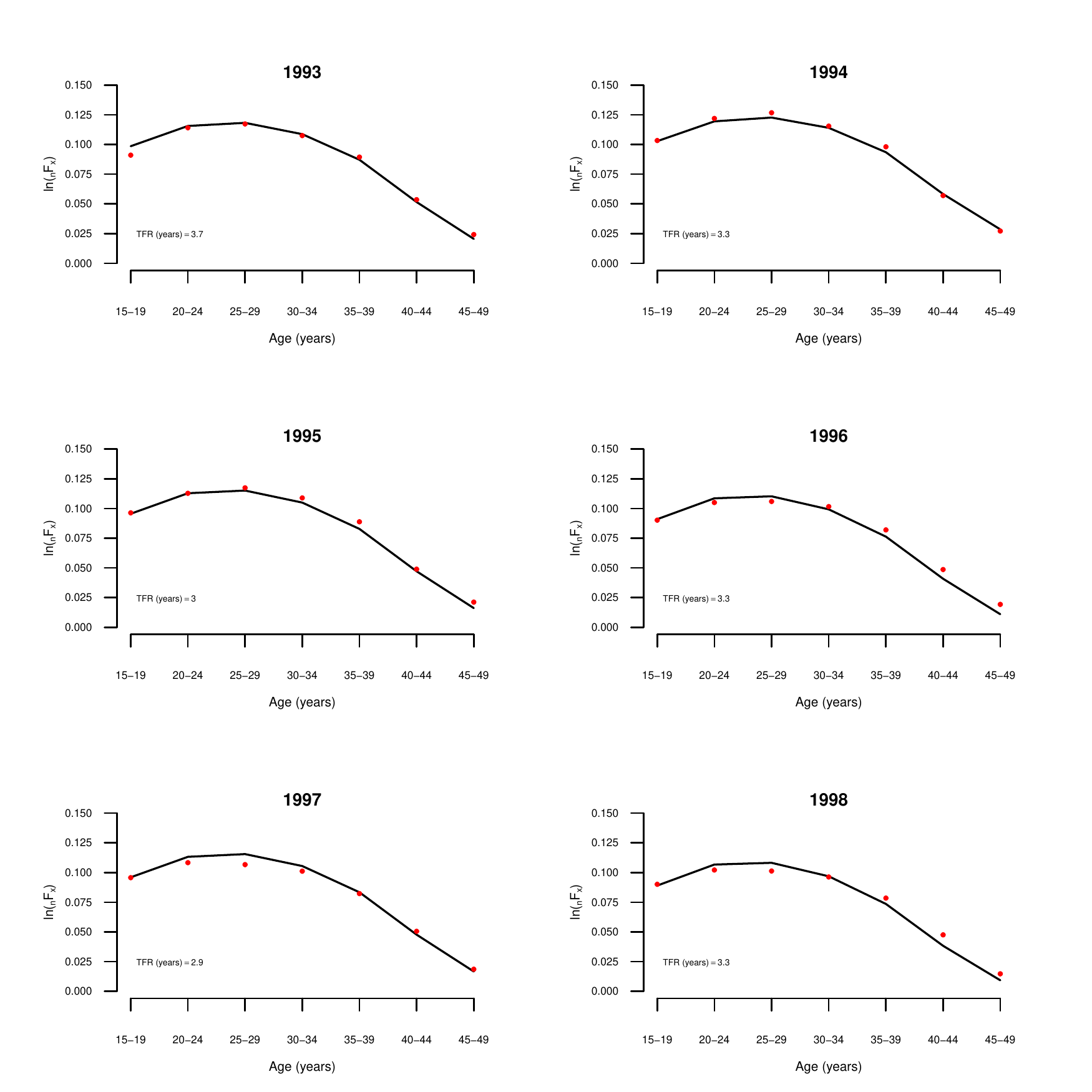}
}
\captionof{figure}{{\bf Age-specific Fertility Rates Predicted as a Function of the Total Fertility Rate, 1993--1998.} The red dots are the data, and the solid black line indicates predicted values.}
\label{fig:aginFertPredsNatScale1}
\end{minipage}

\noindent
\begin{minipage}{\linewidth}
\captionsetup{format=plain,font=small,margin=1cm,justification=justified}
\makebox[\linewidth]{
\includegraphics[width=1.0\textwidth]{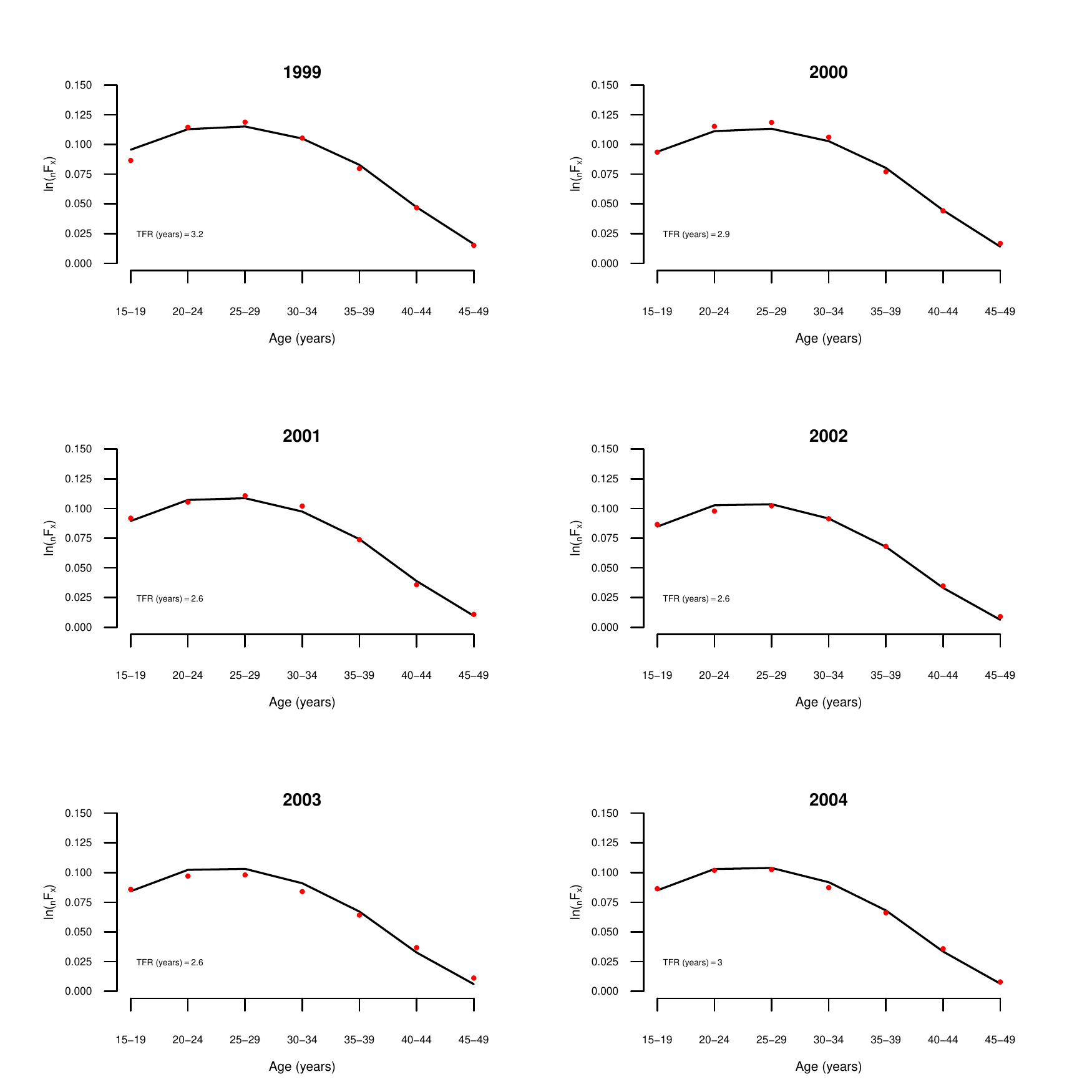}
}
\captionof{figure}{{\bf Age-specific Fertility Rates Predicted as a Function of the Total Fertility Rate, 1999--2004.} The red dots are the data, and the solid black line indicates predicted values.}
\label{fig:aginFertPredsNatScale2}
\end{minipage}

\noindent
\begin{minipage}{\linewidth}
\captionsetup{format=plain,font=small,margin=1cm,justification=justified}
\makebox[\linewidth]{
\includegraphics[width=1.0\textwidth]{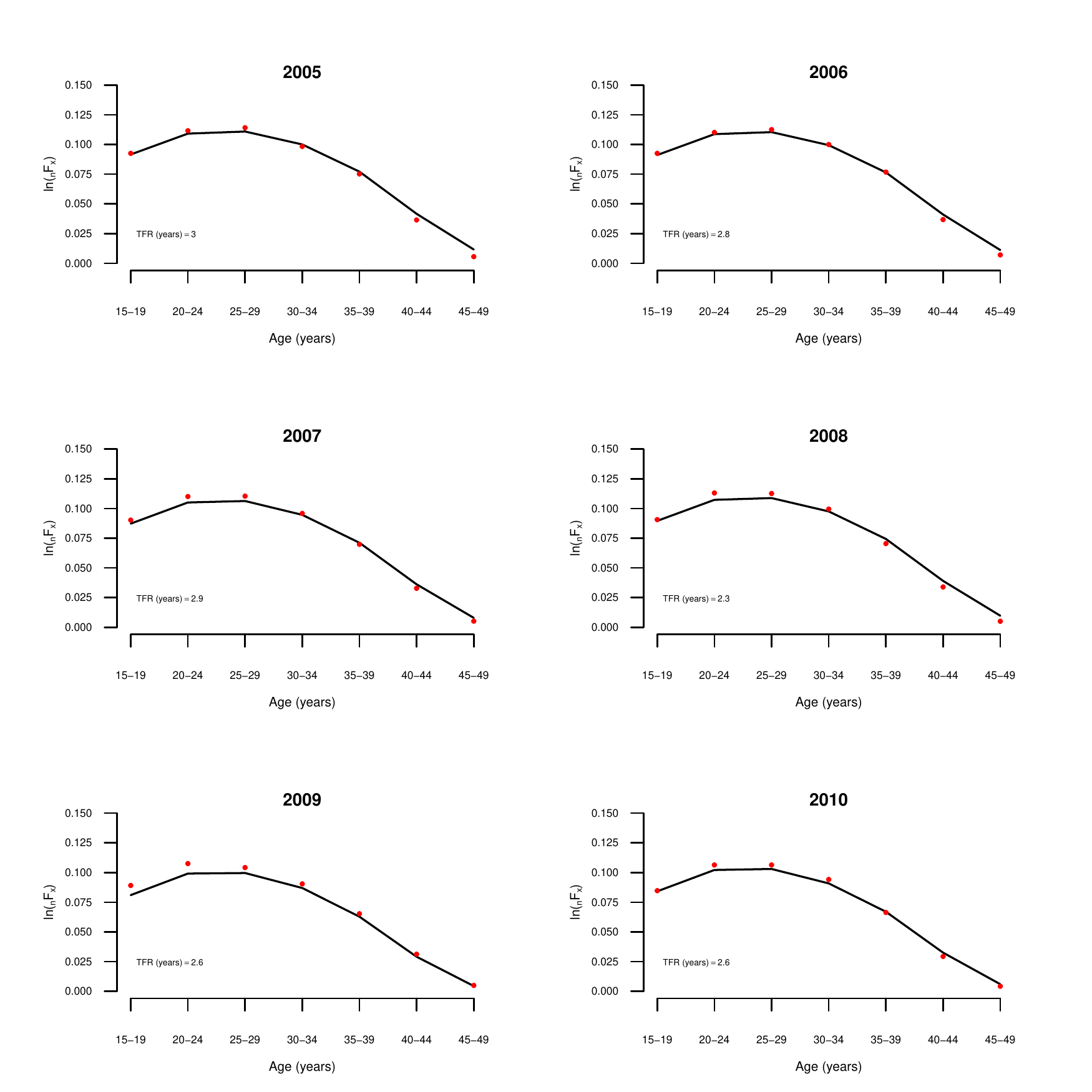}
}
\captionof{figure}{{\bf Age-specific Fertility Rates Predicted as a Function of the Total Fertility Rate, 2005--2010.} The red dots are the data, and the solid black line indicates predicted values.}
\label{fig:aginFertPredsNatScale3}
\end{minipage}

\noindent
\begin{minipage}{\linewidth}
\captionsetup{format=plain,font=small,margin=1cm,justification=justified}
\makebox[\linewidth]{
\includegraphics[width=1.0\textwidth]{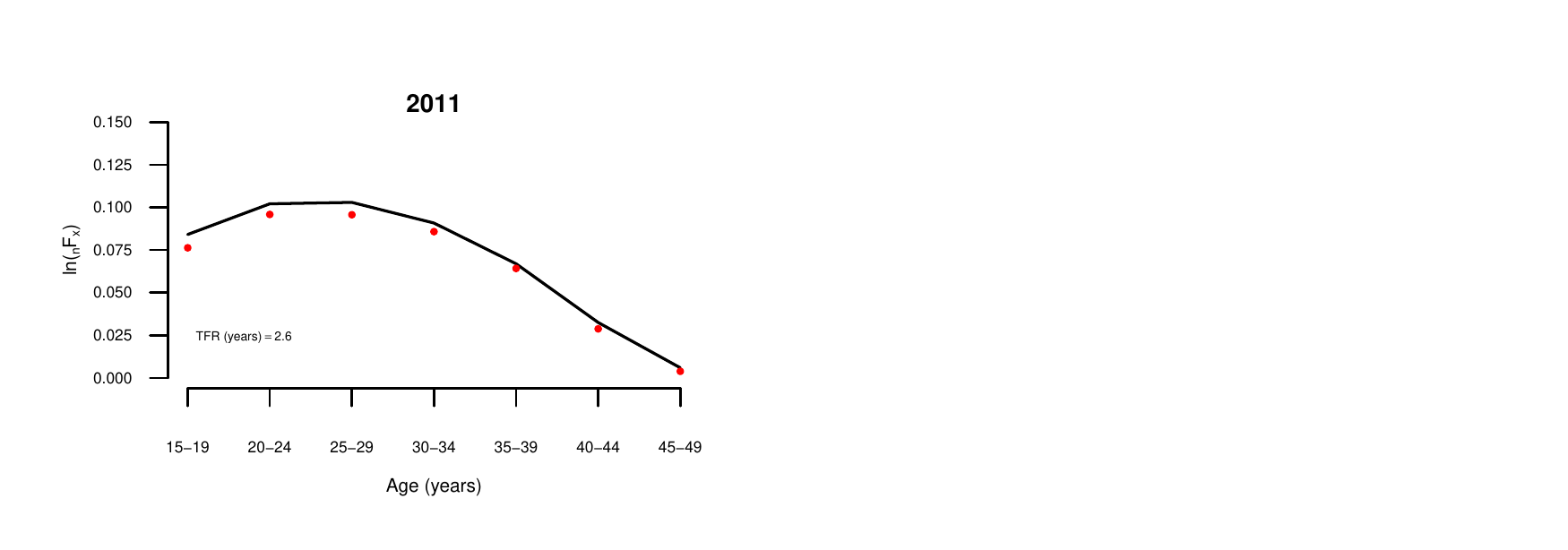}
}
\captionof{figure}{{\bf Agincourt Age-specific Fertility Rates Predicted as a Function of the Total Fertility Rate, 2011.} The red dots are the data, and the solid black line indicates predicted values.}
\label{fig:aginFertPredsNatScale4}
\end{minipage}

\end{document}